\documentclass[aps,preprint,floats,nofootinbib,12pt]{revtex4}
\usepackage{mathrsfs}
\usepackage{amsfonts}
\usepackage{graphicx}
\usepackage{epsfig}
\usepackage{amsmath}
\usepackage{array}
\usepackage{epstopdf}
\usepackage{amstext}
\usepackage{amssymb}
\usepackage{cancel}
\usepackage{multirow}
\usepackage{float}
\usepackage{amsmath}
\allowdisplaybreaks[4]

\def\beq{\begin{eqnarray}}
\def\eeq{\end{eqnarray}}
\def\non{\nonumber}
\def\lqcd{\Lambda_{\rm QCD}}

\makeatletter

\newcommand{\Rmnum}[1]{\expandafter\@slowromancap\romannumeral #1@}
\makeatother

\newcommand{\tabincell}[2]{\begin{tabular}{@{}#1@{}}#2\end{tabular}}

\begin{document}

\title{ The semi-leptonic and non-leptonic weak decays of $\Lambda_b^0$ }

\vspace{1cm}

\author{Jie Zhu$^1$\footnote{zhujllwl@mail.nankai.edu.cn},
 Zheng-Tao Wei$^1$\footnote{weizt@nankai.edu.cn}, and
 Hong-Wei Ke$^{2}$\footnote{khw020056@hotmail.com} }

\affiliation{  $^{1}$ School of Physics, Nankai University, Tianjin 300071, China \\
   $^{2}$ School of Science, Tianjin University, Tianjin 300072, China }

\vspace{12cm}

\begin{abstract}

The recent experimental developments require a more precise theoretical study of weak decays of heavy baryon $\Lambda_b^0$. In this work, we provide an updated and systematic analysis of both the semi-leptonic and nonleptonic decays of $\Lambda^0_b$ into baryons $\Lambda^+_c$, $\Lambda$, $p$, and $n$. The diquark approximation is adopted so that the methods developed in the $B$ meson system can be extended into the baryon system. The baryon-to-baryon transition form factors are calculated in the framework of a covariant light-front quark model. The form factors $f_3, ~g_3$ can be extracted and are found to be non-negligible. The semi-leptonic processes of $\Lambda^0_b\to \Lambda^+_c(p)l^-\bar\nu_l$ are calculated and the results are consistent with the experiment. We study the non-leptonic processes within the QCD factorization approach. The decay amplitudes are calculated at the next-to-leading order in strong coupling constant $\alpha_s$. We calculate the non-leptonic decays of $\Lambda^0_b$ into a baryon and a s-wave meson (pseudoscalar or vector) including 44 processes in total. The branching ratios and direct CP asymmetries are predicted. The numerical results are compared to the experimental data and those in the other theoretical approaches. Our results show validity of the diquark approximation and application of QCD factorization approach into the heavy baryon system.

\end{abstract}

\maketitle

\section{Introduction}

The weak decays of heavy baryon $\Lambda_b^0$ provide an important place to extract the Cabibbo-Kobayashi-Maskawa (CKM) matrix elements, explore CP violation and study different theoretical models of hard interaction. Recently, a lot of experimental developments were made, and many processes were observed or seen \cite{Patrignani:2016xqp}. For the exclusive semi-leptonic processes, the branching fraction of $\Lambda_c^+ l^-\bar\nu_l$ mode is the biggest, at the order of 10\%. The decay rate of $p\mu^-\bar\nu_\mu$ is about $10^{-4}$. For the nonleptonic two-body processes, the charmful decays of $\Lambda_c^+\pi^-(K^-, D^-, D_s^-)$ are observed and their branching ratios are at the order of $10^{-3}$ or $10^{-4}$. The charmonium mode $\Lambda J/\psi$ has fraction of order of $10^{-4}$. The charmless processes with final states $p\pi^-(K^-)$ are observed to be of order of $10^{-6}$. The pentaquark is observed in $\Lambda_b^0\to J/\psi p K^-$ process. The $\Lambda_b^0\to \Lambda \phi$ is observed with a final vector meson $\phi$ and the fraction is of $ 10^{-6}$ \cite{Aaij:2016zhm}. The mode $\Lambda\mu^+\mu^-$ is observed at the order of $10^{-6}$.  The LHC run II \cite{Piucci:2017kih} and the possible future upgrade of LHC will accumulate more data than ever, we expect that the study of $\Lambda_b^0$ will enter into a precise era.

Theoretical interests on $\Lambda_b^0$ decays were increased recently, such as light-front quark model \cite{Ke:2007tg,Wei:2009np}, QCD factorization (QCDF) approach \cite{Zhu:2016bra}, generalized factorization approach (GFA) \cite{Hsiao:2014mua,Geng:2016gul,Hsiao:2017tif}, light-cone sum rules \cite{Khodjamirian:2011jp}, lattice QCD method \cite{Detmold:2015aaa}, soft-collinear-effective-theory (SCET) approach \cite{Feldmann:2011xf}, perturbative QCD (pQCD) approach \cite{Lu:2009cm}, SU(3) symmetry relations \cite{He:2015fwa}, etc.. In the previous works \cite{Ke:2007tg,Wei:2009np,Zhu:2016bra}, we have calculated the weak decay of $\Lambda_b^0$ with the light-front quark model, diquark approximation and factorization assumption. For the charmful processes, the theory predictions within the heavy quark limit for the four processes of $\Lambda_c^+\pi^-(K^-,D^-,D_s^-)$ are well consistent with the data. The consistency shows effectiveness of the diquark approximation and factorization assumption. For the charmless processes, some inconsistencies are found when the data become precise. The theory predictions of the semi-leptonic decays of $p l^-\bar\nu_l$ modes are smaller than the data. For the charmless non-leptonic processes, it is known from the $B$ meson study that the naive factorization is insufficient to explain the experiment. The strong penguin effects are important and even dominant in many decay modes. In \cite{Wei:2009np}, only the tree operators are considered. Although the penguin effects are included in \cite{Zhu:2016bra}, the discussion is only restricted to one process of $p K^-$. Thus, the experimental improvements require the theory developments to compete.

From the theoretical point of view, one difficult thing is to evaluate the transition form factors between two baryons. The method we will use is a relativistic quark model in the light-front form. The basic ingredient is the hadron light-front wave function which is explicitly Lorentz invariant. The conventional form, in which the constitute quarks are on mass shell, has been applied to obtain many meson decay constants and weak form factors \cite{Jaus:1989au, Jaus:1991cy, Ji:1992yf, Jaus:1996np, Cheng:1996if}. In \cite{Ke:2007tg,Wei:2009np}, the conventional light-front quark model is employed into the $\Lambda_b^0$ decays. The baryon-to-baryon transition form factors are derived from a particular plus component of the corresponding current operator in a specific Lorentz frame, e.g. the transverse frame with $q^{+}=0$. Among the six form factors, only four quantities can be calculated in this way. While the form factors $f_3$ and $g_3$ are not obtained. For the transitions of $\Lambda_b^0$ to light baryons such as $p,\Lambda,n$, there is no reasonable argument to guarantee that they are small. It is necessary to estimate their effects. In \cite{Jaus:1999zv}, a covariant light-front quark model is constructed to render the hadron transition matrix elements covariant. This approach has been applied to many meson processes \cite{Cheng:2003sm}. In this study, we will use the covariant approach to derive all the form factors including $f_3$ and $g_3$.  Then, we give the numerical predictions for the semi-leptonic decays.

For the non-leptonic processes, the QCD dynamics is more complicated than the semi-leptonic one. Theory treatment relies on different factorization approaches which developed for the $B$ meson system. In this study, we will work within a framework of QCD factorization (QCDF) approach  \cite{Beneke:1999br, Beneke:2000ry, Beneke:2001ev, Beneke:2003zv}. In the heavy quark limit, the decay amplitudes are expressed by a factorizable form which separates the perturbative contribution from the non-perturbative part. The naive factorization is its lowest order approximation. The non-factorzaible contributions can be systematically calculated in strong coupling constant $\alpha_s$ order by order in leading power of $1/m_b$. Under the diquark approximation, a baryon is similar to a meson. We might expect that the QCDF approach can be applied into the heavy baryon decays. In this study, we extend the QCDF method to the non-leptonic two-body decays of $\Lambda_b^0$ and give a systematic study for decays of $\Lambda^0_b$ into final states containing a baryon and a s-wave meson (pseudoscalar or vector).

The paper is organized as follows: In Section \Rmnum{2}, we give formulations of the covariant light-front approach, and derive the six transition form factors ($f_i$ and $g_i$ with i=1,2,3) of  $\Lambda^0_b\rightarrow \Lambda^+_c (p,\Lambda, n)$ transitions. In Section \Rmnum{3}, the expressions for the semi-leptonic processes are given. In Section \Rmnum{4}, we discuss the nonleptonic decays in QCD factorization approach. In Section \Rmnum{5}, we discuss the input phenomenological parameters, and then give the numerical results for the weak transition form factors. In Section \Rmnum{6}, the numerical results for the semi-leptonic processes are given. In Section \Rmnum{7}, the numerical results for the non-leptonic are presented. The theory predictions are compared with the experimental data and other theory approaches. In the last section \Rmnum{8}, the discussions and conclusions are given.

\section{$\Lambda^0_b\rightarrow H(\Lambda^+_c, p,\Lambda, n)$ transition form factors in the covariant light-front approach}

At first, we discuss the diquark hypothesis. A diquark is a two-quark correlation  \cite{Anselmino:1992vg}. The interaction of two quark can be attractive if they are antisymmetric in color space. This is a special characteristic of QCD, unlike the QED case where the interaction between two like-charged particle is repulsive. The diquark is not a fundamental particle, because it contains color and can only  exist in a hadron containing more than two quarks. The size of the diquark should should be larger than that of a quark and smaller than a hadron. In phenomenology, the size is usually neglected. Thus the diquark is considered as a point-like object.

Since the diquark is composed of two quarks with spin one-half, the spin of the diquark can be 0 and 1. According to spin, the diquark system is classified into scalar and vector diquark. The spin of a scalar diquark is 0, and the two quarks are anti-symmetric in spin space in order to satisfy the Pauli principle. As a result, the two quarks in the diquark are anti-triplet states in both the color and spin spaces. The scalar diquark contains smaller mass than a vector one. One can expect that a hadron with the scalar diquark is lower in mass than a hadron with the vector diquark.

A baryon is composed of three quarks in the conventional quark model. Within the constituent quark model, it is a complicated three-body problem. The treatment is usually difficult. Under the diquark approximation, three-quark picture is changed to a quark-diquark picture, and the three-body problem is turned to a two-body one. This change will cause a great simplification in technic. For the low energy hadron reactions, the diquark hypothesis is tested to be workable \cite{Anselmino:1992vg}. The success of the diquark hypothesis in phenomenology indicates that the contributions from two correlated quarks are dominant. For a hadron with more than three quarks, the diquark approximation is even inevitable. The concept of diquark has been applied to many hadron phenomenology, e.g. the new exotic \cite{Jaffe:2003sg,Cheng:2004cc}.

For a light baryon, any two quarks may be correlated. But for a heavy baryon, such as $\Lambda_b^0$, the case is different. $b$ quark is heavy and will decay. The system of a diquark with a heavy quark and a light quark must break firstly and then decay. While for the two light quarks, they act as spectator. They are more likely to be correlated and unchanged during the weak interaction. Thus, a heavy baryon is considered to be composed of one heavy quark and a light diquark. For the ground state $\Lambda_b$ or $\Lambda_c$ which is an iso-singlet state, the light diquark is a scalar. As a spectator, the diquark in the light baryon, such as $p, n, \Lambda$, is also the scalar \cite{Korner:1994nh}. Thus, the baryons considered in this study ($\Lambda^0_b$, $\Lambda^+_c$, $\Lambda$, $p$, $n$) are composed of one quark ($b$, $c$, $s$, $u$, $d$) and a light diquark $[ud]$. The diquark is in a $0^+$ scalar state ($s=0, ~l=0$) and the orbital angular momentum between the quark and the diquark  is also  zero, i.e. $L=l=0$.

Under the diquark approximation, a baryon is similar to a meson. We call this phenomenon as meson-baryon similarity. The meson-baryon similarity has been noticed for a long time. In this study, we will see more examples and applications.

\subsection{Notations and conventions}

At first, we give our notations and conventions in the covariant light-front quark model. About the conventional light-front approach used in the previous works \cite{Ke:2007tg,Wei:2009np}, we collect their formulations in the Appendix A for reference. For a covariant four-component momentum denoted by $p$, it can be written with the light-front components as
\begin{eqnarray}
p=(p^-,p^+,p_{\perp}), ~~~~p^{\pm}=p^0\pm p^3.
\end{eqnarray}
The momentum square is $p^2=p^+p^--p_{\perp}^2$.

\begin{figure}[!ht]
\centering
\includegraphics[scale=0.8]{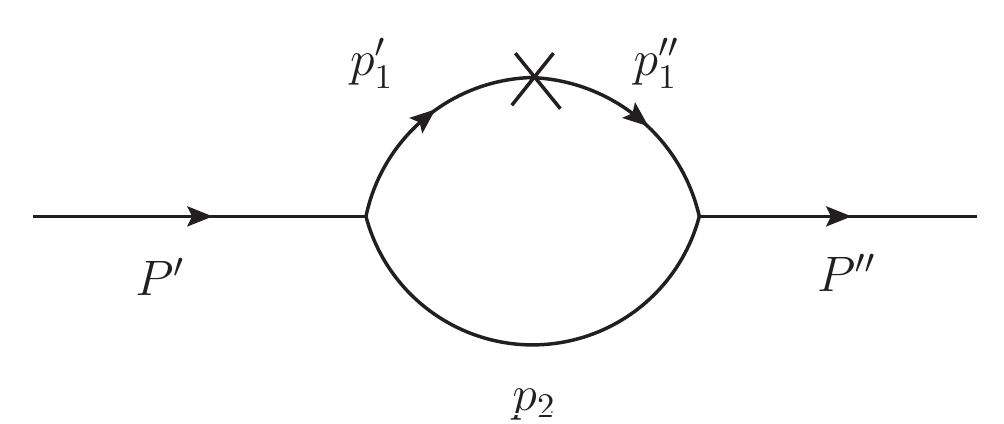}
\caption{Feynman diagram for the baryon to baryon transition amplitudes. The "$\times$" denotes the corresponding $V-A$ current vertex.}\label{fig:feyn}
\end{figure}

The Feynman diagram for the baryon to baryon transition are given by a one-loop graph shown in Fig. \ref{fig:feyn}. At each vertex where quarks and diquarks are off-shell, the four-component momentum is conserved. The momentum of the baryon is equal to the sum of the momenta of its constitutes. Thus, the incoming (outgoing) baryon has the momentum
\begin{eqnarray}
P^{\prime(\prime\prime)}=p^{\prime(\prime\prime)}_1+p_2.
\end{eqnarray}
where $P^{\prime(\prime\prime)}$ is the initial (final) baryon momentum, and $p^{\prime(\prime\prime)}_1$ and $p_2$ are momenta of the off-shell quark and diquark, respectively. The associated constituent masses are denoted by $m^{\prime(\prime\prime)}_1$ and $m_2$. The momentum transfer is $q=P'-P''$.  In order to describe the kinematics of the constituents in a baryon, it is convenient to introduce two intrinsic variable ($x_i,~p'_{\perp}$) where $x_i$ is the light-front momentum faction of the i-th constituent $i=1,2$  and $p'_{\perp}$ the relative transverse momentum between the quark and diquark. They are defined through
\begin{eqnarray}
p'^{+}_{1,2}=x_{1,2}P'^{+}, ~~~~~~ p'_{1,2\perp}=x_{1,2}P'_{\perp}\pm p'_{\perp}.
\end{eqnarray}
with $x_1+x_2=1$. The reason that $x_i,~p'_{\perp}$ are called by the intrinsic variable is that they are independent of the total momentum of the baryon and are invariant under the external Lorentz boost. Thus, the hadron wave function $\Psi(x_i,p'_{\perp})$ is explicitly Lorentz invariant. This is one advantage of the light-front framework.

In the purely longitudinal frame where $q_{\perp}=0$, the so-called Z-diagram contribution occurs and should be taken into account. But it is difficult to treat such contribution. So, we don't consider this frame in this study. As in \cite{Ke:2007tg,Wei:2009np}, we choose the transverse frame where $q^+=0$ and $q^2=-q_{\perp}^2$. The relation $x_2'=x_2''=x_2$ is satisfied in this particular frame. Some useful quantities are given below:
\begin{eqnarray}
M'^2_0&=&(e'_1+e_2)^2=\frac{p'^2_{\perp}+m'^2_1}{x_1}+\frac{p'^2_{\perp}+m^2_2}{x_2},\nonumber\\
M''^2_0&=&(e''_1+e_2)^2=\frac{p''^2_{\perp}+m''^2_1}{x_1}+\frac{p''^2_{\perp}+m^2_2}{x_2},
 \nonumber\\
e^{\prime}_i&=&\sqrt{m^{\prime2}_i+p'^2_{\perp}+p'^2_{z}},\nonumber\\
p'_z&=&\frac{x_2M'_0}{2}-\frac{m_2^2+p'^2_z}{2x_2M'_0},\nonumber\\
p''_{\perp}&=&p'_{\perp}-x_2q_{\perp}.
\end{eqnarray}

\subsection{Baryon-to-baryon transition matrix elements}

For the baryon transition $H_Q \rightarrow H_{Q^{\prime}}$ ($Q$, $Q'$ denote the incoming and outgoing quarks, respectively) depicted in Fig. 1, the amplitude can be expressed as
\begin{eqnarray}\label{eq:amu}
A_{\mu}=-i^3\frac{N_c}{(2\pi)^4}\int d^4p_1'\frac{H' H''}{N'_1 N''_1 N_2}s_{\mu},
\end{eqnarray}
where $H',H''$ are the vertex functions of the baryon-quark-diquark. Their explicit forms will be given below. The $s_\mu$ is
\begin{eqnarray}\label{eq:smu}
s_{\mu}=\bar{u}_{H_{Q^{\prime}}}(P'',S''_{z})[(\cancel{p}''_1+m''_1)
\gamma_{\mu}(1-\gamma_5)(\cancel{p}'_1+m'_1)]u_{H_Q}(P',S'_z).
\end{eqnarray}
where $u_{H_Q}(P',S'_z)$ is the baryon $H_Q$ spinor, $N'_1=p'^2_1-m'^2_1+i\varepsilon$, $N''_1=p''^2_1-m''^2_1+i\varepsilon$ and $N_2=p^2_2-m^2_2+i\varepsilon$. Obvious, the above equations are covariant.

Now, we turn to the light-front treatment. In order to do the integration over the $p'^-_1$ component in $A_{\mu}$ of Eq. (\ref{eq:amu}), we close the contour in the upper complex $p'^-_1$  plane and assuming the vertices $H'$ and $H''$ are analytic. This corresponds to putting the diquark on its mass shell, i.e., $\hat{p}^2_2=m^2_2$. The other momenta can be obtained by momentum conservation, $\hat{p}'_1=P'-\hat{p}_2$ and $\hat{p}''_1=P''-\hat{p}_2$. Note that this is one difference between the covariant approach and the conventional one where the momentum conservation is not satisfied in each vertex. Then, one can do the following replacement:
\begin{eqnarray}
 N'_1 &\rightarrow& \hat{N}'_1=\hat{p}'^{2}_1-m'^2_1=x'_1(M'^2-M'^2_0),\nonumber\\
 N''_1&\rightarrow& \hat{N}''_1=\hat{p}''^{2}_1-m''^2_1=x''_1(M''^2-M''^2_0),\nonumber\\
 H'   &\rightarrow& h',\nonumber\\
 H''  &\rightarrow& h'',\nonumber\\
 \int \frac{ d^4p'_1}{N'_1 N''_1 N_2}H' H''s_{\mu}&\rightarrow& -i\pi
  \int\frac{dx_2 d^2p'_{\bot}}{x_2 \hat{N}'_1\hat{N}''_1}h'h''\hat{s}.
\end{eqnarray}
As in \cite{Cheng:2003sm}, we also find that the factor $(M^{\prime(\prime\prime)2}-M^{\prime(\prime\prime)2}_0)
\sqrt{x^{\prime(\prime\prime)}_1x^{\prime(\prime\prime)}_2}$ cancels out the same expression in the denominator of Eq. (\ref{eq:amu}).

The explicit forms of $h'$ and $h''$ are given by
\begin{eqnarray}
h'&=&(M'^2-M'^2_0)\sqrt{\frac{x'_1x'_2}{N_c}}\frac{1}{\sqrt{2}\tilde M'_0}\varphi',\nonumber\\
h''&=&(M''^2-M''^2_0)\sqrt{\frac{x''_1x''_2}{N_c}}\frac{1}{\sqrt{2}\tilde M''_0}\varphi''.
\end{eqnarray}
where $\tilde M'_0=\sqrt{M'^2_0-(m_1'-m_2)^2}$ and $\tilde M''_0=\sqrt{M''^2_0-(m_1''-m_2)^2}$.
The $\varphi'$ and $\varphi''$ are light-front wave functions for the incoming and outgoing baryons, respectively. We use the Gaussian-type wave function as
\begin{eqnarray}
\varphi'=\varphi'(x_2,p'_{\perp})&=&4\left(\frac{\pi}{\beta^2}\right)^{3/4}\sqrt{\frac{\partial p'_z}{\partial x_2}}\text{exp}\left(-\frac{p'^2_z+p'^2_{\perp}}{2\beta'^2}\right),\nonumber\\
\varphi''=\varphi''(x_2,p''_{\perp})&=&4\left(\frac{\pi}{\beta^2}\right)^{3/4}\sqrt{\frac{\partial p''_z}{\partial x_2}}\text{exp}\left(-\frac{p''^2_z+p''^2_{\perp}}{2\beta''^2}\right),
\end{eqnarray}
with
\begin{eqnarray}
\frac{\partial p'_z}{\partial x_2}=\frac{e'_1e_2}{x_1x_2M'_0},~~~~~~
\frac{\partial p''_z}{\partial x_2}=\frac{e''_1e_2}{x_1x_2M''_0}.
\end{eqnarray}
The baryon parameter $\beta$ is the essential phenomenological input of the light-front quark model. In principle, it is at the order of the confinement scale.

\subsection{Formulations for the baryon-to-baryon transition form factors}

The form factors for the weak transition $H_Q\rightarrow H_{Q^{\prime}}$ are defined in the standard way as
\begin{eqnarray}\label{formfactor}
A_{\mu}&=&\langle H_{Q^{\prime}}(P'',S'',S''_z)|\bar{Q}'\gamma_{\mu}
 (1-\gamma_5)Q|H_Q(P',S',S'_z)\rangle\nonumber\\
&=&\bar{u}_{H_{Q^{\prime}}}(P'',S''_z)\left[\gamma_{\mu}f_1(q^2)
 +i\sigma_{\mu\nu}\frac{q^{\nu}}{M_{H_Q}}f_2(q^2)+\frac{q_{\mu}}
 {M_{H_Q}}f_3(q^2)\right]u_{H_Q}(P',S'_z)\nonumber\\
&&-\bar{u}_{H_{Q^{\prime}}}(P'',S''_z)\left[\gamma_{\mu}g_1(q^2)
 +i\sigma_{\mu\nu}\frac{q^{\nu}}{M_{H_Q}}g_2(q^2)
 +\frac{q_{\mu}}{M_{H_Q}}g_3(q^2)\right]\gamma_5u_{H_Q}(P',S'_z).
\end{eqnarray}
where $u_{H_Q}$ and $u_{H_{Q^{\prime}}}$ are Dirac spinors of the initial and final baryons $H_Q$, $H_{Q^{\prime}}$, respectively. There are six form factors in total. For the heavy-to-heavy $\Lambda_b^0\to\Lambda_c^+$ transitions, there is a well-known symmetry: the heavy quark symmetry in the infinite quark mass limit. The flavor and spin symmetries provide model-independent relations for form factors:
\begin{eqnarray}
f_1=g_1, ~~~~f_2=g_2=f_3=g_3=0.
\end{eqnarray}
Thus, $f_1$ and $g_1$ are dominant and other form factors are higher powers in $1/m_b$.
For the heavy-to-light transitions $\Lambda_b^0\to p(\Lambda)$,  the above relations are still
valid in the large energy limit for the large recoil region \cite{Mannel:2011xg}.

After the replacements in the covariant approach, the amplitude $A_{\mu}$ in the transition $H_Q \rightarrow H_{Q^{\prime}}$ given in the above subsection is expressed by
\begin{eqnarray}\label{amu2}
 A_{\mu} &=& N_{IF}\frac{N_c}{16\pi}\int\frac{dx_2 d^2p'_{\bot}}{x_2 \hat{N}'_1\hat{N}''_1}
 h'h''\bar{u}_{H_{Q^{\prime}}}(P'',S''_{z})\left[(\hat{\cancel{p}}''_1+m''_1)\gamma_{\mu}\right. \non\\
 && \left.~~\times(1-\gamma_5)(\hat{\cancel{p}}'_1+m'_1)\right]u_{H_Q}(P',S'_z).
\end{eqnarray}
where $N_{IF}$ is a flavor-spin factor which will be given for different processes later.

In principle, the six form factors can be extracted out by comparing Eqs. (\ref{formfactor}) and (\ref{amu2}). But, the initial and final baryon spinors produce some difficulties. Our treatment is to use the familiar spin sum relation of the Dirac spinors $\sum\limits_{S'_z}\bar{u}_{H_Q} (P',S'_z) u_{H_Q}(P',S'_z)=\cancel{P}'+M'$. To proceed, we multiply $\sum\limits_{S'_z,S''_z}\bar{u}_{H_Q}(P',S'_z) ~u_{H_{Q^{\prime}}}(P'',S''_{z})~P^{\mu}$, $\sum\limits_{S'_z,S''_z} \bar{u}_{H_Q}(P',S'_z)
~u_{H_{Q^{\prime}}}(P'',S''_{z})~q^{\mu}$ and $\sum\limits_{S'_z,S''_z}\bar{u}_{H_Q}(P',S'_z)
~\gamma^{\mu}~u_{H_{Q^{\prime}}}(P'',S''_{z})$ onto the right side of Eqs. (\ref{formfactor}) and (\ref{amu2}). According to the equality of the two equations, we obtain three independent equations. From these equations, the three physical quantities $f_1$, $f_2$ and $f_3$ can be solved. Because there are more terms occurred than the meson case, our method is different from the treatment in \cite{Cheng:2003sm}. After a lengthy calculation and with help of the computer program, we obtain the analytic formulae for the form factors $f_1$, $f_2$ and $f_3$ as
\begin{eqnarray}
f_1(q^2)&=& N_{IF}\int\frac{dx_2 d^2p'_{\perp}}{16\pi^3}\frac{\varphi_{H_Q}(x'_2,p'_{\perp})\varphi_{H_{Q'}}
 (x''_2,p''_{\perp})}{\sqrt{[(m'_1+x'_1M'_0)^2+p'^2_{\perp}][(m''_1+x''_1M''_0)^2+p''^2_{\perp}]}}\non\\
 &&\frac{1}{(M'+M'')^2-q^2}\left\{A^{(1)}_1[(M''+M')^2-q^2](2m'_1M''+2m''_1M'-q^2)\right.\nonumber \\
 &&+A^{(1)}_2q^2[(M'+M'')^2-q^2]+2A^{(2)}_1[(M'+M'')^2-q^2]\nonumber\\
 &&+A^{(2)}_2[2(M'+M'')^4-4M'M''q^2-q^4]\nonumber\\
 &&+2A^{(2)}_3(M'-M'')(M'+M'')^3+A^{(2)}_4q^4\nonumber\\
 &&+m'_1m''_1[(M'+M'')^2-q^2]-[x_1(M'^2-M'^2_0)+m'^2_1](M'+M'')^2\Big\},
\end{eqnarray}

\begin{eqnarray}
f_2(q^2)&&=N_{IF}\int\frac{dx_2 d^2p'_{\perp}M}{16\pi^3}\frac{\varphi_{H_Q}(x'_2,p'_{\perp})
\varphi_{H_{Q'}}(x''_2,p''_{\perp})}{\sqrt{[(m'_1+x'_1M'_0)^2+p'^2_{\perp}]
 [(m''_1+x''_1M''_0)^2+p''^2_{\perp}]}}\non\\
 &&\frac{1}{(M'+M'')^2-q^2} \left\{A^{(1)}_1[(M''+M')^2-q^2](m'_1+m''_1-2M')\right.\nonumber \\
 &&+A^{(1)}_2(m'_1-m''_1)[(M'+M'')^2-q^2]+4A^{(2)}_1[(M'+M'')]\nonumber\\
 &&+A^{(2)}_2(M'+M'')[4M'^2+4M'M''+4M''^2-3q^2]\nonumber\\
 &&+2A^{(2)}_3(M'-M'')[2(M'+M'')^2-q^2]+A^{(2)}_4q^2(M'+M'')\nonumber\\
 &&-m'_1[(M'+M'')^2-q^2]-[x_1(M'^2-M'^2_0)+m'^2_1](M'+M'')\Big\},
\end{eqnarray}

\begin{eqnarray}
f_3(q^2)&&=N_{IF}\int\frac{dx_2 d^2p'_{\perp}M}{16\pi^3}\frac{\varphi_{H_Q}(x'_2,p'_{\perp})
\varphi_{H_{Q'}}(x''_2,p''_{\perp})}{\sqrt{[(m'_1+x'_1M'_0)^2+p'^2_{\perp}]
[(m''_1+x''_1M''_0)^2+p''^2_{\perp}]}} \non\\
&&\frac{1}{(M'+M'')^2-q^2}\left\{A^{(1)}_1[(M''+M')^2-q^2](m'_1-m''_1-2M'')\right.\nonumber \\
&&+A^{(1)}_2(m'_1+m''_1-2M'+M'')[(M'+M'')^2-q^2]-4A^{(2)}_1[(M'-M'')]\nonumber\\
&&-A^{(2)}_2(M'-M'')(2M'^2+2M''^2-q^2)\nonumber\\
&&+2A^{(2)}_3(M'+M'')(4M'M''-q^2)+A^{(2)}_4(M'-M'')[2(M'+M'')^2-3q^2]\nonumber\\
&&-m'_1[(M'+M'')^2-q^2]+[x_1(M'^2-M'^2_0)+m'^2_1](M'-M'')\Big\}£¬
\end{eqnarray}
where $A^{(i)}_j$ are functions of $x_2$, $p'^2_{\perp}$, $p'_{\perp}\cdot q_{\perp}$ and $q^2$. Their explicit expressions are \cite{Jaus:1999zv}
\beq
&&A^{(1)}_1=\frac{x_1}{2}, \qquad \qquad  ~~~~A^{(1)}_2=A^{(1)}_1-\frac{p'_{\perp}\cdot q_{\perp}}{q^2},
 \non\\
&&A^{(2)}_1=-p'^2_{\perp}-\frac{(p'_{\perp}\cdot q_{\perp})^2}{q^2}, \qquad
 A^{(2)}_2=(A^{(1)}_1)^2, \non\\
&&A^{(2)}_3=A^{(1)}_1 A^{(1)}_2, ~~~~\qquad A^{(2)}_4=(A^{(1)}_2)^2-\frac{A^{(2)}_1}{q^2}.
\eeq

The other three form factors $g_1$, $g_2$ and $g_3$ can be obtained in a similar way. A $\gamma_5$ matrix is needed to insert into the spinors. We multiply $\sum\limits_{S'_z,S''_z} \bar{u}_{H_Q}(P',S'_z) \gamma^{5} u_{H_{Q^{\prime}}}(P'',S''_{z})P^{\mu}$, $\sum\limits_{S'_z,S_z''} \bar{u}_{H_Q}(P',S'_z) \gamma^{5} u_{H_{Q^{\prime}}}(P'',S''_{z})q^{\mu}$  and $\sum\limits_{S'_z,S_z''} \bar{u}_{H_Q}(P',S'_z) \gamma^{\mu}\gamma_5 u_{H_{Q^{\prime}}}(P'',S''_{z})$  onto the right side of Eqs. (\ref{formfactor}) and (\ref{amu2}). Then by solving another three equations, the form
factors $g_1$, $g_2$ and $g_3$ are obtained as
\begin{eqnarray}
g_1(q^2)&&=N_{IF}\int\frac{dx_2 d^2p'_{\perp}}{16\pi^3}\frac{\varphi_{H_Q}(x'_2,p'_{\perp})\varphi_{H_{Q'}}(x''_2,p''_{\perp})}
{\sqrt{[(m'_1+x'_1M'_0)^2+p'^2_{\perp}][(m''_1+x''_1M''_0)^2+p''^2_{\perp}]}}\non\\
&&\frac{1}{(M'-M'')^2-q^2}\left\{A^{(1)}_1[(M'-M¡®')^2-q^2](2m'_1M''+2m''_1M'+q^2)\right.\nonumber \\
&&-A^{(1)}_2q^2[(M'-M'')^2-q^2]-2A^{(2)}_1[(M'-M'')^2+q^2]\nonumber\\
&&-A^{(2)}_2[2(M'-M'')^4+4M'M''q^2-q^4]\nonumber\\
&&-2A^{(2)}_3(M'+M'')(M'-M'')^3-A^{(2)}_4q^4\nonumber\\
&&+m'_1m''_1[(M'-M'')^2-q^2]+[x_1(M'^2-M'^2_0)+m'^2_1](M'-M'')^2\Big\},
\end{eqnarray}

\begin{eqnarray}
g_2(q^2)&&=N_{IF}\int\frac{dx_2 d^2p'_{\perp}M}{16\pi^3}\frac{\varphi_{H_Q}(x'_2,p'_{\perp})
\varphi_{H_{Q'}}(x''_2,p''_{\perp})}{\sqrt{[(m'_1+x'_1M'_0)^2+p'^2_{\perp}]
[(m''_1+x''_1M''_0)^2+p''^2_{\perp}]}}\non\\
&&\frac{1}{(M'-M'')^2-q^2}\left\{A^{(1)}_1[(M'-M'')^2-q^2](m'_1-m''_1-2M')\right.\nonumber \\
&&+A^{(1)}_2(m'_1+m''_1)[(M'-M'')^2-q^2]+4A^{(2)}_1[(M'-M'')]\nonumber\\
&&+A^{(2)}_2(M'-M'')[4M'^2-4M'M''+4M''^2-3q^2]\nonumber\\
&&+2A^{(2)}_3(M'+M'')[2(M'-M'')^2-q^2]+A^{(2)}_4q^2(M'-M'')\nonumber\\
&&-m'_1[(M'-M'')^2-q^2]-[x_1(M'^2-M'^2_0)+m'^2_1](M'-M'')\Big\},
\end{eqnarray}

\begin{eqnarray}
g_3(q^2)&&=N_{IF}\int\frac{dx_2 d^2p'_{\perp}M}{16\pi^3}\frac{\varphi_{H_Q}(x'_2,p'_{\perp})\varphi_{H_{Q'}}(x''_2,p''_{\perp})}
{\sqrt{[(m'_1+x'_1M'_0)^2+p'^2_{\perp}][(m''_1+x''_1M''_0)^2+p''^2_{\perp}]}}\non\\
&&\frac{1}{(M'-M'')^2-q^2}\left\{A^{(1)}_1[(M'-M'')^2-q^2](m'_1+m''_1+2M'')\right.\nonumber \\
&&+A^{(1)}_2(m'_1-m''_1-2M'-2M'')[(M'-M'')^2-q^2]-4A^{(2)}_1[(M'+M'')]\nonumber\\
&&-A^{(2)}_2(M'+M'')(2M'^2+2M''^2-q^2)\nonumber\\
&&-2A^{(2)}_3(M'-M'')(4M'M''+q^2)+A^{(2)}_4(M'+M'')[2(M'-M'')^2-3q^2]\nonumber\\
&&-m'_1[(M'-M'')^2-q^2]+[x_1(M'^2-M'^2_0)+m'^2_1](M'+M'')\Big\}.
\end{eqnarray}
One can find that the formulations for $f_i$ and $g_i$ are quite similar except for some sign difference.

From \cite{Singleton:1990ye}, the spin-flavor factors $N_{IF}$ for different transitions are given by
 \beq \label{sff}
 N_{\Lambda^0_b\Lambda^+_c}=1,
 \qquad N_{\Lambda^0_bp}=\frac{1}{\sqrt{2}}, \qquad
 N_{\Lambda^0_b\Lambda}=\frac{1}{\sqrt{3}}.
 \eeq
These factors are necessary to obtain the correct theory predictions. Without them, the $\Lambda_b^0\to p$ process will be increased by a factor of 2 and the $\Lambda_b^0\to \Lambda$ process will be increased by a factor of 3. In \cite{Singleton:1990ye}, these factors are derived in the three-quark picture. In the quark-diquark picture, the the spin-flavor factors remain the same and it is easier to obtain them. The heavy baryon flavor and spin wave functions are
 \beq
 |\Lambda_b^0\rangle=b[ud]\chi_A,     \qquad  \qquad  |\Lambda_c^+\rangle=c[ud]\chi_A,
 \eeq
where $[ud]$ is the scalar diquark with $[ud]=\frac{ud-du}{\sqrt{2}}$ and $\chi_A$ is the spin function which is anti-symmetric for the diquark. For the light baryons $p$ and $\Lambda$,
 \beq
 &&|p\rangle=\frac{1}{\sqrt{2}}\left( u[ud]\chi_A+\phi_{S}\chi_S\right), \non\\
 &&|\Lambda\rangle=\frac{1}{\sqrt{2}}\frac{1}{\sqrt{6}}\left(2[ud]s\chi_A+[ds]u\chi_A
 +[su]d\chi_A+\phi_{S}\chi_S\right).
 \eeq
The $\phi_{S}$ and $\chi_S$ are mixed symmetric flavor and spin wave functions. Their explicit forms are irrelevant because the diquark in the final baryon comes from the scalar diquark in the initial heavy baryon which is flavor and spin anti-symmetric. The factor $\frac{1}{\sqrt{2}}$ comes from the equal components of the mixed symmetric and mixed anti-symmetric flavor wave functions of the baryon SU(3) octets. By comparing the coefficients of the diquark $[ud]$ for each baryon, we obtain the same spin-flavor factors as  Eq. (\ref{sff}). It is noted that the authors in \cite{Cheng:1995fe} use a totally antisymmetric flavor wave function for $\Lambda$ which is not correct for a ground state baryon. But their results are correct.

\section{ Semi-leptonic decays of $\Lambda^0_b\rightarrow \Lambda^+_c\ (p) l^-\bar{\nu}_l$ }

In this section, we provide formulations for the rates and some asymmetries of the semi-leptonic processes.
In order to study the semi-leptonic decays, another parametrization of the transition form factors adopted in \cite{Faustov:2016pal} is useful. It is given by
\begin{eqnarray} \label{formfactor2}
&&\langle H_{Q^{\prime}}(P'',S'',S''_z)|V_{\mu}|H_Q(P',S',S'_z)\rangle\nonumber\\
&&=\bar{u}_{H_{Q^{\prime}}}(P'',S''_z)\left[\gamma_{\mu}F_1(q^2)+\frac{P'_{\mu}}{M_{H_Q}}
F_2(q^2)+\frac{P''_{\mu}}{M_{H_{Q'}}}F_3(q^2)\right]u_{H_Q}(P',S'_z)\nonumber\\
&&\langle H_{Q^{\prime}}(P'',S'',S''_z)|A_{\mu}|H_Q(P',S',S'_z)\rangle\nonumber\\
&&=\bar{u}_{H_{Q^{\prime}}}(P'',S''_z)\left[\gamma_{\mu}G_1(q^2)+\frac{P'_{\mu}}
{M_{H_Q}}G_2(q^2)+\frac{P''_{\mu}}{M_{H_{Q'}}}G_3(q^2)\right]\gamma_5u_{H_Q}(P',S'_z).
\end{eqnarray}

The two parametrization forms of Eqs. (\ref{formfactor}) and (\ref{formfactor2}) are related by
\begin{eqnarray}
F_1(q^2)&=&f_1(q^2)-(M_{H_Q}+M_{H_{Q^{\prime}}})\frac{f_2(q^2)}{M_{H_Q}},\nonumber\\
F_2(q^2)&=&f_3(q^2)+f_2(q^2),\nonumber\\
F_3(q^2)&=&-\frac{M_{H_{Q^{\prime}}}}{M_{H_Q}}\left[f_3(q^2)-f_2(q^2)\right],\nonumber\\
G_1(q^2)&=&g_1(q^2)+(M_{H_Q}-M_{H_{Q^{\prime}}})\frac{g_2(q^2)}{M_{H_Q}},\nonumber\\
G_2(q^2)&=&g_3(q^2)+g_2(q^2),\nonumber\\
G_3(q^2)&=&-\frac{M_{H_{Q^{\prime}}}}{M_{H_Q}}\left[g_3(q^2)-g_2(q^2)\right].
\end{eqnarray}

Following \cite{Faustov:2016pal, Bialas:1992ny}, it is necessary to define the helicity amplitudes which are expressed in terms of the weak form factors. The different helicity amplitudes are defined by
\begin{eqnarray}
H^V_{+1/2,0}&=&\frac{1}{\sqrt{q^2}}\sqrt{2M_{H_Q}M_{H_{Q^{\prime}}}(\omega-1)}~
\left[(M_{H_Q}+M_{H_{Q^{\prime}}})F_1(q^2)+M_{H_{Q}^{\prime}}(\omega+1)F_2(q^2)\right.\nonumber\\
&&+M_{H_{Q}}(\omega+1)F_3(q^2)\Big ],\nonumber\\
H^A_{+1/2,0}&=&\frac{1}{q^2}\sqrt{2M_{H_Q}M_{H_{Q^{\prime}}}(\omega+1)}~
\left[(M_{H_Q}-M_{H_{Q^{\prime}}})F_1(q^2)-M_{H_{Q}^{\prime}}(\omega-1)F_2(q^2)\right.\nonumber\\
&&-M_{H_{Q}}(\omega-1)F_3(q^2)\Big ],\nonumber\\
H^V_{+1/2,1}&=&-2\sqrt{M_{H_Q}M_{H_{Q^{\prime}}}(\omega-1)}~F_1(q^2),\nonumber\\
H^A_{+1/2,1}&=&-2\sqrt{M_{H_Q}M_{H_{Q^{\prime}}}(\omega+1)}~G_1(q^2),\nonumber\\
H^V_{+1/2,t}&=&\frac{1}{\sqrt{q^2}}\sqrt{2M_{H_Q}M_{H_{Q^{\prime}}}(\omega+1)}~
\left[(M_{H_Q}-M_{H_{Q^{\prime}}})F_1(q^2)+(M_{H_Q}-M_{H_{Q^{\prime}}}\omega)F_2(q^2)\right.\nonumber\\
&&+(M_{H_Q}\omega-M_{H_{Q^{\prime}}})F_3(q^2)\Big ],\nonumber\\
H^A_{+1/2,t}&=&\frac{1}{\sqrt{q^2}}\sqrt{2M_{H_Q}M_{H_{Q^{\prime}}}(\omega-1)}~
\left[(M_{H_Q}+M_{H_{Q^{\prime}}})G_1(q^2)-(M_{H_Q}-M_{H_{Q^{\prime}}}\omega)G_2(q^2)\right.\nonumber\\
&&-(M_{H_Q}\omega-M_{H_{Q^{\prime}}})G_3(q^2)\Big ].
\end{eqnarray}
where
\begin{eqnarray}
\omega=\frac{M^2_{H_Q}+M^2_{H_{Q^{\prime}}}-q^2}{2M_{H_Q}M_{H_{Q^{\prime}}}}.
\end{eqnarray}
The helicity amplitudes $H^{V,A}_{\lambda',\lambda_{W}}$ where $\lambda'$ and $\lambda_W$ are the helicities of the final baryon and the virtual $W$-boson, are the amplitudes for vector ($V$) and axial ($A$) vector currents, respectively. Because of the $V-A$ structure of the charged current weak interaction, the total helicity amplitudes are obtained as
\begin{eqnarray}
H_{\lambda',\lambda_W}=H^V_{\lambda',\lambda_W}-H^A_{\lambda',\lambda_W}.
\end{eqnarray}
The helicity amplitudes for the negative values of the helicities satisfy the relations
\begin{eqnarray}
H^V_{-\lambda',-\lambda_W}=+H^V_{\lambda',\lambda_W},\qquad
H^A_{-\lambda',-\lambda_W}=-H^A_{\lambda',\lambda_W}.
\end{eqnarray}

For the semi-leptonic process of $H_Q\rightarrow H_{Q^{\prime}} W^-(\rightarrow l^- \bar{\nu}_l)$,
the twofold angular distribution can be derived to be
\begin{eqnarray}\label{eq:dG1}
\frac{d\Gamma(H_Q\rightarrow H_{Q^{\prime}} l^- \bar{\nu}_l)}{dq^2d\cos\theta}=\frac{G_F^2}{(2\pi)^3}\mid V_{Q'Q}\mid^2\frac{\lambda(q^2-m_l^2)}{48M^3_{H_Q}q^2}W(\theta,q^2).
\end{eqnarray}
where
\begin{eqnarray} \label{eq:H}
W(\theta,q^2)&=&\frac{3}{8}\Big\{(1+\cos^2\theta)H_U{q^2}-2\cos\theta H_P(q^2)+2\sin^2\theta H_L(q^2)\nonumber\\
&&+\frac{m_l^2}{q^2}\left[2H_S(q^2)+\sin^2\theta H_U(q^2)+2\cos^2\theta H_L(q^2)-4\cos\theta H_{SL}(q^2) \right]\Big\},
\eeq
and
\beq
\lambda\equiv\lambda(M^2_{H_Q},M^2_{H_{Q'}},q^2)=M^4_{H_Q}
+M^4_{H_{Q'}}+q^4-2(M^2_{H_Q}M^2_{H_{Q'}}+M^2_{H_Q}q^2+M^2_{H_{Q'}}q^2).
\eeq
The $V_{Q'Q}$ is the CKM matrix elements, $G_F$ the Fermi constant. $m_l$ is the lepton mass $(l=e,\mu,\tau)$, and $\theta$ is the angle between the lepton $l$ and $W$ momenta.

In Eq. (\ref{eq:H}), there are several amplitudes $H_i$ which are given in terms of the helicity amplitudes.
The relevant parity conserving helicity amplitudes are given by
\begin{eqnarray}
H_U(q^2)&=&\mid H_{+1/2,+1}\mid^2+\mid H_{-1/2,-1}\mid^2,\nonumber\\
H_L(q^2)&=&\mid H_{+1/2,0}\mid^2+\mid H_{-1/2,0}\mid^2,\nonumber\\
H_S(q^2)&=&\mid H_{+1/2,t}\mid^2+\mid H_{-1/2,t}\mid^2,\nonumber\\
H_{SL}(q^2)&=&{\rm Re}( H_{+1/2,0}H^{\dag}_{+1/2,t}+H_{-1/2,0}H^{\dag}_{-1/2,t}),
\end{eqnarray}
and the parity violating helicity amplitudes are
\begin{eqnarray}
H_P(q^2)&=&\mid H_{+1/2,+1}\mid^2-\mid H_{-1/2,-1}\mid^2,\nonumber\\
H_{LP}(q^2)&=&\mid H_{+1/2,0}\mid^2-\mid H_{-1/2,0}\mid^2,\nonumber\\
H_{SP}(q^2)&=&\mid H_{+1/2,t}\mid^2-\mid H_{-1/2,t}\mid^2.
\end{eqnarray}

By integrating over $\cos\theta$ of Eq. (\ref{eq:dG1}), we obtain the transverse momentum $q^2$-dependent differential decay as
\begin{eqnarray}
\frac{d\Gamma(H_Q\rightarrow H_{Q'}l\bar{\nu_l})}{dq^2}=\frac{G_F^2}{(2\pi)^3}\mid V_{Q'Q}\mid^2\frac{\lambda(q^2-m_l^2)}{48M^3_{H_Q}q^2}H_{tot}(q^2).
\end{eqnarray}
where
\begin{eqnarray}
H_{tot}(q^2)=[H_U(q^2)+H_L(q^2)]\left(1+\frac{m_l^2}{2q^2}\right)+\frac{3m_l^2}{2q^2}H_S(q^2).
\end{eqnarray}

The forward-backward asymmetry is an important observable quantity. From Eq. (\ref{eq:dG1}), the $q^2$-dependent forward-backward asymmetry of the charged lepton is given by
\begin{eqnarray}
A_{FB}(q^2)=\frac{\frac{d\Gamma}{dq^2}(\text{forward})-\frac{d\Gamma}{dq^2}
(\text{backward})}{\frac{d\Gamma}{dq^2}}=-\frac{3}{4}\frac{H_P(q^2)
+2\frac{m_l^2}{q^2}H_{SL}(q^2)}{H_{tot}(q^2)}.
\end{eqnarray}
The integrated forward-backward asymmetry is obtained as
\begin{eqnarray}
A_{FB}&=&\frac{\int_{m_l^2}^{(M_{H_Q}-M_{H_{Q'}})^2}\frac{d\Gamma}{dq^2}
(\text{forward})-\int_{m_l^2}^{(M_{H_Q}-M_{H_{Q'}})^2}\frac{d\Gamma}{dq^2}
(\text{backward})}{\int_{m_l^2}^{(M_{H_Q}-M_{H_{Q'}})^2}\frac{d\Gamma}{dq^2}},\nonumber\\
&=&-\frac{3}{4}\frac{\int_{m_l^2}^{(M_{H_Q}-M_{H_{Q'}})^2}dq^2[H_P(q^2)
+2\frac{m_l^2}{q^2}H_{SL}(q^2)]}{\int_{m_l^2}^{(M_{H_Q}-M_{H_{Q'}})^2}dq^2[H_{tot}(q^2)]}.
\end{eqnarray}

Similary, the $q^2$-dependent longitudinal polarization of the final baryon $H_{Q'}$ is
\begin{eqnarray}
P_L(q^2)=\frac{[H_P(q^2)+H_{LP}(q^2)]\left(1+\frac{m_l^2}{2q^2}\right)+3\frac{m_l^2}
{2q^2}H_{SP}(q^2)}{H_{tot}(q^2)}.
\end{eqnarray}
The integrated longitudinal polarization of the final baryon $H_{Q'}$ is
\begin{eqnarray}
P_L=\frac{\int_{m_l^2}^{(M_{H_Q}-M_{H_{Q'}})^2}dq^2\left\{\left[H_P(q^2)+H_{LP}(q^2)\right]
\left(1+\frac{m_l^2}{2q^2}\right)+3\frac{m_l^2}{2q^2}H_{SP}(q^2)\right\}}
{\int_{m_l^2}^{(M_{H_Q}-M_{H_{Q'}})^2}dq^2H_{tot}(q^2)}.
\end{eqnarray}

\section{Nonleptonic decays of $\Lambda^0_b\rightarrow H+M$ in QCD factorization approach}

In this section, we study the exclusive nonleptonic decays $\Lambda^0_b\rightarrow H + M$ where H represents baryon ($\Lambda^+_c$, $p$, $n$, $\Lambda$) and $M$ represents a meson. For the meson $M$, we restrict our discussions for the ground state, i.e. pseudoscalar (P) or vector (V) meson in this study.

\subsection{Classification}

At first, we discuss the classification of the $\Lambda_b^0$ decays. In the B meson case, it is usually classified by the charmful and charmless processes according to the charm quark component of the final mesons. This classification can be done for the heavy baryon. But it may not be most convenient. The heavy baryon $\Lambda_b^0$ decays have one property: the spectator can only enter into the baryon. This argument is valid under the diquark assumption. Without the diquark approximation, one spectator quark can enter into the final meson. While for the meson case, the spectator quark is possible to enter into either of the two final mesons. This difference makes us to choose a more convenient classification method. The $\Lambda_b^0$ decays are classified by the final baryon. According to this classification rule, the $\Lambda_b^0$ decays are classified into four classes: (1) $\Lambda_b^0\to \Lambda_c^+ + M$, ~(2) $\Lambda_b^0\to p + M$, ~(3) $\Lambda_b^0\to \Lambda + M$, ~(4) $\Lambda_b^0\to n + M$. For each class, the decay modes are collected as following. We only write the final state to represent each decay mode.

\vspace{0.5cm}

(1)~~ $\Lambda_b^0\to \Lambda_c^+ + M$ (8 modes)
 \begin{eqnarray*}
 &&\Lambda_c^+ \pi^-, \qquad ~\Lambda_c^+ \rho^-, \qquad ~~\Lambda_c^+ K^-, \qquad \Lambda_c^+ K^{*-}, \\
 &&\Lambda_c^+ D^-, \qquad \Lambda_c^+ D^{*-}, \qquad \Lambda_c^+ D_s^-, \qquad \Lambda_c^+ D_s^{*-}.
 \end{eqnarray*}
Since the initial and final baryons are $\Lambda_b^0$ and $\Lambda_c^+$,  the final meson $M$ must be negative charged because of the charge conservation. The negative charged quark-antiquark pair combined by $u,d,c,s$ quarks can be: $\bar u d$, $\bar u s$, $\bar c d$, $\bar c s$. Correspondingly, the ground state mesons are: $\pi^-$, $\rho^-$, $K^-$, $K^{*-}$, $D^-$, $D^{*-}$, $D_s^-$, $D_s^{*-}$.

\vspace{0.5cm}

(2)~~ $\Lambda_b^0\to p + M$ (8 modes)
 \begin{eqnarray*}
 &&p\pi^-, \qquad ~p\rho^-, \qquad ~~p K^-, \qquad p K^{*-}, \\
 &&p D^-, \qquad p D^{*-}, \qquad p D_s^-, \qquad p D_s^{*-}.
 \end{eqnarray*}
Similar discussions follow from the above arguments, and the final meson $M$ can be: $\pi^-$, $\rho^-$, $K^-$, $K^{*-}$, $D^-$, $D^{*-}$, $D_s^-$, $D_s^{*-}$.

\vspace{0.5cm}

(3)~~ $\Lambda_b^0\to \Lambda + M$ (14 modes)
 \begin{eqnarray*}
 &&\Lambda\pi^0, \qquad ~\Lambda\rho^0, \qquad ~\Lambda K^0,     \qquad \Lambda K^{*0}, \\
 &&\Lambda\eta, \qquad ~~~\Lambda\eta', \qquad ~\Lambda\omega,   \qquad ~~~\Lambda\phi, \\
 &&\Lambda D^0, \qquad \Lambda D^{*0},  \qquad \Lambda \bar D^0, \qquad \Lambda \bar D^{*0}, \\
 &&\Lambda \eta_c, \qquad ~\Lambda J/\psi.
 \end{eqnarray*}
The final meson $M$ must be neutral charged according to the charge conservation. Among all the neutral charged mesons, the two states of $\bar K^{(*)0}$ are not allowed. It is because the states $\Lambda \bar K^{(*)0}$ contain two $s$ quarks. They cannot be produced by the tree or penguin operators of the weak effective interactions to be given below. The neutral charged quark-antiquark pair combined by $u,d,c,s$ quarks can be: $\bar u u$, $\bar d d$, $\bar s s$, $\bar s d$, $\bar d s$, $\bar u c$, $\bar c u$, $\bar c c$. Correspondingly, except $\bar K^{(*)0}$, the neutral ground state mesons include: $\pi^0$, $\rho^0$, $K^0$, $K^{*0}$, $\eta$, $\eta'$, $\omega$, $\phi$, $D^0$, $D^{*0}$, $\bar D^0$, $\bar D^{*0}$, $\eta_c$, $J/\psi$.

\vspace{0.5cm}

(4)~~ $\Lambda_b^0\to n + M$ (14 modes)
 \begin{eqnarray*}
 &&n\pi^0, \qquad ~n\rho^0,  \qquad ~n\bar K^0, \qquad ~n\bar K^{*0}, \\
 &&n\eta,  \qquad ~~~n\eta', \qquad ~n\omega,   \qquad ~~~n\phi, \\
 &&n D^0,  \qquad n D^{*0},  \qquad n \bar D^0, \qquad n \bar D^{*0}, \\
 &&n \eta_c, \qquad ~n J/\psi.
 \end{eqnarray*}
The final meson $M$ must be neutral charged due to the charge conservation. Among all the neutral charged mesons, $K^{(*)0}$ are not allowed. It is because $n K^{(*)0}$ contains one $\bar s$ quark which cannot be produced by the tree or penguin operators.

There are 44 decay modes in total. We will discuss these modes in the part of numerical results in detail.

\subsection{The effective Hamiltonian and QCD factorization approach}

There are three separate energy scales in $\Lambda_b^0$ weak decays: $M_W\gg m_b\gg\lqcd$. One convenient method is the effective field theory. By integrating out the high energy degree of freedom and performing the operator product expansion, the interactions are expressed as a series of local effective operators. The information of high energy is encoded in the Wilson coefficients. In this study, the effective Hamiltonian $H_{eff}$ for $b\rightarrow s$ transitions ($b\to d$ transitions are done by the replacement of $s\to d$) can be written by \cite{Buras:1998raa}:
 \begin{eqnarray}
  \mathcal{H}_{eff}=\frac{G_F}{\sqrt{2}}\sum\limits_{q=u,c}v_q\left(C_1O^{q}_1+
  C_2O^{q}_2+\sum\limits^{10}_{i=3}C_{i}O_{i}+C_{7\gamma}O_{7\gamma}+C_{8g}O_{8g}\right),
 \end{eqnarray}
where $v_q=V_{qb}V^{*}_{qs}$. The $C_i$ are Wilson coefficients evaluated at the renormalization scale $\mu$. The current-current operators $O_1^{u}$ and $O_2^{u}$ are
 \begin{eqnarray}
 O^{u}_1=\bar{s}_{\alpha}\gamma^{\mu}L u_{\alpha}\cdot\bar{u}_{\beta}\gamma_{\mu}L b_{\beta}, ~~~~~ &O^{u}_2=\bar{s}_{\alpha}\gamma^{\mu}L u_{\beta}\cdot\bar{u}_{\beta}\gamma_{\mu}L b_{\alpha}.
 \end{eqnarray}
where $\alpha$ and $\beta$ are the SU(3) color indices, and $L$ and $R$ are the left- and right-handed projection operators with $L=1-\gamma_5$ and $R=1+\gamma_5$, respectively.

The usual tree-level W-exchange contribution in the effective theory corresponds to $O_1$ and $O_2$ emerges due to the QCD corrections. The operators $O_3-O_6$ are
 \begin{eqnarray}
 O_3=\bar{s}_{\alpha}\gamma^{\mu}L b_{\alpha}\cdot\sum\limits_{q'}\bar{q}'_{\beta}\gamma_{\mu}Lq'_{\beta}, ~~~~~ & O_4=\bar{s}_{\alpha} \gamma^{\mu}L b_{\beta}\cdot\sum\limits_{q'}\bar{q}'_{\beta}\gamma_{\mu}L
  q'_{\alpha}, \non\\
 O_5=\bar{s}_{\alpha}\gamma^{\mu}L b_{\alpha}\cdot\sum\limits_{q'}\bar{q}'_{\beta}\gamma_{\mu}R
  q'_{\beta}, ~~~~~ &
 O_6=\bar{s}_{\alpha}\gamma^{\mu}L b_{\beta}\cdot\sum\limits_{q'}\bar{q}'_{\beta}\gamma_{\mu}R
  q'_{\alpha}.
\end{eqnarray}
They arise from the QCD penguin diagrams which contribute in order $\alpha_s$ through the initial values of the Wilson coefficients at $\mu\approx M_W$ and operator mixing due to the QCD corrections. The sum over $q'$ runs over the quark fields that are active at the scale $\mu=O(m_b)$, i.e. $q'={u,d,s,c}$. The operators $O_7,\ldots,O_{10}$ which arise from the electroweak-penguin diagrams are given by
 \begin{eqnarray}
 \nonumber O_7=\frac{3}{2} \bar{s}_{\alpha}\gamma^{\mu}L b_{\alpha}\cdot\sum\limits_{q'}e_{q'}\bar{q}'_{\beta}\gamma_{\mu}R q'_{\beta}, ~~~~~
 &&O_8=\frac{3}{2} \bar{s}_{\alpha}\gamma^{\mu}L b_{\beta}\cdot\sum\limits_{q'}e_{q'}\bar{q}'_{\beta}\gamma_{\mu}R q'_{\alpha},\\[8pt]
 O_9=\frac{3}{2} \bar{s}_{\alpha}\gamma^{\mu}L b_{\alpha}\cdot\sum\limits_{q'}e_{q'}\bar{q}'_{\beta}\gamma_{\mu}L q'_{\beta}, ~~~~~
 &&O_{10}=\frac{3}{2} \bar{s}_{\alpha}\gamma^{\mu}L b_{\beta}\cdot\sum\limits_{q'}e_{q'}\bar{q}'_{\beta}\gamma_{\mu}L q'_{\alpha},
 \end{eqnarray}
The last two operators $O_{7\gamma}$ and $O_{8g}$ are
 \begin{eqnarray}
 O_{7\gamma}=\frac{-e}{8\pi^2}m_b\bar{s}\sigma_{\mu\nu}(1+\gamma_5)F^{\mu\nu}b, ~~~~~
 &&O_{8g}=\frac{-g_s}{8\pi^2}m_b\bar{s}\sigma^{\mu\nu}RG^{\mu\nu}b.
 \end{eqnarray}
where $G^{\mu\nu}$ denotes the gluon field strength tensor. The $O_{7\gamma}$ and $O_{8g}$ are
the electromagnetic and chromomagnetic dipole operators, respectively.

In phenomenology, it is more convenient to use the coefficients $a_i$ which are obtained from the Wilson coefficients $C_j$. Without QCD corrections, $a_i$ are given by
 \beq
 a_i=C_i+\frac{1}{N_c}C_{i+1}~~(i={\rm odd});  \qquad
 a_i=\frac{1}{N_c}C_{i-1}+C_i~~(i={\rm even}).
 \eeq
where $i=1,...,10$. With QCD corrections, all the dynamical information is encoded in coefficients $a_i$.

\subsection{The QCD factorization approach}

For the nonleptonic decays, there are at least three hadrons in one system. How to calculate the hadronic matrix elements of the local operators given in the effective Hamilatonian is a notorious difficult problem. The factorization hypothesis is proposed to simplify the hadronic matrix elements. The original idea is called by the naive factorization \cite{Bauer:1986bm}. Take the $B\to M_1M_2$ decay as an example. The recoiled $M_1$ denotes the meson which picks up the light spectator quark. Another meson $M_2$ is called the emitted meson which is created from one current. The assumption of factorization is that the emitted $M_2$ decouple from the remained $BM_1$ system. This assumption corresponds to vacuum insertion approximation. Under this approximation, the three meson matrix element is simplified into product of a decay constant and form factor. The naive factorization is tested to work well for the color-allowed tree dominated processes. But it fails to explain the color-suppressed and penguin dominated processes. In these processes, the non-factorizable QCD corrections between $M_2$ and $BM_1$ are important. The generalized factorization approach solves the renormalization scale and scheme dependence problem in the naive factorization \cite{Ali:1998eb}. But it is not a systematic method because it introduces a phenomenological color number to account for the non-factorizable contributions.

The QCD factorization approach is a rigourous theoretical method within which the non-factorizable QCD corrections can be systematically calculated \cite{Beneke:1999br,Beneke:2000ry,Beneke:2001ev,Beneke:2003zv}. It states that in the heavy quark limit, the transition matrix element of an operator $O_i$ in the weak decays $B\to M_1M_2$ can be
factorized into a convolution of hard scattering kernel and meson distribution amplitude as
 \beq
 \langle M_1M_2|O_i|B\rangle&=&\sum\limits_{j}F_j^{BM_1}(m_2^2)\int_0^1 dx ~T^I_{ij}(x)\Phi_{M_2}(x)
  +(M_1\leftrightarrow M_2) \non\\
  &&+\int_0^1 d\xi dxdy ~T^{II}(\xi,x,y)\Phi_B(\xi)\Phi_{M_1}(y)\Phi_{M_2}(x).
 \eeq
The term in the second line is the hard spectator scattering contribution. When $M_1$ is heavy and $M_2$ is light, Only the first term in the first line has contribution.  The hard scattering kernels $T^I$ and $T^{II}$ can be perturbatively calculated order by order in $\alpha_s$. The $\Phi_M(x)$ is the meson light-cone distribution amplitude which is universal and process independent. In QCDF, the factorization means the separation of perturbative contribution from the non-perturbative part. It is proved that the factorization is valid for final states containing two light mesons or the case with one heavy and one light mesons.

Under the diquark approximation, a baryon is similar to the meson. This similarity makes the application of QCDF into the heavy baryon decays possible. But one need to be cautious about the hard spectator scattering. When a hard gluon interacts with a diquark, the loosely bounded diquark may be broken and the diquark approximation is invalid. This case occurs for a light final baryon, such as $p$ where the two quarks in the diquark are both energetic. In this case, one has to return to the three-quark picture and use the  perturbative method, e.g. \cite{Lu:2009cm}. However, the interactions with two hard gluon exchanges are suppressed by $\alpha_s^2$.  Another possibility is that the diquark remains unbroken and it interacts with the hard gluons like a point particle. As we know, the diquark is not a fundamental particle. One needs to introduce a form factor to compensate for its structure. The form factor can not be calculated from first principles. A decay constant for a baryon is also required to be introduced. Due to these technical difficulty and the theory uncertainties, we will not consider the hard spectator scattering in this study.

Without the hard spectator interaction contribution, QCD factorization can be extended to the $\Lambda^0_b\rightarrow H + M$ decays when the emitted meson $M$ is light. In the rest frame of $\Lambda^0_b$, the light meson is energetic. It is a compact object and has small transverse size. The soft gluons decouple from the light meson $M$. This is statement of color transparency \cite{Bjorken:1988kk}. The $\Lambda^0_b\rightarrow H$
transitions are soft dominated and the form factors are evaluated in the covariant light-front quark model.
The QCD interactions between $M$ and $\Lambda^0_b H$ are mediated by the hard gluon exchange and perturbatively calculable. Thus, we have a factorized form for the the decay $\Lambda^0_b\rightarrow H + M$ as
 \beq
 \langle H M|O_i|\Lambda^0_b\rangle&=&\sum\limits_{j}F_j^{\Lambda_b H}(M^2)\int_0^1 dx ~T^I_{ij}(x)\Phi_{M}(x).
 \eeq
where $F_j^{\Lambda_b H}$ denote the $\Lambda^0_b\rightarrow H$ form factors and $\Phi_{M}(x)$ is the light-cone distribution amplitude of the meson $M$.

\begin{figure}[!ht]
\centering
\includegraphics[scale=1.0]{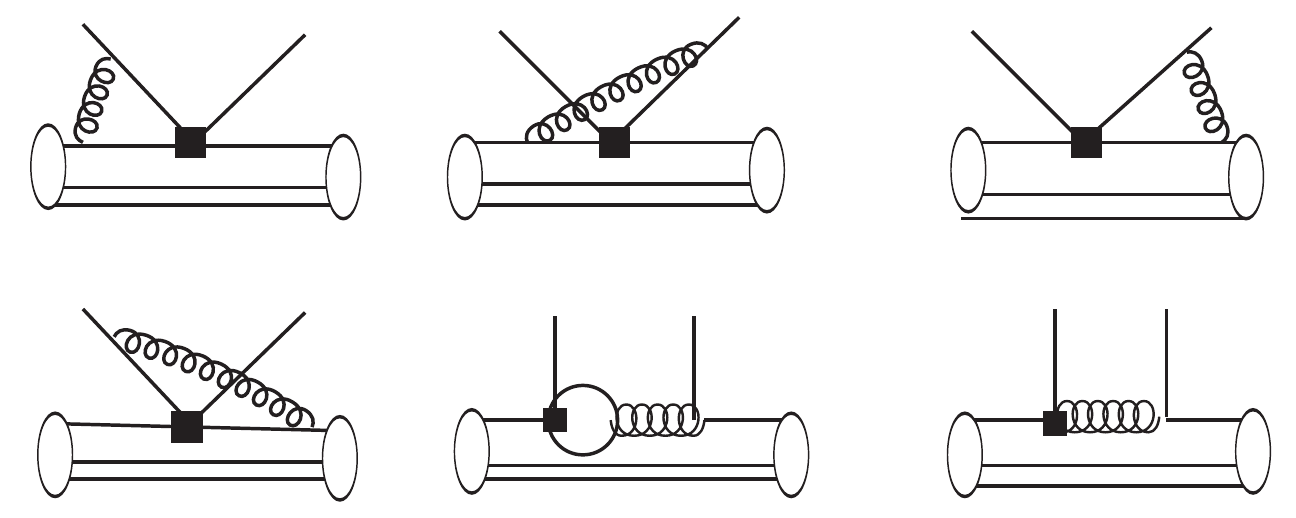}
\caption{Feynman diagrams in the QCD factorization approach.}\label{fig:QCDF}
\end{figure}

\begin{table}
\caption{Numerical values of the coefficients $a_i$.}\label{tab:ai}
\begin{center}
\begin{tabular}{ |c |c | c |c|}\hline\hline
~~~~$a_i$~~~~ &$\mu=m_b/2$                            &$\mu=m_b$ &$\mu=2m_b$\\\hline
$a_1$      &$1.096+0.037i$                            &$1.067+0.020i$                              &$1.046+0.011i$\\
$a_2$      &$0.200-0.114i$                            &$0.200-0.084i$                              &$0.200-0.067i$\\
$a_3$      &$(9.293+3.665i)\times10^{-3}$             &$(7.007+2.041i)\times10^{-3}$
 &$(5.044+1.187i)\times10^{-3}$ \\
$a^u_4$    &$(-2.157-2.059i)\times10^{-2}$            &$(-2.290-1.623i)\times10^{-2}$
 &$(-2.290-1.350i)\times10^{-2}$ \\
$a^c_4$    &~~$(-2.949-0.924i)\times10^{-2}~~$        &~~$(-2.875-0.785i)\times10^{-2}~~$
 &~~$(-2.755-0.684i)\times10^{-2}¡¤~$ \\
$a_5$      &$(-6.681-5.112i)\times10^{-3}$            &$(-5.106-2.570i)\times10^{-3}$
 &$(-3.494-1.374i)\times10^{-3}$ \\
$a^u_6$    &$(-4.611-1.891i)\times10^{-2}$            &$(-3.561-1.535i)\times10^{-2}$
 &$(-2.974-1.301i)\times10^{-2}$ \\
$a^c_6$    &$(-5.069-0.685i)\times10^{-2}$            &$(-3.899-0.644i)\times10^{-2}$
 &$(-3.243-0.593i)\times10^{-2}$ \\
$a_7$      &$(1.58+3.17i)\times10^{-5}$             &$(7.43+1.60i)\times10^{-5}$     &$1.91\times10^{-4}$ \\
$a^u_8$    &$3.98\times10^{-4}$  &$(2.62-0.56i)\times10^{-4}$     &$(1.59-0.96i)\times10^{-4}$  \\
$a^c_8$    &$3.98\times10^{-4}$  &$(2.52-0.30i)\times10^{-4}$     &$(1.40-0.50i)\times10^{-4}$ \\
$a_9$      &$(-9.21-0.29i)\times10^{-3}$  &$(-8.93-0.16i)\times10^{-3}$  &$(-8.63+0.09i)\times 10^{-3}$\\
$a^u_{10}$ &$(1.06+0.95i)\times10^{-3}$   &$(5.99+6.48i)\times10^{-4}$   &$(1.62+4.57i)\times10^{-4}$ \\
$a^c_{10}$ &$(1.06+0.95i)\times10^{-3}$   &$(5.82+6.73i)\times10^{-4} $  &$(1.32+4.50i)\times10^{-4}$ \\\hline\hline
\end{tabular}
\end{center}
\end{table}

At the $\alpha_s$ order, the QCD corrections can be shown in Fig. \ref{fig:QCDF}. The four diagrams (the three in the first line and the first one in the second line) are vertex corrections. The second diagram in the second line is penguin diagram and the third diagram is the chromomagnetic dipole diagram. Their formulations are presented in Appendix B. All the QCD corrections are included in the coefficients $a_i$ which are obtained from the Wilson coefficients $C_j$ given in the effective Hamiltonian. The coefficients $a_i$ is calculated up to $\alpha_s$ order, including the one-loop vertex corrections and penguin contributions. The terms of $a_6$ and $a_8$ contains the chirally enhanced twist-3 contributions since they are numerically important. For the other coefficients $a_i$, only the leading twist contributions are considered and the asymptotic form of the twist-2 meson distribution amplitude is adopted. About the coefficient $a_2$, its value is small considering the vertex corrections and penguin contributions. It is insufficient to explain the experimental data for the color suppressed processes. The hard spectator scattering contribution is important for the coefficient $a_2$. After taking the hard spectator scattering contribution into account, the real part of $a_2$ is $0.2$ and nearly independent of the renormalization scale $\mu$ \cite{Beneke:2001ev}. We use this value to partly compensate the neglected hard spectator scattering contributions. The numerical results for the coefficients $a_i$ are given in Table \ref{tab:ai}.

When the meson in $\Lambda^0_b\rightarrow H + M$ decays is heavy, such as $D$ or $D^*$, the color transparency argument is not valid. The QCD factorization is considered to be inapplicable for this type processes. According to this criteria, about half of the 44 processes can not be analyzed. In order to study these processes, we prefer to adopt a more phenomenological point of view at the cost of losing some theoretical rigorousness. Assuming $m_c\ll m_b$ so that $D$ and $D^*$ mesons are considered to be light. Under this assumption, the QCDF approach can be applied to all the 44 processes listed in the subsection of Classification. From the previous study \cite{Ke:2007tg},  the naive factorization works very well for the color-allowed processes with two heavy final states. One needs to worry about the color-suppressed processes. We make a crude estimate that the uncertainties caused by the approximation is estimated to be order of $m_c/m_b$, about 30\% at the amplitude level. In \cite{Beneke:2000ry}, the authors calculated $a_2$ in $B\to \pi D$ process. By choosing a very asymmetric distribution amplitude for the $D$ meson, they obtain $a_2\approx 0.22 e^{-i41^{\rm o}}$ which is not far from the value of $a_2$ given in Table \ref{tab:ai}.

About the processes containing the final state of charmonium $\eta_c$ or $J/\psi$, QCD factorization is still applicable due the the small transverse size of the charmonium in the heavy quark limit \cite{Cheng:2000kt}.
A combined coefficient $\bar a_2$ extracted from the experiment data of $B\to J/\psi K$ is $|\bar a_2|_{\rm expt}=0.26$ is close to the value of $a_2$ given in Table \ref{tab:ai}.

\subsection{The decay rate and direct CP asymmetry}

Under the factorization assumption, the transition amplitude of $\Lambda^0_b\rightarrow H M$ can be written generally by
\begin{eqnarray}
\mathcal{M}(\Lambda^0_b\rightarrow H P)&&=\bar{u}_{H}(A+B\gamma_5)u_{\Lambda^0_b},\nonumber\\
\mathcal{M}(\Lambda^0_b\rightarrow H V)&&=\bar{u}_{H}\epsilon^{*\mu}[A_1\gamma_{\mu}\gamma_5+A_2(p_{H})_{\mu}\gamma_5
+B_1\gamma_{\mu}+B_2(p_{H})_{\mu}]u_{\Lambda^0_b},
\end{eqnarray}
with
\begin{eqnarray} \label{AB}
A&=&\lambda \left[(M_{\Lambda^0_b}-M_{H})f_1(M^2)+q^2\frac{f_3(M^2)}{M_{\Lambda^0_b}}\right],
 \nonumber\\
B&=&\lambda \left[(M_{\Lambda^0_b}+M_{H})g_1(M^2)-q^2\frac{g_3(M^2)}{M_{\Lambda^0_b}}\right],
 \nonumber\\
A_1&=&-\lambda~  M\left[g_1(M^2)+g_2(M^2)\frac{M_{\Lambda^0_b}-M_{H}}{M_{\Lambda^0_b}}\right],
\nonumber\\
A_2&=&-2\lambda~ M\frac{g_2(M^2)}{M_{\Lambda^0_b}},\nonumber\\
B_1&=&\lambda~  M\left[f_1(M^2)-f_2(M^2)\frac{M_{\Lambda^0_b}+M_{H}}{M_{\Lambda^0_b}}\right],
\nonumber\\
B_2&=&2\lambda~ M\frac{f_2(M^2)}{M_{\Lambda^0_b}},
\end{eqnarray}
where $M$ represents the meson mass and $q^2=M^2$. The function $\lambda$ is an essential quantity in the decay amplitude. Note that the function $\lambda$ given here is different from the Wolfenstein parameter $\lambda$ in the CKM elements. In order to avoid confusion, we change the Wolfenstein parameter $\lambda$ to
$\lambda_W$. Except the baryon-to-baryon form factors, all the other quantities, such as the meson decay constant, Fermi constant, CKM matrix elements, Wilson coefficients, the non-factorizable corrections are contained in $\lambda$. The explicit forms of $\lambda$ for different processes are collected in the Appendix \ref{function}.

The decay rates of $\Lambda^0_b\rightarrow H P$ and the up-down asymmetries are
\begin{eqnarray}
 \Gamma&=&\frac{p_c}{8\pi}\left[\frac{(M_{\Lambda^0_b}+M_{H})^2-M^2}{M^2_{\Lambda^0_b}}\mid A\mid^2+\frac{(M_{\Lambda^0_b}-M_{H})^2-M^2}{M^2_{\Lambda^0_b}}\mid B\mid^2\right],\nonumber\\
 \alpha&=&-\frac{2\kappa \text{Re}(A^{*}B)}{\mid A\mid^2+\kappa^2\mid B\mid^2}.
\end{eqnarray}
where $p_c$ is the momentum of the final baryon $H$ in the rest frame of $\Lambda^0_b$ and $\kappa=\frac{p_c}{E_{H}+M_{H}}$. For $\Lambda^0_b\rightarrow H V$ decays, the decay rates and up-down asymmetries are
\begin{eqnarray}
\Gamma&=&\frac{p_c(E_{H}+M_{H})}{8\pi M_{\Lambda^0_b}}\left[2\left(\mid S\mid^2+\mid P_2\mid^2\right)+\frac{E^2}{M^2}\left(\mid S+D\mid^2+\mid P_1\mid^2\right)\right],\nonumber\\
\alpha&=&\frac{4M^2 \text{Re}(S^*P_2)+2E^2\text{Re}(S+D)^*P_1}{2M^2(\mid S\mid^2+\mid P_2\mid^2)+E^2(\mid S+D\mid^2+\mid P_1\mid^2)},
\end{eqnarray}
where $E$ is the energy of the vector meson, and
\begin{eqnarray}
S&=&-A_1,\nonumber\\
P_1&=&-\frac{p_c}{E}\left(\frac{M_{\Lambda^0_b}+M_H}{E_H+M_H}B_1+B_2\right),\nonumber\\
P_2&=&\frac{p_c}{E_H+M_H}B_1,\nonumber\\
D&=&-\frac{p^2_c}{E(E_H+M_H)}(A_1-A_2).
\end{eqnarray}

The direct CP asymmetry of decay $\Lambda_b^0\rightarrow HM$ is defined by
 \beq
 A_{CP}\equiv\frac{{\cal B}(\Lambda_b^0\rightarrow HM)-{\cal B}(\bar \Lambda_b^0\rightarrow \bar H~ \bar M)}
 {{\cal B}(\Lambda_b^0\rightarrow HM)+{\cal B}(\bar \Lambda_b^0\rightarrow \bar H~ \bar M)}.
 \eeq
At the quark level, the CP violation is represented by $b$ quark decay rate minus the $\bar b$ anti-quark which follows the standard convention. In order to produce CP violation, it requires both the weak and strong phase differences.
Only the tree diagram contribution cannot satisfy the condition. Usually, the direct CP asymmetry arises from the interference of tree and penguin contributions. It is also possible for the processes which contain pure penguin contributions. This is due to the interference between the virtual $u$ and $c$ quark exchanges in the penguin loop diagrams.

The weak phases are contained in the CKM matrix elements. The strong phases come from the the diagrams where the virtual quarks or gluons become on-shell. In QCDF approach, it has two origins: (1) In the penguin contributions, the quark-antiquark loop produces an imaginary part. This is usually called the BSS mechanism \cite{Bander:1979px}.
(2) In the vertex corrections, the hard gluon exchange between the final two hadrons can also produces an imaginary part. These two origins of strong phase are perturbative.

\subsection{Chirally enhanced contributions}

When the final meson is a pseudoscalar, the penguin operators from $O_5$ to $O_8$ with (V+A) current will give non-zero contributions. We take the process of $\Lambda_b^0\to p\pi^-$ as an example to illustrate. Considering the operator $O_5$, the matrix element is
 \beq \label{chiral}
 &&\langle p\pi^-|(\bar d b)_{\rm V-A}(\bar u u)_{\rm V+A} |\Lambda_b^0\rangle \non\\
 &&=(-2)\langle p\pi^-|\bar d_\alpha(1+\gamma_5)u_\beta \bar u_\beta(1-\gamma_5)b_\alpha|
  \Lambda_b^0\rangle \non\\
 &&=\frac{1}{N_c}R_{\pi}\langle \pi^-|(\bar d u)_{\rm V-A}|0\rangle \langle p|
  (\bar u b)_{\rm V+A}|\Lambda_b^0\rangle,
 \eeq
where
 \beq
 R_\pi=\frac{2m_\pi^2}{m_b(m_d+m_u)}.
 \eeq
In the above equation, we have used the Fierz transformation, factoriztion and the equations of motion.
From the power counting, the operator $O_5$ contribution belongs to power correction in $1/m_b$. However, the small masses of the $u,d$ current quarks make the factor $R_\pi$ numerically large, and $R_\pi$ is nearly about $1$ for the realistic $b$ quark mass. So, this term is usually called the "chirally enhanced" contribution. It is important in the penguin dominated processes. We include this term in the calculations.

The occurrence of (V+A) current in the matrix element of Eq. (\ref{chiral}) causes one complication which is special for the baryon decay. For the meson case, only the vector current contribute to $B\to P$ transition form factor and only the axial-vector current contribute to $B\to V$ transition (the vector current part vanishes when couples to the pseudoscalar momentum). The (V+A) current can be changed to (V-A) current and relative minus sign is required for $B\to PP$ and $B\to VP$. In particular, for $\bar B^0\to \pi^+\pi^-$ and $\bar B^0\to \rho^+\pi^-$, they have the same quark component.  Their decay amplitudes are
 \beq \label{eq:pipi}
 M(\bar B^0\to \pi^+\pi^-)&=&-i\frac{G_F}{\sqrt{2}}f_{\pi}F_0^{B\pi}(m_\pi^2)(m_B^2-m_\pi^2)
  \left[V_{ub}V^{\ast}_{ud}a_1+V_{ub}V^{\ast}_{ud}(a_4^u+a_{10}^u)\right. \non\\
 &&+V_{cb}V^{\ast}_{cd}(a_4^c+a_{10}^c)+R_{\pi}\left.(V_{ub}V^{\ast}_{ud}(a_6^u+a_{8}^u)
  +V_{cb}V^{\ast}_{cd}(a_6^c+a_{8}^c))\right],
 \eeq
and
 \beq \label{eq:rhopi}
 M(\bar B^0\to \rho^+\pi^-)&=&\sqrt{2}G_F f_{\pi}A_0^{B\rho}(m_\pi^2)m_\rho(\epsilon\cdot p_\pi)
  \left[V_{ub}V^{\ast}_{ud}a_1+V_{ub}V^{\ast}_{ud}(a_4^u+a_{10}^u)\right. \non\\
 &&+V_{cb}V^{\ast}_{cd}(a_4^c+a_{10}^c)-R_{\pi}\left.(V_{ub}V^{\ast}_{ud}(a_6^u+a_{8}^u)
  +V_{cb}V^{\ast}_{cd}(a_6^c+a_{8}^c))\right].
 \eeq
One can see that the $a_6$ and $a_8$ contributions in $\bar B^0\to \pi^+\pi^-$ and $\bar B^0\to \rho^+\pi^-$ decays are opposite in sign. Neglecting the small difference in the Wilson coefficients $a_i^u$ and $a_i^c$ and using the unitarity of the CKM matrix elements, the above formulae are same as the expressions given in \cite{Ali:1998eb}.

But for baryon case, the vector and axial-vector currents both contribute to the baryon-to-baryon form factors. The operators $O_{5-8}$ contribute to (V-A)$\otimes$(V+A) while other operators contribute to (V-A)$\otimes$(V-A). These two contributions from different types of current have to be treated differently. Our method is to divide the vector current and axial vector current parts and absorb them into $A$ and $B$ terms of the Eq. (\ref{AB}). Here, we give formulae of the $\lambda$ function in $\Lambda^0_b\rightarrow p \pi^-$ process. For the other processes, their forms are collected in the Appendix \ref{function}. In $\Lambda^0_b\rightarrow p \pi^-$ process, the $\lambda$ function for A term is:
\begin{eqnarray} \label{eq:ppiA}
\lambda&=&\frac{G_F}{\sqrt{2}}f_{\pi}\left[V_{ub}V^{\ast}_{ud}a_1+V_{ub}V^{\ast}_{ud}
(a_4^u+a_{10}^u)+V_{cb}V^{\ast}_{cd}(a_4^c+a_{10}^c)\right. \non\\
&&+R_{\pi}\left.(V_{ub}V^{\ast}_{ud}(a_6^u+a_{8}^u)+V_{cb}V^{\ast}_{cd}(a_6^c+a_{8}^c))\right],
\end{eqnarray}
and for B term is:
\begin{eqnarray} \label{eq:ppiB}
\lambda&=&\frac{G_F}{\sqrt{2}}f_{\pi}\left[V_{ub}V^{\ast}_{ud}a_1+V_{ub}V^{\ast}_{ud}
(a_4^u+a_{10}^u)+V_{cb}V^{\ast}_{cd}(a_4^c+a_{10}^c)\right. \non \\
&&-R_{\pi}\left.(V_{ub}V^{\ast}_{ud}(a_6^u+a_{8}^u)+V_{cb}V^{\ast}_{cd}(a_6^c+a_{8}^c))\right].
\end{eqnarray}
There is only one difference: a relative minus sign for $a_6$ and $a_8$ contributions in A and B terms. We find a relation: the term in square bracket of Eq. (\ref{eq:pipi}) is same as the corresponding one of Eq. (\ref{eq:ppiA}); and the term in square bracket of Eq. (\ref{eq:rhopi}) is same as the corresponding one of Eq. (\ref{eq:ppiB}). The complication caused by the (V-A)$\otimes$(V+A) current structure is one difference between the baryon and meson. The authors in \cite{Lu:2009cm} observed this phenomenon earlier. While this point is not realized in the previous work \cite{Zhu:2016bra}. We correct this error in this study.

\subsection{ Similarity of meson and baryon }

Under the diquark approximation, the baryon is similar to a meson. We may use this similarity to obtain some information for the $\Lambda_b^0$ decays by using the correponding $B$ meson decays. Consider $\Lambda^0_b\rightarrow \Lambda\phi$ decay as an example. If we change the diquark $[ud]$ by a antiquark $\bar d$, we have the meson decay $\bar{B}^0\rightarrow \bar{K}^0 \phi$. If the meson-baryon similarity is rigorous, we expect that the two processes have the same QCD dynamics at the quark level. We prove this assumption below.

The decay amplitude of the process $\bar{B}^0\rightarrow \bar{K}^0 \phi$ is written by
 \beq \label{eq:phi1}
 M(\bar{B}^0\rightarrow \bar{K}^0 \phi)&=&\sqrt{2}G_Ff_{\phi}F_1^{B K}
 (m_{\phi}^2)m_{\phi}(\epsilon\cdot p_K)\left[V_{ub}V^{\ast}_{us}\left(a_3+a^u_4+a_5-\frac{1}{2}a_7
 \right.\right.\non\\
 &&\left.-\frac{1}{2}a_9-\frac{1}{2}a^u_{10}\right)
 +\left.V_{cb}V^{\ast}_{cs}\left(a_3+a^c_4+a_5-\frac{1}{2}a_7-\frac{1}{2}a_9
 -\frac{1}{2}a^c_{10}\right)\right]\non\\
 &=&-\sqrt{2}G_Ff_{\phi}F_1^{B K}
 (m_{\phi}^2)m_{\phi}(\epsilon\cdot p_K)V_{tb}V^{\ast}_{ts}~\bar a,
 \eeq
where the factor $\bar a$ is
 \beq
 \bar a&=&\frac{-1}{V_{tb}V^{\ast}_{ts}}\left[V_{ub}V^{\ast}_{us}\left(a_3+a^u_4+a_5-\frac{1}{2}a_7
 -\frac{1}{2}a_9-\frac{1}{2}a^u_{10}\right)\right.\non\\
 &&~~~~~~~+\left.V_{cb}V^{\ast}_{cs}\left(a_3+a^c_4+a_5-\frac{1}{2}a_7-\frac{1}{2}a_9
 -\frac{1}{2}a^c_{10}\right)\right].
 \eeq
The $\bar a$ is a combined coefficient where all the QCD corrections are included. In fact, $\bar a$ can be simplified into a familiar form. Neglecting the difference of $a_i^u$ and $a_i^c$, and using the unitarity relation $V_{ub}V^{\ast}_{us}+V_{cb}V^{\ast}_{cs}=-V_{tb}V^{\ast}_{ts}$, the factor $\bar a$ can be rewritten
by
 \beq
 \bar a=a_3+a_4+a_5-\frac{1}{2}(a_7+a_9+a_{10}).
 \eeq
With this $\bar a$, the formula of Eq. (\ref{eq:phi1}) reproduces the result in \cite{Ali:1998eb}.

For the $\Lambda^0_b\rightarrow \Lambda\phi$ decay, what we need is the $\lambda$ function. It is
 \beq\label{eq:phi2}
 \lambda&=&\frac{G_F}{\sqrt{2}}f_{\phi}\left[V_{ub}V^{\ast}_{us}\left(a_3+a^u_4+a_5-\frac{1}{2}a_7
 -\frac{1}{2}a_9-\frac{1}{2}a^u_{10}\right)\right.\non\\
 &&+\left.V_{cb}V^{\ast}_{cs}\left(a_3+a^c_4+a_5-\frac{1}{2}a_7-\frac{1}{2}a_9
 -\frac{1}{2}a^c_{10}\right)\right]\non\\
 &=&-\frac{G_F}{\sqrt{2}}f_{\phi}V_{tb}V^{\ast}_{ts}~ \bar a.
 \eeq

Comparing the Eqs. (\ref{eq:phi1}) and (\ref{eq:phi2}), we find that the baryon and meson decay amplitudes have the same factor $\bar a$. That means,
 \beq \label{eq:relation}
 \bar a(\Lambda^0_b\rightarrow \Lambda\phi)=\bar a(\bar{B}^0\rightarrow \bar{K}^0 \phi).
 \eeq
Since $\bar a$ encodes the QCD dynamics, we can say that the baryon and meson decays have the same QCD dynamics at the quark level.  This is a rigorous relation obtained from the meson-baryon similarity.

The meson-baryon similarity has important meaning and applications. The calculation of the QCD dynamics in $\Lambda_b^0$ decays depends on the theory approach and contains large hadron uncertainties. At present, the B meson data is very precise. The meson-baryon similarity permits us to give a model-independent prediction. In particular, we can extract $\bar a$ from the data of the meson decay $\bar{B}^0\rightarrow \bar{K}^0 \phi$, and then use it to predict the baryon decay $\Lambda^0_b\rightarrow \Lambda\phi$. Using the meson data to predict baryon decay $\Lambda^0_b\rightarrow p K^-$ has been done in \cite{Zhu:2016bra}. It is shown that this model-independent prediction accords with the experiment very well.

\section{Input parameters and numerical results of the form factors}

In this section, we first present the input parameters. Then we use them to calculate the  baryon-to-baryon
transition form factors in the covariant light-front approach.

\subsection{Input parameters}

In the calculations, the baryon masses are $M_{\Lambda^0_b}=5.619\ \text{GeV}$, $M_{\Lambda_c}=2.285\ \text{GeV}$, $M_{\Lambda}=1.116\ \text{GeV}$ and $M_p=0.938\ \text{GeV}$ \cite{Patrignani:2016xqp}.

The quark mass appeared in the light-front quark model is the constituent mass. Its value should be process independent. So we can use the quark masses determined from the meson process. The quark masses are taken from the previous works \cite{Ke:2007tg,Wei:2009np}:
\begin{eqnarray}
m_b=4.4~{\rm GeV}, ~~~~ m_c=1.3~{\rm GeV}, ~~~~ m_s=0.45~{\rm GeV}, ~~~~ m_u=m_d=0.3~{\rm GeV}.
\end{eqnarray}
The $[ud]$ diquark mass is not well determined. From \cite{Cheng:2004cc}, it is assumed that mass of a $[ud]$ diquark is close to the constituent strange quark mass. In the literature, the mass of the constituent light scalar diquark $m_{[ud]}$ is rather arbitrary, ranging from 400-800 MeV. In \cite{Ke:2007tg}, $m_{[ud]}=500~{\rm MeV}$ is fitted from the process of $\Lambda_b^0\to\Lambda_c^+ l^-\bar\nu_l$ when other parameters are fixed. We also use this value for our calculations and adjust it when necessary.

The quark in the QCDF approach and the equations of motion is the current quark, and the mass is current mass. The values for the three light current quarks are
 \beq
 m_u=2.3~{\rm MeV}, \qquad  m_d=4.8~{\rm MeV}, \qquad m_s=95~{\rm MeV}.
 \eeq
For the heavy quark mass, the values are chosen the same as those given in the constituent mass.

The baryon parameter $\beta$ in the Gaussian-type wave function is at the order of the QCD scale $\Lambda_{\rm QCD}$ and needs to be specified. For the meson case, the parameter $\beta$ can be determined from the decay constant which is measured by experiment. But this method cannot be applied to the baryon. The flavor symmetry
can provide some helpful relations. In the heavy quark limit, the heavy quark symmetry gives  $\beta_{\Lambda_b}=\beta_{\Lambda_c}$. From the light quark SU(3) symmetry, $\beta_{\Lambda}=\beta_p$. Isospin symmetry gives $\beta_p=\beta_n$. The $\beta$ parameters are determined by fitting the theory prediction to the data. For example, the parameters $\beta_{\Lambda_b}$ and $\beta_{\Lambda_c}$ are fixed by data of $\Lambda_b^0\rightarrow \Lambda_c^+ l^-\bar\nu_l$ and $\Lambda_b^0\rightarrow \Lambda_c^+\pi^-$ processes. From these two process, the $\beta_{\Lambda_b}$ and $\beta_{\Lambda_c}$ are chosen to be $\beta_{\Lambda_b}=0.40$ GeV and $\beta_{\Lambda_c}=0.34$ GeV. The value of $\beta_{\Lambda_c}$ is slightly smaller than $\beta_{\Lambda_b}$.
The proton parameter $\beta_p$ is fixed from $\Lambda_b^0\rightarrow p l^-\bar\nu_l$ process. The fitted value is $\beta_p=0.38$ GeV. The values of $\beta_p$ is nearly equal to $\beta_{\Lambda_b}$. The choice of a large value for $\beta_p=0.38$ GeV is forced by the experimental data. The previous chosen $\beta_p=0.3$ GeV in \cite{Wei:2009np} gives predictions of ${\cal B}(\Lambda_b^0\rightarrow p l^-\bar\nu_l)=2.54\times 10^{-4}$ and ${\cal B}(\Lambda_b^0\rightarrow p\pi^-)=3.15\times 10^{-6}$. These predictions are insufficient to explain the present data of ${\cal B}(\Lambda_b^0\rightarrow p \mu^-\bar\nu_\mu)=(4.1\pm 1.0)\times 10^{-4}$ and ${\cal B}(\Lambda_b^0\rightarrow p\pi^-)=(4.2\pm 0.8)\times 10^{-6}$.  So we have to choose a large value for $\beta_p$. The leptonic decay of $\Lambda_b^0\to \Lambda\mu^+\mu^-$ is a flavor-changing-neutral-current process. Its discussion is beyond the scope of this study. So, it is difficult to determine $\beta_{\Lambda}$ from the experiment. We use the light quark SU(3) symmetry relation $\beta_{\Lambda}=\beta_p$ and neglect the SU(3) breaking effect. In fact, the theory results are not sensitive to
the variation of $\beta_{\Lambda}$. Neglecting SU(3) breaking in this case is reasonable. The input parameters of the constituent quark masses and the $\beta$ parameters are collected in Table \ref{tab:mass}.

\begin{table}
\caption{Input parameters in the covariant light-front approach (in units of GeV).}\label{tab:mass}
\begin{ruledtabular}
\begin{tabular}{cccccccccc}
$m_b$ & $m_c$ & $m_s$ & $m_u$ & $m_{[ud]}$ & $\beta_{\Lambda_b}$ & $\beta_{\Lambda_c}$ & $\beta_{\Lambda}$ & $\beta_p$ & $\beta_n$  \\\hline
4.4 & 1.3 & 0.45 & 0.3 & 0.5 & 0.40 & 0.34 & 0.38 & 0.38 & 0.38
\end{tabular}
\end{ruledtabular}
\end{table}

For the $\omega$ and $\phi$ mesons, the ideal mixing is assumed so that the quark component of the two
mesons are $\omega=\frac{1}{\sqrt 2}(u\bar u+d\bar d)$ and $\phi=s\bar s$. For the $\eta$ and $\eta'$ mesons,
both of them require two decay constants.  We adopt the Feldmann-Kroll-Stech scheme \cite{Feldmann:1998vh} for the $\eta-\eta'$ mixing. The mesons $\eta$ and $\eta'$ are superposition of the non-strange and strange flavor bases as
 \beq
 \left(\begin{array}{c} \eta \\  \eta' \end{array}\right)=
 \left(\begin{array}{cc} {\rm cos}\phi & -{\sin}\phi  \\
  {\rm sin}\phi & {\rm cos}\phi  \end{array}\right)
 \left(\begin{array}{c} \eta_n \\ \eta_s \end{array}\right),
 \eeq
where
 \beq
 \eta_n=\frac{u\bar u+d\bar d}{\sqrt 2}=n\bar n, \qquad \eta_s=s\bar s.
 \eeq
The mixing angle $\phi=39.3^{\rm o}\pm 1.0^{\rm o}$. In this mixing scheme, only two decay constants
$f_n~(n=u,d)$ and $f_s$ are needed \cite{Lu:2007sg}:
\begin{eqnarray}
   \langle 0|\bar n\gamma_\mu\gamma_5 n|\eta_n(P)\rangle
   &=& \frac{i}{\sqrt2}\,f_n\,P_\mu \;,\nonumber \\
   \langle 0|\bar s\gamma_\mu\gamma_5 s|\eta_s(P)\rangle
   &=& i f_s\,P_\mu \;.\label{deffq}
\end{eqnarray}
This is based on the assumption that the intrinsic $\bar nn(\bar ss)$ component is absent in the
$\eta_s(\eta_n)$ meson. These decay constants have been determined from the related exclusive processes \cite{Feldmann:1999uf}. Their values are
 \beq
  f_n=(1.07\pm 0.02)f_\pi, \qquad
  f_s=(1.34\pm 0.06)f_\pi.
 \eeq

The decay constants of $\eta$ and $\eta'$ are defined by
 \beq
 &&\langle 0|\bar u\gamma_\mu \gamma_5 u|\eta(P)\rangle=if_\eta^u P_\mu, \qquad
 ~\langle 0|\bar s\gamma_\mu \gamma_5 s|\eta(P)\rangle=if_\eta^s P_\mu, \non\\
 &&\langle 0|\bar u\gamma_\mu \gamma_5 u|\eta'(P)\rangle=if_{\eta'}^u P_\mu, \qquad
 \langle 0|\bar s\gamma_\mu \gamma_5 s|\eta'(P)\rangle=if_{\eta'}^s P_\mu.
 \eeq
Then, we have
 \beq
 &&f_{\eta}^u=f_{\eta}^d=54~{\rm MeV}, \qquad  ~f_{\eta}^s=-111~{\rm MeV},  \non\\
 &&f_{\eta'}^u=f_{\eta'}^d=44~{\rm MeV}, \qquad  f_{\eta'}^s=136~{\rm MeV}.
 \eeq

\begin{table}
\caption{Meson decay constants  $f_M$ (in units of MeV). }\label{tab:fp}
\begin{ruledtabular}
\begin{tabular}{ccccccccc}
  Meson & $\pi$ & $\rho$ & $K$  & $K^{*}$ & $D$  & $D^{*}$ & $D_s$  & $D_s^{*}$ \\ \hline
  $f$   & 131   & 216    & 160  & 210     & 200  & 220     & 230    & 230 \\ \hline\hline
  Meson & $\omega$ & $\phi$ & $\eta^u$ & $\eta^s$ & $\eta'^u$ & $\eta'^s$ & $\eta_c$ & $J/\psi$  \\ \hline
  $f$   & 195      & 233    & 54       & -111     & 44        & 136       &   335    & 395
\end{tabular}
\end{ruledtabular}
\end{table}

The meson decay constants used in this study are collected in the Table \ref{tab:fp}. The $\eta_c$ decay constant is taken from \cite{Hwang:2006cua,Yang:2009kq}.

The CKM matrix elements are taken from \cite{Patrignani:2016xqp}
\begin{eqnarray}
&&V_{ud}=1-\lambda_W^2/2, \qquad  \qquad ~~ V_{us}=\lambda_W, \qquad ~~~ \qquad V_{ub}=A\lambda_W^3(\rho-i\eta),  \non \\
&&V_{cd}=-\lambda_W, \qquad \qquad \qquad ~~~~ V_{cs}=1-\lambda_W^2/2, \qquad V_{cb}=A\lambda_W^2,  \non \\
&&V_{td}=A\lambda_W^3(1-\rho-i\eta),\qquad  V_{ts}=-A\lambda_W^2, \qquad ~~~ V_{tb}=1.
\end{eqnarray}
where the Wolfenstein parameters are $\lambda_W=0.225$, $A=0.823$, $\rho=0.141$ and $\eta=0.349$. Here we use the symbol $\lambda_W$ to replace the familiar form $\lambda$ in order to avoid confusion with the $\lambda$ function given in the decay amplitude.

\subsection{Numerical results for the form factors}

The form factors are evaluated in the frame $q^+=0$ where $q^2\leq 0$. The calculated form factors are in the space-like momentum region. In order to obtain the physical form factors, we need an analytic extrapolation from the space-like to the time-like region. Following \cite{Wei:2009np}, the form factors are parameterized in a three-parameter form as
\begin{eqnarray}
F(q^2)=\frac{r_1}{(1-\frac{q^2}{M^2_{\text{fit}}})}+\frac{r_2}{(1-\frac{q^2}{M^2_{\text{fit}}})^2}
\end{eqnarray}
where $F$ represents the form factors $f_{1,2,3}$ and $g_{1,2,3}$. The parameters $r_1$, $r_2$, and $M_{\text{fit}}$ are fixed by performing a three-parameter fit to the form factors in the space-like region and then extrapolate to the physical regions. Because there is no singularity for the obtained form factors at $q^2<M^2_{\Lambda_b}$, the analytic extrapolation is reasonable. The fitted values of $r_1$, $r_2$, and $M_{\text{fit}}$  for different form factors $f_{1,2,3}$ and $g_{1,2,3}$ are given in Tables \ref{tab:ff1}, \ref{tab:ff2}, \ref{tab:ff3} and \ref{tab:ff4} .

\begin{table}
\caption{The $\Lambda_b \rightarrow \Lambda_c$ form factors in the covariant light-front approach.}
\begin{center}\label{tab:ff1}
\begin{tabular}{c@{\hspace{60pt}}c@{\hspace{60pt}}c@{\hspace{60pt}}c@{\hspace{60pt}}c}\hline\hline
 $F$   & $r_1$ & $r_2$  & $M_{\text{fit}}$(GeV)&F(0) \\ \hline
 $f_1$ & -3.22 & 3.72   & 13.9 & 0.500 \\
 $f_2$ & 0.736 & -0.834 & 13.9 &-0.098 \\
 $f_3$ & 0.063 & -0.071 & 13.9 &-0.009 \\
 $g_1$ & -3.30 & 3.82   & 13.9 &0.509  \\
 $g_2$ & 0.131 & -0.146 & 13.9 &-0.015 \\
 $g_3$ & 0.573 & -0.657 & 13.9 &-0.085 \\ \hline\hline
\end{tabular}
\end{center}
\end{table}

\begin{table}
\caption{The $\Lambda_b \rightarrow p$ form factors in the covariant light-front approach.}
\begin{center}\label{tab:ff2}
\begin{tabular}{c@{\hspace{60pt}}c@{\hspace{60pt}}c@{\hspace{60pt}}c@{\hspace{60pt}}c}\hline\hline
$F$&$r_1$&$r_2$&$M_{\text{fit}}$(GeV)&F(0)\\\hline
$f_1$ &-0.078 &0.206   &6.0 &0.128\\
$f_2$ &0.055  &-0.110  &6.0 &-0.056\\
$f_3$ &0.036  &-0.073 &6.0 &-0.037\\
$g_1$ &-0.078 &0.207   &6.0 &0.129\\
$g_2$ &0.032  &-0.065 &6.0 &-0.033\\
$g_3$ &0.086  &-0.121  &6.0 &-0.062\\\hline\hline
\end{tabular}
\end{center}
\end{table}

\begin{table}
\caption{The $\Lambda_b \rightarrow \Lambda$ form factors in the covariant light-front approach.}
\begin{center}\label{tab:ff3}
\begin{tabular}{c@{\hspace{60pt}}c@{\hspace{60pt}}c@{\hspace{60pt}}c@{\hspace{60pt}}c}\hline\hline
$F$&$r_1$&$r_2$&$M_{\text{fit}}$(GeV)&F(0)\\\hline
$f_1$ &-0.091 &0.222   &6.2 &0.131\\
$f_2$ &0.051  &-0.098 &6.2 &-0.048\\
$f_3$ &0.028  &-0.055 &6.2 &-0.027\\
$g_1$ &-0.092 &0.224   &6.2 &0.132\\
$g_2$ &0.026  &-0.050 &6.2 &-0.023\\
$g_3$ &0.053  &-0.105  &6.2 &-0.052\\\hline\hline
\end{tabular}
\end{center}
\end{table}

\begin{table}
\caption{The $\Lambda_b \rightarrow n$ form factors in the covariant light-front approach.}
\begin{center}\label{tab:ff4}
\begin{tabular}{c@{\hspace{60pt}}c@{\hspace{60pt}}c@{\hspace{60pt}}c@{\hspace{60pt}}c}\hline\hline
$F$&$r_1$&$r_2$&$M_{\text{fit}}$(GeV)&F(0)\\\hline
$f_1$ &-0.078 &0.207   &6.0 &0.128\\
$f_2$ &0.055  &-0.110  &6.0 &-0.056\\
$f_3$ &0.036  &-0.073  &6.0 &-0.037\\
$g_1$ &-0.078 &0.207   &6.0 &0.129\\
$g_2$ &0.032  &-0.065  &6.0 &-0.033\\
$g_3$ &0.059  &-0.121  &6.0 &-0.062\\\hline\hline
\end{tabular}
\end{center}
\end{table}

For the heavy-to-heavy transitions $\Lambda_b \rightarrow \Lambda_c$, the numerical results of the form factors are presented in Table \ref{tab:ff1}. The form factors $f_1, ~g_1$ are positive and of the order of 1. They are nearly equal, i.e. $f_1\approx g_1$ which satisfies the heavy quark symmetry. The other four form factors $f_2, g_2, f_3, g_3$ are all negative. At $q^2=0$, $f_2\approx g_3$, and they are about 20\% of $f_1(g_1)$. The quantities $f_3, g_2$ are the smallest, $f_3\sim g_2\approx 0$, and they can be neglected. The numerical results show the validity of heavy quark symmetry and the power corrections are at the order of 20\%.

For the heavy-to-light transitions $\Lambda_b\to p(\Lambda,n)$, the numerical results of the form factors are presented in Tables \ref{tab:ff2},  \ref{tab:ff3} and \ref{tab:ff4}. The form factors $f_1, ~g_1$ are the largest, but their values are only about 0.1. This form factor suppression comes from the large momentum transfer to the final baryon. Similar to heavy-to-heavy transitions, the other form factors are negative.
At the large recoil point $q^2=0$, $f_2\approx g_3$, and they are about 50\% of $f_1(g_1)$. That means the large energy limit relations are broken significantly. The quantities $f_3, g_2$ are small but not negligible, about 10-20\% of $f_1(g_1)$. Comparing Tables \ref{tab:ff2} and \ref{tab:ff3}, one can find that the corresponding form factors in the $\Lambda_b\to p$ and $\Lambda_b\to \Lambda$ two processes are nearly equal. This is due to the light quark flavor symmetry. $\Lambda_b\to n$ form factors are same as $\Lambda_b\to p$ due to isospin symmetry.

\begin{figure}
\centering
\includegraphics[scale=0.55]{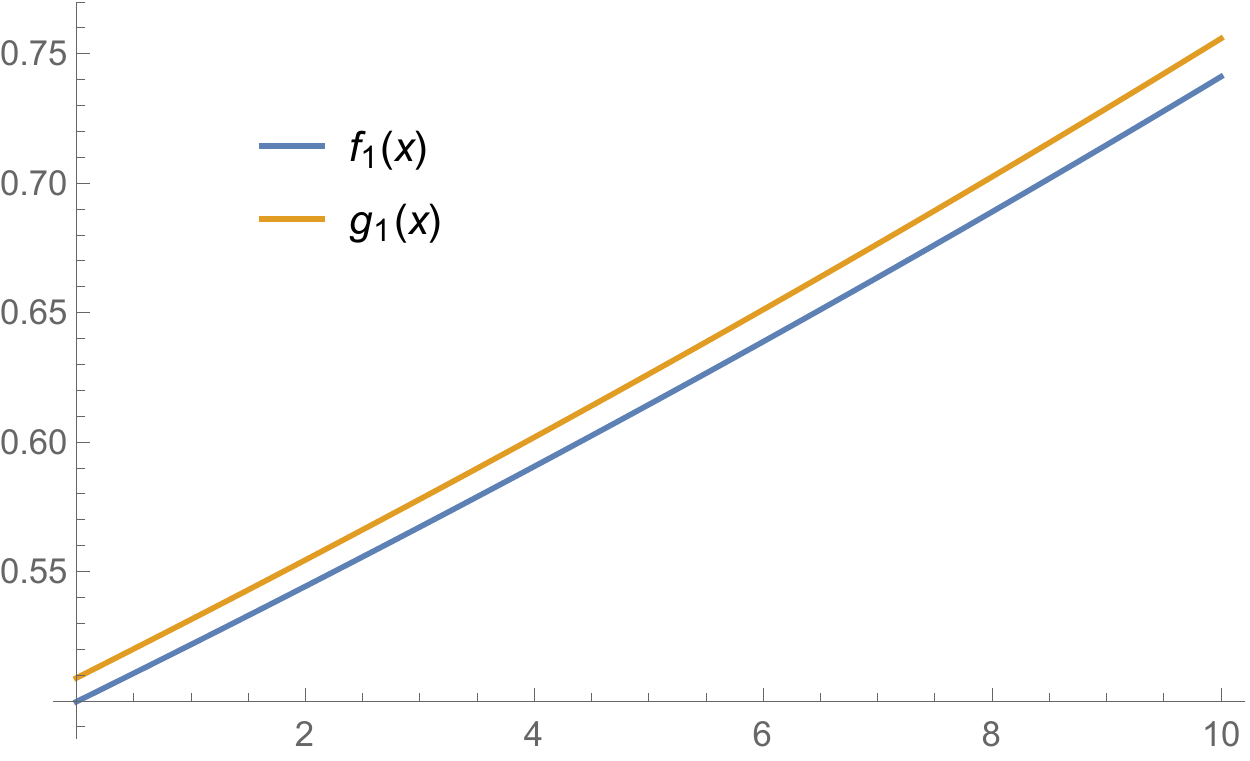}\includegraphics[scale=0.55]{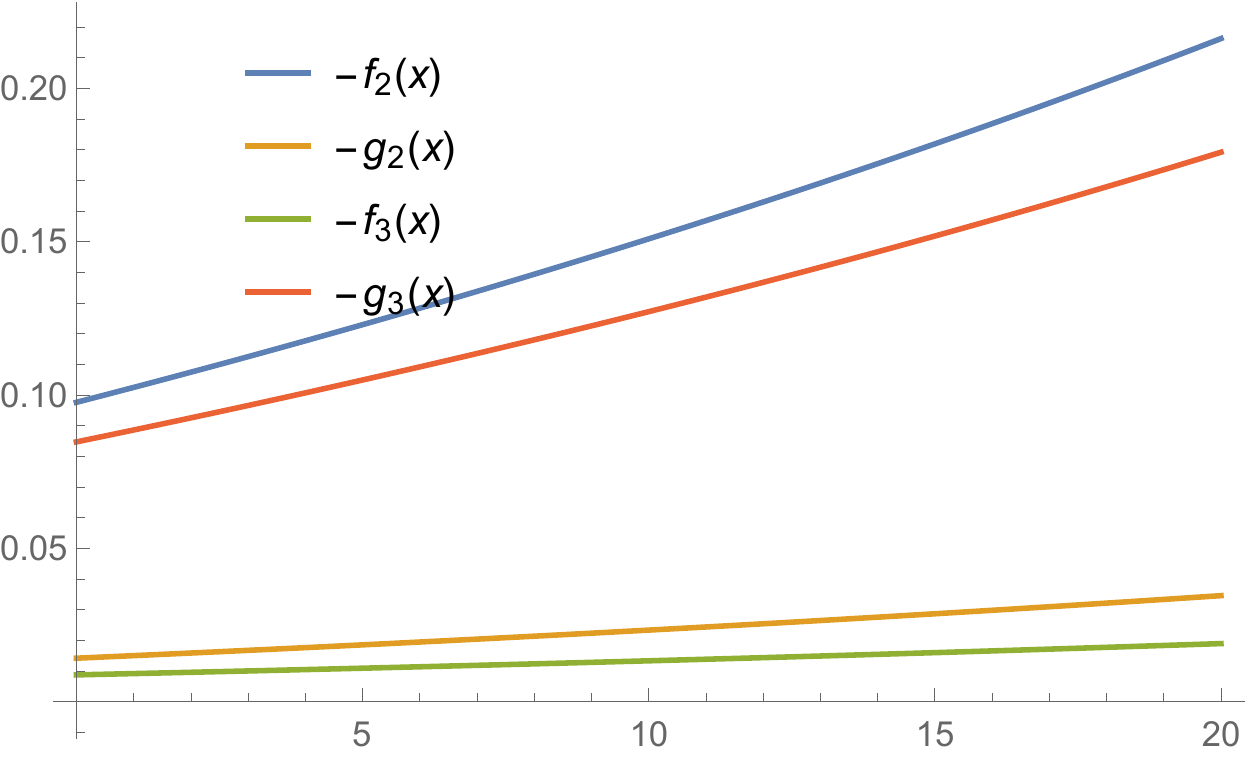}
\caption{The $q^2$-dependence of the $\Lambda_b \rightarrow \Lambda_c$ transition form factors. The horizontal $q^2$ variable is given in units of ${\rm GeV}^2$. }\label{fig:q21}
\end{figure}

\begin{figure}
\centering
\includegraphics[scale=0.55]{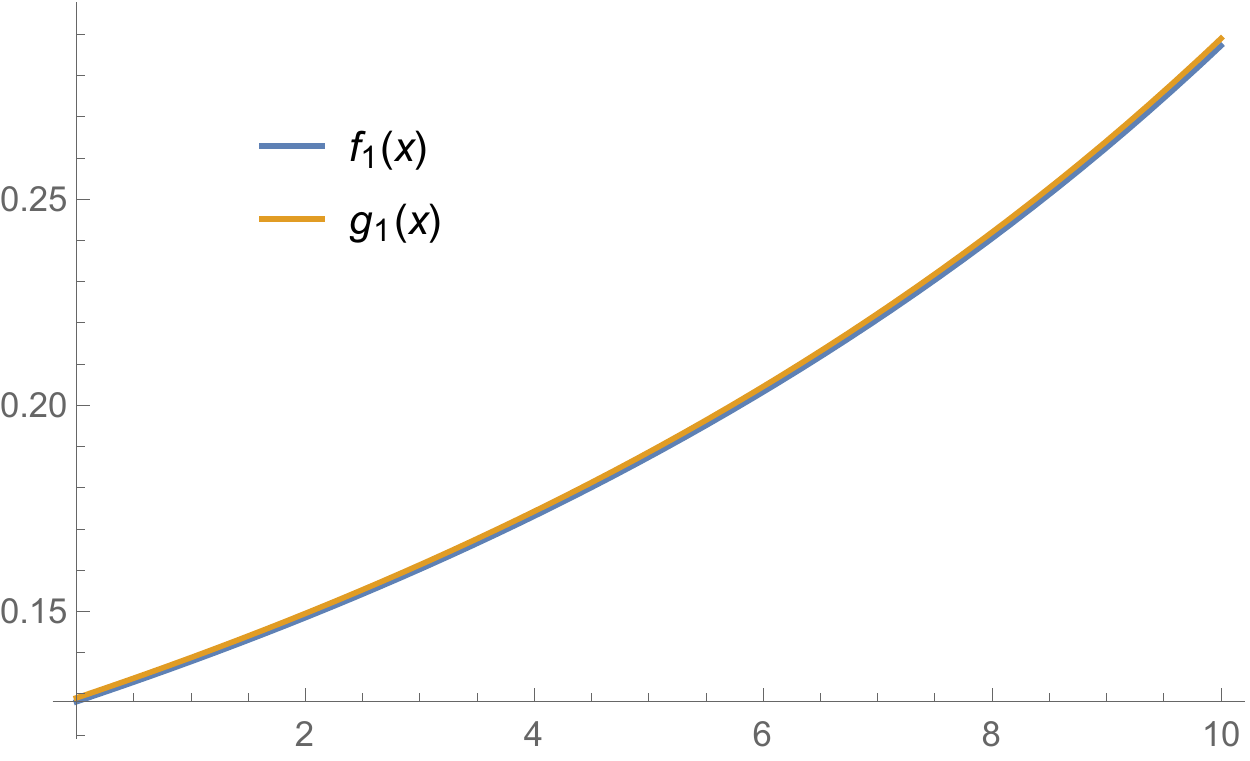}\includegraphics[scale=0.55]{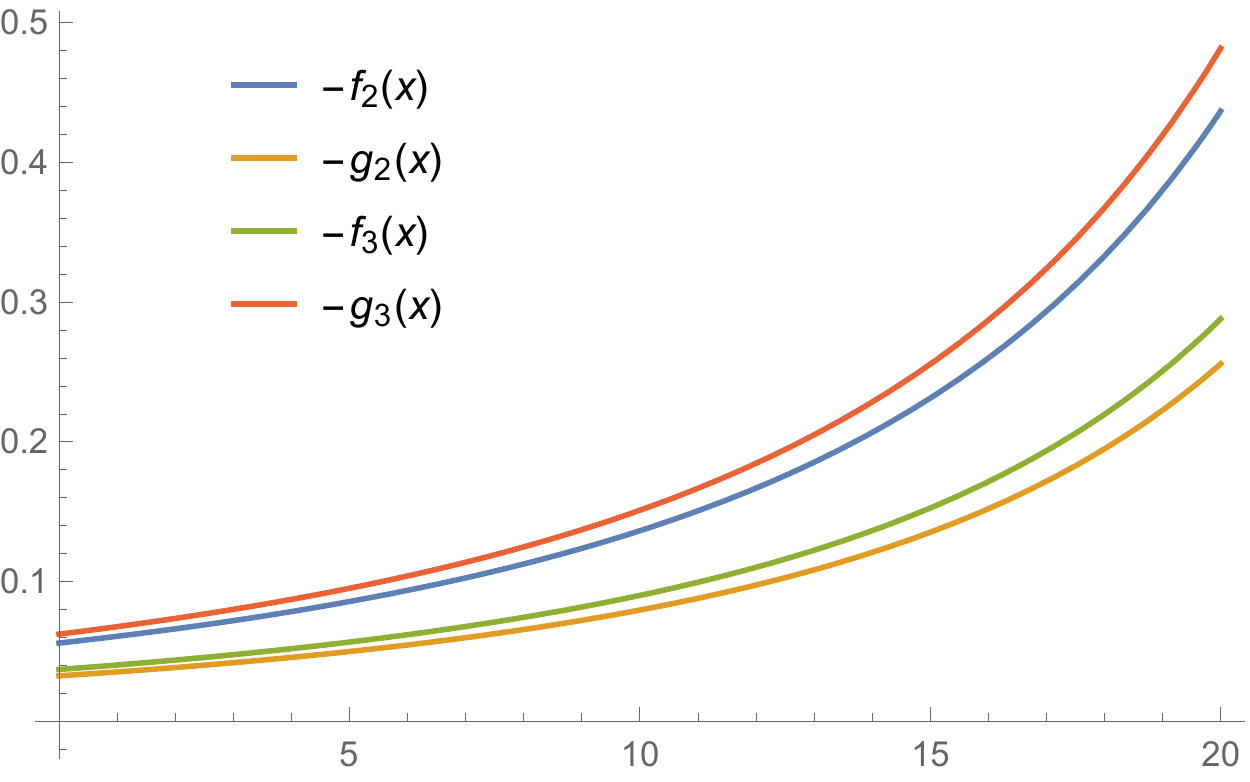}
\caption{The $q^2$-dependence of the $\Lambda_b \rightarrow p$ transition form factors. The horizontal $q^2$ variable is given in units of ${\rm GeV}^2$. }\label{fig:q22}
\end{figure}

\begin{figure}
\centering
\includegraphics[scale=0.55]{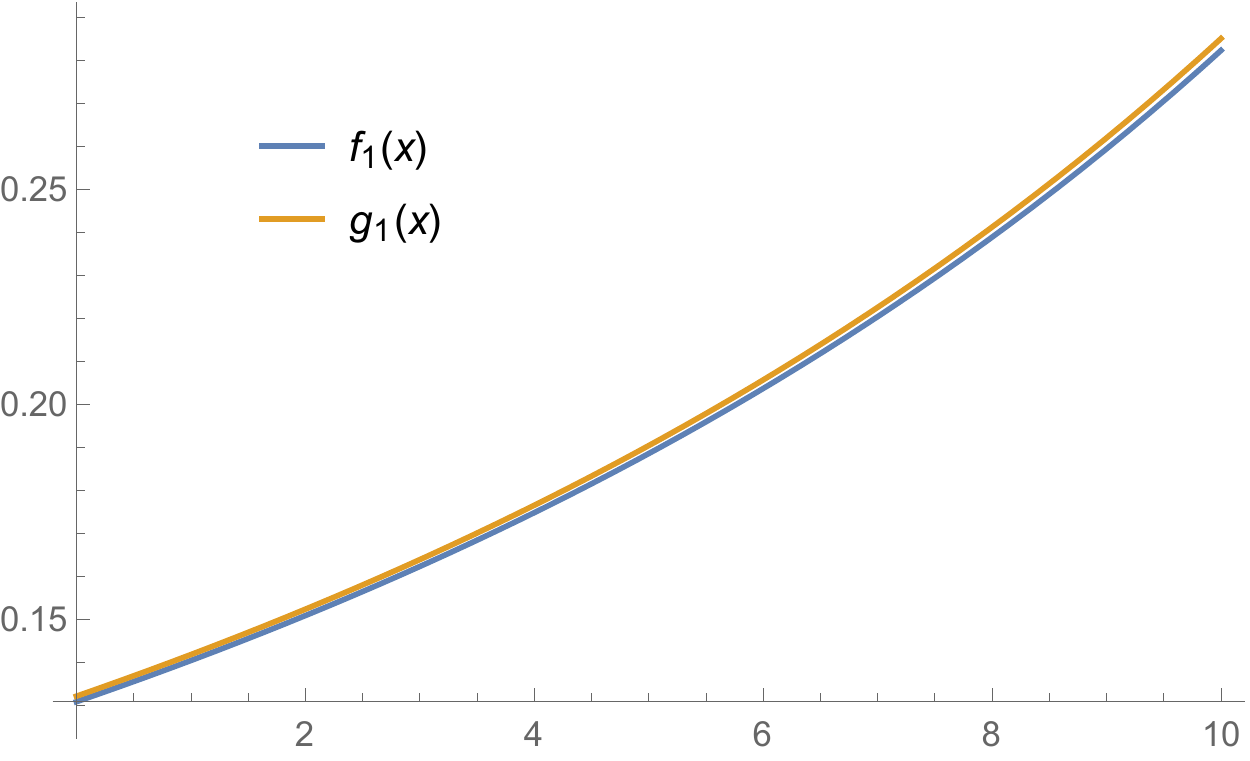} \includegraphics[scale=0.55]{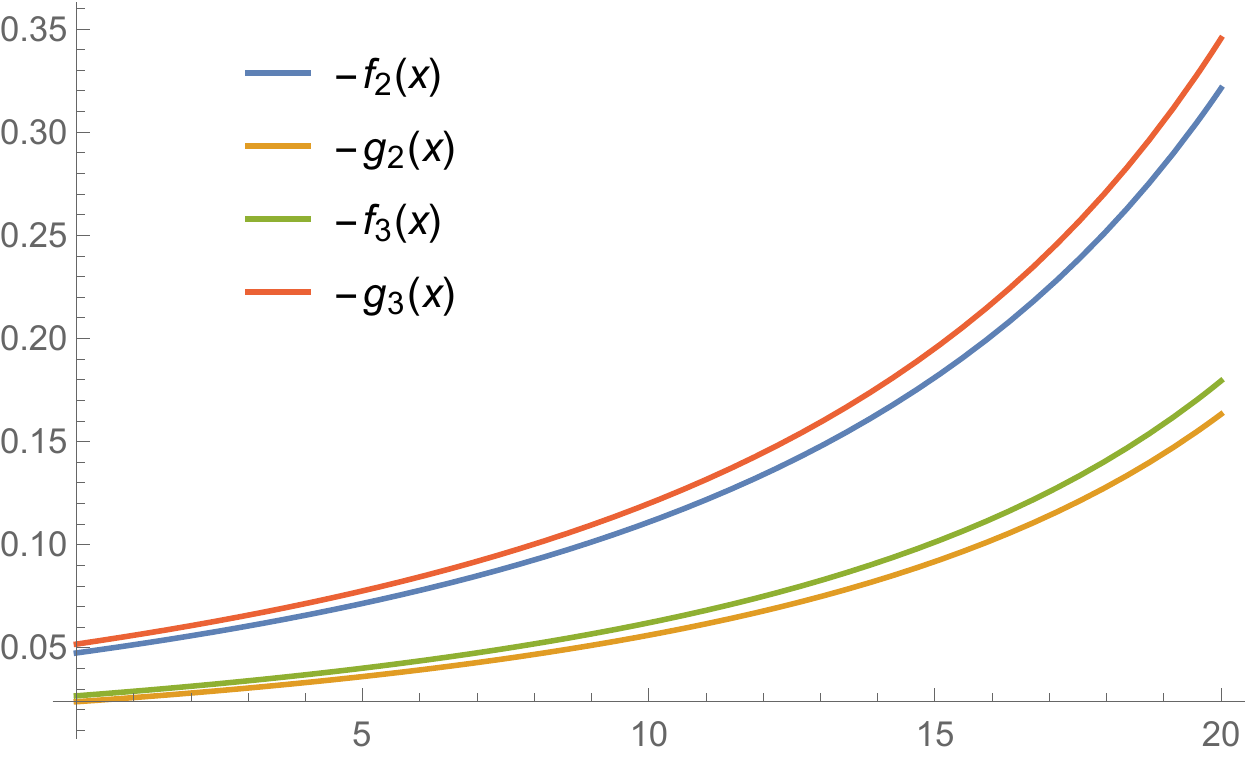}
\caption{The $q^2$-dependence of the $\Lambda_b \rightarrow \Lambda$ transition form factors. The horizontal $q^2$ variable is given in units of ${\rm GeV}^2$. }\label{fig:q23}
\end{figure}

\begin{figure}
\centering
\includegraphics[scale=0.55]{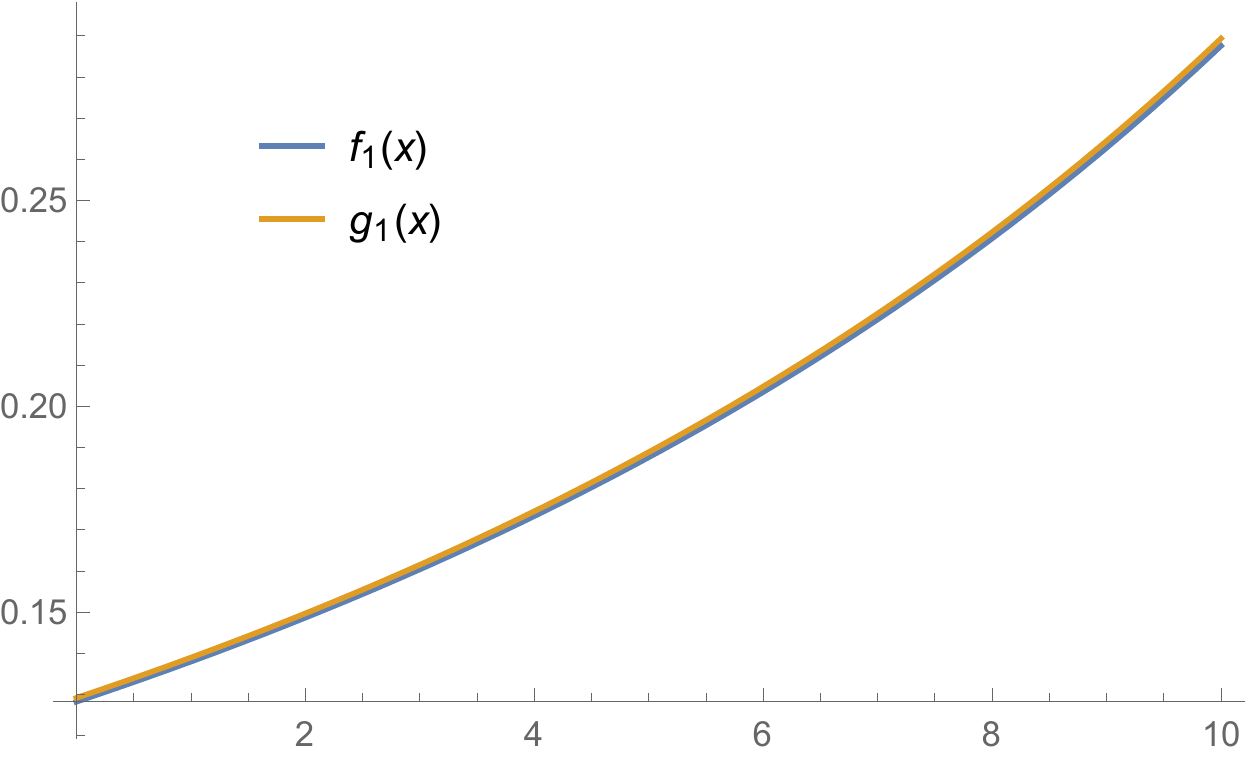} \includegraphics[scale=0.55]{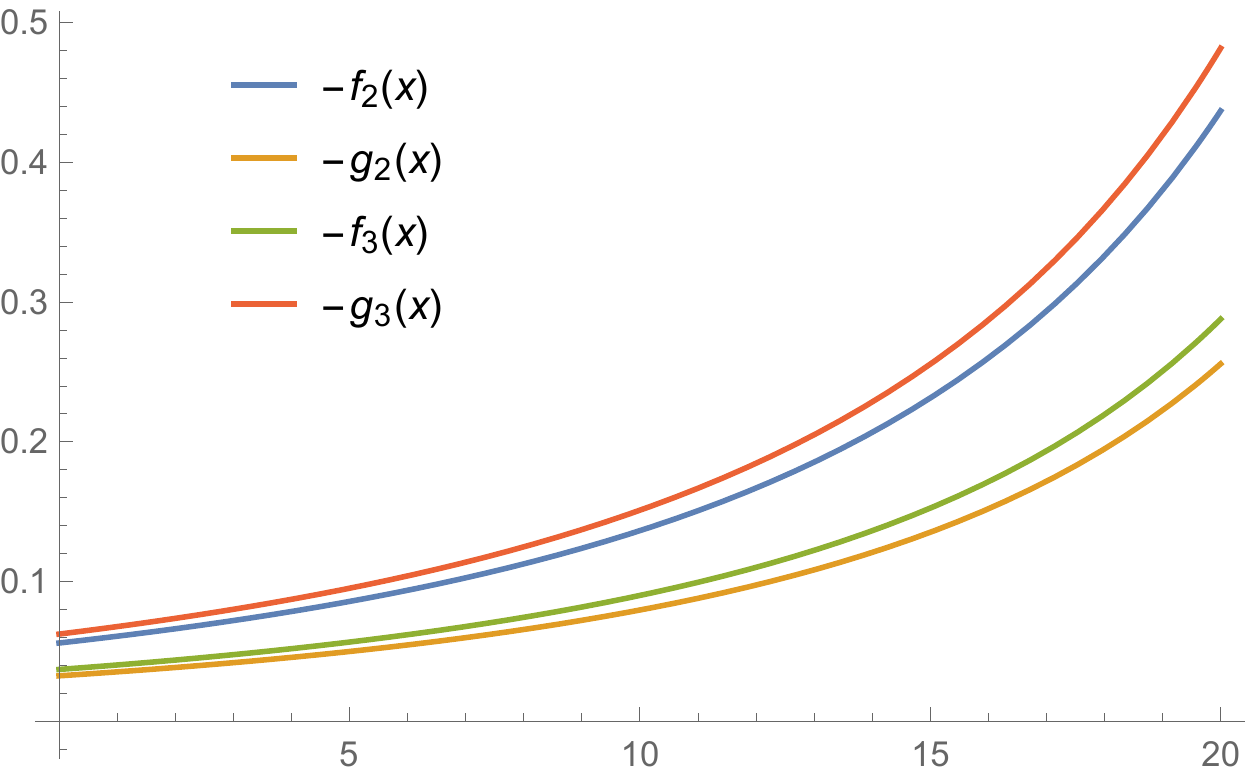}
\caption{The $q^2$-dependence of the $\Lambda_b \rightarrow n$ transition form factors. The horizontal $q^2$ variable is given in units of ${\rm GeV}^2$. }\label{fig:q24}
\end{figure}

The $q^2$-dependence of the $\Lambda_b\to \Lambda_c(p,\Lambda, n)$ form factors are plotted Figs. \ref{fig:q21}, \ref{fig:q22},  \ref{fig:q23} and \ref{fig:q24}. In all the four cases, the absolute values of the six form factors are increasing function of $q^2$. The dependence of form factors on $q^2$ is smooth. The $q^2$-dependence is crucial for the behavior of the differential decay width of the semi-leptonic processes and also has effects on the non-leptonic processes.

The baryon-to-baryon form factors are dominated by the non-pertubative QCD dynamics. The calculation of the transition form factors are model dependent and the theory uncertainties are difficult to estimate. In \cite{Faustov:2016pal}, the authors compare the predictions of the $\Lambda_b\to \Lambda_c,~p$ form factors in different theory models. They obtain a conclusion: there is reasonable agreement between predictions of significant different approaches for calculating the baryon form factors.

\section{Numerical results for semi-leptonic decays of $\Lambda^0_b\rightarrow \Lambda^+_c(p)\ l^-\bar{\nu}_l$}

Now, we are able to calculate the branching ratios and various asymmetries of the semi-leptonic decays $\Lambda^0_b\rightarrow \Lambda^+_c(p) \ l^- \bar{\nu}_l$. The numerical results of our model predictions in the covariant light-front approach are presented in Table \ref{tab:semi}.

\begin{table}
\caption{The branching ratios and asymmetries of the semi-leptonic decays.}\label{tab:semi}
\begin{center}
\begin{tabular}{l@{\hspace{20pt}}c@{\hspace{15pt}}c@{\hspace{25pt}}c@{\hspace{25pt}}c} \hline\hline
 ~~~~~Mode & ${\cal B}$ &  $A_{FB}$ & $P_L$ & \\ \hline
 $\Lambda^0_b\rightarrow \Lambda^+_c e^- \bar{\nu}_e$ & $5.59\times 10^{-2}$ & $-0.03$ &  $-0.96$ & \multirow{2}*{\tabincell{c} {covariant approach \\ (this work)}} \\
 $\Lambda^0_b\rightarrow \Lambda^+_c \mu^- \bar{\nu}_\mu$ & $5.57\times 10^{-2}$ & $-0.07$ &  $-0.93$ \\
 $\Lambda^0_b\rightarrow \Lambda^+_c \tau^- \bar{\nu}_\tau$ & $1.54\times 10^{-2}$ & $-0.13$ &  $-0.79$ \\
 $\Lambda^0_b\rightarrow p e^- \bar{\nu}_e$ & $4.02\times10^{-4}$ & 0.12 & $-0.97$ \\
 $\Lambda^0_b\rightarrow p \mu^- \bar{\nu}_\mu$& $4.02\times10^{-4}$ & 0.18 & $-0.95$ \\
 $\Lambda^0_b\rightarrow p \tau^- \bar{\nu}_\tau$ & $2.74\times10^{-4}$ & 0.10 & $-0.94$ \\
 \hline\hline
 $\Lambda^0_b\rightarrow \Lambda^+_c l^-\bar{\nu}_l$ & $6.30\times 10^{-2}$
 & $$ & $-0.80$ & \multirow{2}*{\tabincell{c}{conventional\\approach \cite{Ke:2007tg,Wei:2009np}}} \\
 $\Lambda^0_b\rightarrow p l^- \bar{\nu}_l$ & $2.54\times 10^{-4}$ &  & $-0.97$ \\\hline\hline
 $\Lambda^0_b\rightarrow \Lambda^+_c l^- \bar{\nu}_l$ &$(6.2^{+1.4}_{-1.3})\times 10^{-2}$
 & & & Experiment \cite{Patrignani:2016xqp}\\
 $\Lambda^0_b\rightarrow p \mu^- \bar{\nu}_\mu$ & $(4.1\pm 1.0)\times 10^{-4}$ & & & \\\hline\hline
\end{tabular}
\end{center}
\end{table}

\begin{table}
\caption{Predictions for the semi-leptonic decays in \cite{Faustov:2016pal}.}\label{tab:semi2}
\begin{center}
\begin{tabular}{l@{\hspace{30pt}}c@{\hspace{30pt}}c@{\hspace{30pt}}c} \hline\hline
 ~~~~~~Mode~~~~~~~~~ & ~~~~~~~~${\cal B}$~~~~~~~~ &  ~~~~~~~$A_{FB}$~~~~~~~ & ~~~~~$P_L$~~~~~ \\\hline
 $\Lambda^0_b\rightarrow \Lambda^+_c e^- \bar{\nu}_e$ & $6.48\times 10^{-2}$ & $0.20$ &  $-0.80$ \\
 $\Lambda^0_b\rightarrow \Lambda^+_c \mu^- \bar{\nu}_\mu$ & $6.46\times 10^{-2}$ & $0.19$ &  $-0.80$ \\
 $\Lambda^0_b\rightarrow \Lambda^+_c \tau^- \bar{\nu}_\tau$ & $2.03\times 10^{-2}$ & $-0.02$ &  $-0.91$ \\
 $\Lambda^0_b\rightarrow p e^- \bar{\nu}_e$ & $4.5\times10^{-4}$ & 0.35 & $-0.91$ \\
 $\Lambda^0_b\rightarrow p \mu^- \bar{\nu}_\mu$& $4.5\times10^{-4}$ & 0.34 & $-0.91$ \\
 $\Lambda^0_b\rightarrow p \tau^- \bar{\nu}_\tau$ & $2.9\times10^{-4}$ & -0.19 & $-0.89$ \\\hline\hline
\end{tabular}
\end{center}
\end{table}

The semi-leptonic decays $\Lambda^0_b\rightarrow \Lambda^+_c l^- \bar{\nu}_l$ decays where the final lepton is electron or muon are observed with a large branching ratio $(6.2^{+1.4}_{-1.3})\times 10^{-2}$. At present, the experimental error is still large. At the quark level, it is $b\to cl^- \bar{\nu}_l$ transition and the involved CKM matrix element is $V_{cb}$. Theory prediction for the electron process is $5.59\times 10^{-2}$. The ratio for the muon process is nearly equal to electron mode. That means that the mass of the light lepton can be neglected for the branching ratios. But it can't be neglected for the forward-backward asymmetry and the the longitudinal polarization. Our theory prediction for the ratio of the process $\Lambda^0_b\rightarrow \Lambda^+_c \ l^- \bar{\nu}_l$ is slightly smaller than the central value of the data, and consistent with the data within the experimental error. This result is obtained based upon taking account of both the semi-leptonic and non-leptonic processes. The data of the non-leptonic processes $\Lambda^0_b\rightarrow \Lambda^+_c+M$ given in the next section is more precise than the ones of the semi-leptonic processes. If we choose the parameters $\beta_{\Lambda_b}$ and $\beta_{\Lambda_c}$ to fit the central value of the data of $\Lambda^0_b\rightarrow \Lambda^+_c \ l^- \bar{\nu}_l$, the predictions for the non-leptonic processes of $\Lambda^0_b\rightarrow \Lambda^+_c+M$ will be found to be inconsistent with the data. Besides the ratios of the absolute ratios of the semi-leptonic and non-leptonic processes, one also needs to consider the relative ratio of semi-leptonic to non-leptonic decays, such as $\frac{{\cal B}(\Lambda^0_b\rightarrow \Lambda^+_c  l^- \bar{\nu}_l)}{{\cal B}(\Lambda^0_b\rightarrow \Lambda^+_c \pi)}$. We will discuss this ratio later.

For the tau lepton process $\Lambda^0_b\rightarrow \Lambda^+_c \tau^- \bar{\nu}_\tau$, it is not observed by experiment. The theory prediction for the branching ratio is $1.54\times 10^{-2}$, which is smaller than the ratio of the light lepton process but at the same order. We expect the tau lepton process to be observed in a near future. A discrepancy is observed in $B\to D^{(*)}$ semi-leptonic processes. The Standard Model (SM) prediction for the ratio of the heavy tau lepton to the light lepton processes is not consistent with the data. It is necessary to test whether the discrepancy exists  in the baryon case. Following \cite{Faustov:2016pal}, we define a ratio as
 \beq
 R^{\tau l}_{\Lambda_c}=\frac{{\cal B}(\Lambda^0_b\rightarrow \Lambda^+_c \tau^- \bar{\nu}_\tau)}{{\cal B}(\Lambda^0_b\rightarrow \Lambda^+_c l^- \bar{\nu}_l)}.
 \eeq
Our theory prediction is $R^{\tau l}_{\Lambda_c}=0.28$ which agrees with the result $0.31$ in \cite{Faustov:2016pal}.

About the forward-backward asymmetry $A_{FB}$, our predictions for the processes $\Lambda^0_b\rightarrow \Lambda^+_c e^- \bar{\nu}_e$ and $\Lambda^0_b\rightarrow \Lambda^+_c \mu^- \bar{\nu}_\mu$ are quite small, only several percent. For the tau lepton process, the asymmetry is about 10\%, but the detection efficiency of tau is low. So, it is difficult to measure the forward-backward asymmetry for the semi-leptonic processes of the $\Lambda^0_b\rightarrow \Lambda^+_c l^- \bar{\nu}_l$ in experiment. The longitudinal polarization $P_L$ is close to 1 which represents the longitudinal polarization dominance.

For the semi-leptonic decays of $\Lambda^0_b\rightarrow p$ transitions, only the process involving the muon lepton $\Lambda^0_b\rightarrow p \mu^- \bar{\nu}_\mu$ is reported. At the quark level, it is $b\to ul^- \bar{\nu}_l$ transition and the CKM matrix element is $V_{ub}$. Because $\frac{|V_{ub}|}{|V_{cb}|}\sim 0.1$, the measured ratio of the decay $\Lambda^0_b\rightarrow p \mu^- \bar{\nu}_l$ is two orders smaller than the ratio of $\Lambda^0_b\rightarrow \Lambda^+_c l^- \bar{\nu}_l$. Theory prediction agrees with the data as it should be, since we use the semi-leptonic decay $\Lambda^0_b\rightarrow p \mu^- \bar\nu_\mu$ to determine the proton parameter $\beta_p$. For the decay $\Lambda^0_b\rightarrow p l^- \bar{\nu}_l$, it is also longitudinal polarization dominant. The forward-backward asymmetry is at the order of 10-20\%, which is difficult to measure due to its suppressed rate. Similar to $R^{\tau l}_{\Lambda_c}$, we can define $R^{\tau l}_p$ by
 \beq
 R^{\tau l}_p=\frac{{\cal B}(\Lambda^0_b\rightarrow p \tau^- \bar{\nu}_\tau)}{{\cal B}(\Lambda^0_b\rightarrow p l^- \bar{\nu}_l)}.
 \eeq
Our model prediction is $R^{\tau l}_p=0.68$ which agrees with the result $0.65$ in \cite{Faustov:2016pal}.

For comparison, we discuss two models in literature. One is the conventional light-front approach used in the previous study \cite{Ke:2007tg,Wei:2009np}. The results have been included in Table \ref{tab:semi}. The previous prediction for the ratio of $\Lambda^0_b\rightarrow p \mu^- \bar{\nu}_\mu$ decay is $2.54\times 10^{-4}$ which is smaller than the data. This is the reason that we choose a large value for $\beta_p$. Another approach is a relativistic quark model given in \cite{Faustov:2016pal}. Their numerical results are listed in Table \ref{tab:semi2}.
One can see that the main difference in theory predictions is the forward-backward asymmetry. The asymmetry is small and sensitive to the details of the models. The measurement of the forward-backward asymmetry can test the different theory approaches.

The LHCb collaboration reported a measurement on the ratio of the heavy-to-heavy and heavy-to-light semi-leptonic decays in the restricted momentum region of $q^2$ \cite{Aaij:2015bfa}. The ratio is defined by
 \beq
 R_{\Lambda_c p}=\frac{\int_{15{\rm GeV^2}}^{q^2_{max}}dq^2\frac{d\Gamma(\Lambda^0_b\to p\mu^-\bar{\nu}_\mu)}
 {dq^2}}{\int_{7{\rm GeV^2}}^{q^2_{max}}dq^2\frac{d\Gamma(\Lambda^0_b\to \Lambda_c^+\mu^-\bar{\nu}_\mu)}{dq^2}}.
\eeq
The measurement of the above ratio permits us to extract the CKM matrix elements $|V_{ub}|/|V_{cb}|$ in the heavy baryon decays. It provides an independent measurement outside of the $B$ meson system and a crosscheck for the CKM matrix elements. In our model, the calculation gives the numerical result as
 \beq
 R_{\Lambda_c p}=1.10~\frac{|V_{ub}|^2}{|V_{cb}|^2}.
 \eeq
The result in \cite{Faustov:2016pal} is $R_{\Lambda_c p}=(0.78\pm 0.08)~\frac{|V_{ub}|^2}{|V_{cb}|^2}$. The lattice calculation gives $R_{\Lambda_c p}=(1.471\pm 0.095\pm 0.109)~\frac{|V_{ub}|^2}{|V_{cb}|^2}$ \cite{Detmold:2015aaa}. Our prediction lies in the middle of them.

By use of the CKM parameters chosen in this study, we obtain $R_{\Lambda_c p}=1.09\times 10^{-2}$.
The experimental measurement from the LHCb collaboration is \cite{Aaij:2015bfa}
 \beq
 R_{\Lambda_c p}=(1.00\pm 0.04\pm 0.08)\times 10^{-2}.
 \eeq
Our model prediction is slightly larger that the central value of the data. They are consistent within the experimental error. Taking into account of the theoretical errors would increase the consistency.

We can also extract the the CKM elements $|V_{ub}|/|V_{cb}|$ from the data by using our model calculations. We obtain
 \beq
 \frac{|V_{ub}|}{|V_{cb}|}=0.091\pm 0.08.
 \eeq
The error comes from the experiment data. At present, the determination of $|V_{cb}|$ is more precise due to the heavy quark symmetry. From PDG \cite{Patrignani:2016xqp}, an average of the experiments gives $|V_{cb}|=(40.5\pm 1.5)\times 10^{-3}$. From the precise value of $|V_{cb}|$, we can extract $|V_{ub}|$ by use of our model as
 \beq
 |V_{ub}|=(3.69\pm 0.3)\times 10^{-3}.
 \eeq
For comparison, we give the values of $|V_{ub}|$ obtained from the inclusive and exclusive determinations  as \cite{Patrignani:2016xqp}
 \beq
 |V_{ub}|&=&(4.49\pm 0.16^{+0.16}_{-0.18})\times 10^{-3}  \qquad ({\rm inclusive}),\non\\
 |V_{ub}|&=&(3.72\pm 0.19)\times 10^{-3} \qquad \qquad ({\rm exclusive}).
 \eeq
and the average is
 \beq
 |V_{ub}|=(4.09\pm 0.39)\times 10^{-3}  \qquad ({\rm average}).
 \eeq
One can see the value of $|V_{ub}|$ extracted from our model agrees with the measurement from the exclusive processes very well. Since our method is adopted for the exclusive processes, the agreement provides a support of our model.

\section{Numerical results for non-leptonic decays of $\Lambda^0_b \rightarrow H + M$}

In this section, we present our numerical predictions for the four types of the non-leptonic decays $\Lambda^0_b \rightarrow H + M$ where $H$ represents $\Lambda^+_c, p, \Lambda, n$. We discuss them case by case.

\subsection{$\Lambda^0_b\rightarrow \Lambda^+_c + M$ decays}

The $\Lambda^0_b\rightarrow \Lambda^+_c + M$ decays have the largest decay ratios in the non-leptonic processes of $\Lambda_b^0$. They belong to charmful processes which are enhanced by the CKM matrix element $V_{cb}$. For the processes with light mesons $\pi,\rho,K,K^*$, they have only the color-allowed tree operator contribution and the Wilson coefficient is $a_1$. For the processes with heavy mesons $D^-,D^{*-},D_s^-,D_s^{*-}$, they contain the $b\to d(s)$ QCD penguin operator contributions which are suppressed by $\alpha_s$. According to the CKM elements, $\Lambda^0_b\rightarrow \Lambda^+_c + M$ decays can be classified into Cabibbo-favored and Cabibbo-suppressed processes. The processes with with mesons $\pi,\rho,D_s,D_s^*$ being the final states are the Cabibbo-favored processes. The corresponding sub-processes are $b\to c\bar u d$ or $b\to c \bar c s$, Their decay ratios are largest, in the region $4\times 10^{-3}$ to $1\times 10^{-2}$.  The processes with mesons $K^-,K^{*-},D^-,D^{*-}$ being the final states are the Cabibbo-suppressed processes. The sub-processes are $b\to c\bar u s$ or $b\to c \bar c d$ which is suppressed by $\lambda=\sin \theta_C\sim 0.22$. Their decay ratios are of order  $(3-5)\times 10^{-4}$. The theory predictions and the experimental data for decay rates of the processes $\Lambda^0_b \rightarrow \Lambda^+_c + M$  are given in Table \ref{tab:non11}. The renormalization scale $\mu$ dependence of the decay rates is small, less than 5\%. For all the observed processes, the theory predictions accord well with the experiment data.  At present, only four processes where the final meson is a pseudoscalar are observed. Because the ratios of the other four processes with the final vector mesons are at the same order, we expect that these vector processes will be measured in the near future.

\begin{table}
\caption{Branching ratios of $\Lambda^0_b\rightarrow \Lambda^+_c + M$ decays.}\label{tab:non11}
\begin{center}
\begin{tabular}{l | c |c | c |c}\hline\hline
~~~Mode &~~~~~$\mu=m_b/2$~~~~~ & ~~~~~$\mu=m_b$~~~~~   & ~~~~~$\mu=2m_b$~~~~~& Experiment \cite{Patrignani:2016xqp} \\\hline
$\Lambda^0_b\rightarrow \Lambda^+_c \pi^-$      & $5.24\times 10^{-3}$  & $4.96\times10^{-3}$
 & $4.76\times 10^{-3}$  & $(4.9\pm0.4)\times 10^{-3}$ \\
$\Lambda^0_b\rightarrow \Lambda^+_c \rho^-$     &$ 9.13\times 10^{-3}$  & $8.65\times10^{-3}$
 & $8.30\times 10^{-3}$  & $-$  \\
$\Lambda^0_b\rightarrow \Lambda^+_c K^- $       & $4.15\times 10^{-4}$  & $3.93\times10^{-4}$
 & $3.77\times 10^{-4}$  & $(3.59\pm0.3)\times 10^{-4}$ \\
$\Lambda^0_b\rightarrow \Lambda^+_c K^{*-}$     & $4.65\times 10^{-4}$  & $4.41\times10^{-4}$
 & $4.23\times 10^{-4}$  & $-$  \\
$\Lambda^0_b\rightarrow \Lambda^+_c D^- $       & $5.52\times 10^{-4}$  & $5.22\times10^{-4}$
 & $5.01\times 10^{-4}$  & $(4.6\pm0.6)\times 10^{-4}$  \\
$\Lambda^0_b\rightarrow \Lambda^+_c D^{*-}$     & $5.51\times 10^{-4}$  & $5.20\times10^{-4}$
 & $4.99\times 10^{-4}$  & $-$  \\
$\Lambda^0_b\rightarrow \Lambda^+_c D^-_{s}$    & $1.31\times 10^{-2}$  & $1.24\times10^{-2}$
 & $1.19\times 10^{-2}$  & $(1.10\pm0.10)\times 10^{-2}$ \\
$\Lambda^0_b\rightarrow \Lambda^+_c D^{*-}_{s}$ & $1.11\times 10^{-2}$  & $1.05\times10^{-2}$
 & $1.01\times 10^{-2}$  & $-$  \\ \hline\hline
\end{tabular}
\end{center}
\end{table}

\begin{table}
\caption{ Up-down and CP asymmetries for $\Lambda^0_b\rightarrow \Lambda^+_c + M$ decays .}
 \label{tab:non12}
\begin{center}
\begin{tabular}{l |c|c|c|c}\hline\hline
 ~~~~Mode &\tabincell{c}{~~~~~~~~~~$\alpha$~~~~~~~~~~} & \multicolumn{3}{c}{$A_{CP}$} \\ \cline{3-5}
 &  & ~~~~$\mu=m_b/2$~~~~ & ~~~~$\mu=m_b$~~~~ & ~~~~$\mu=2m_b$~~~~ \\\hline
$\Lambda^0_b\to \Lambda^+_c \pi^-$     & $-0.998$    & $0$      & $0$        & $0$ \\
$\Lambda^0_b\to \Lambda^+_c \rho^-$    & $-0.888$    & $0$      & $0$        & $0$ \\
$\Lambda^0_b\to \Lambda^+_c K^- $      & $-1.0$      & $0$      & $0$        & $0$ \\
$\Lambda^0_b\to \Lambda^+_c K^{*-}$    & $-0.859$    & $0$      & $0$        & $0$ \\
$\Lambda^0_b\to \Lambda^+_c  D^- $     & $-0.999$    & $1.39\times10^{-2}$   & $1.16\times 10^{-2}$
 & $1.01\times 10^{-2}$  \\
$\Lambda^0_b\to \Lambda^+_c D^{*-} $   & $-0.478$    & $1.26\times10^{-2}$   & $1.04\times 10^{-2}$
 & $8.96\times 10^{-3}$  \\
$\Lambda^0_b\to \Lambda^+_c D^-_{s} $  & $-1.0$      & $-5.71\times10^{-3}$  & $-4.82\times 10^{-3}$
 & $-4.24\times 10^{-3}$ \\
$\Lambda^0_b\to \Lambda^+_c D^{*-}_{s}$ & $-0.439$   & $-6.76\times10^{-4}$  & $-5.58\times 10^{-4}$
 & $-4.81\times 10^{-4}$ \\ \hline\hline
\end{tabular}
\end{center}
\end{table}

The predictions for the up-down and CP asymmetries are given in Table \ref{tab:non12}. Up to now, no up-down and CP asymmetries in $\Lambda^0_b\rightarrow \Lambda^+_c + M$ decays were observed.
All the up-down asymmetries $\alpha$ from theory are negative and the absolute values are about 1 for most processes. Up-down asymmetry reflects parity violation. The parity violation at the order of 1  is due to the
$V-A$ nature of the weak currents which contains the maximal parity violation. For two processes with final states $\Lambda^+_c D^{*-}$ and $\Lambda^+_c D^{*-}_{s}$, the up-down asymmetry is about $0.4$.
This is because more complicated Lorentz structures are entered for the vector final state. All the up-down asymmetry $\alpha$ is nearly independent of $\mu$. For the $\Lambda^0_b\rightarrow p+M$ processes, the $\mu$ dependence will be non-negligible. There is no direct CP violation in the processes of with light mesons $\pi^-,\rho^-,K^-,K^{*-}$ because there is only tree operator contribution with no weak and strong phase difference. The CP asymmetries in the Cabibbo-favored processes $\Lambda_c^+ D^{(*)-}_{s}$ are quite small, about $10^{-3}$ or $10^{-4}$, and it is difficult to detect them in experiment. For the processes with final states $\Lambda_c^+ D^{(*)-}$, the direct CP asymmetries are at the order of $10^{-2}$. But these processes are Cabibbo-suppressed, and also difficult to measure the direct CP asymmetry in them. This "large ratio and small CP violation" phenomenon is familiar in the B meson system. Thus, we can obtain a conclusion that it is nearly impossible to observe the direct CP violation in $\Lambda^0_b\rightarrow \Lambda^+_c + M$ decays. Any observation would be signal of new physics. As will be shown, this conclusion applies to all $\Lambda^0_b$ decays with the final states containing one or two charm quarks.

A ratio of semi-leptonic to non-leptonic fractions is defined by
\beq
 R^{\Lambda_c}_{l\pi}=\frac{{\cal B}(\Lambda^0_b\rightarrow\Lambda^+_c l^-\bar{\nu}_l)}
  {{\cal B}(\Lambda^0_b\rightarrow\Lambda^+_c\pi^-)}.
\eeq
This ratio reduces the theory uncertainties in calculating the baryon-to-baryon form factors. In our model, the semi-leptonic to non-leptonic decay ratio is
 \beq
 R^{\Lambda_c}_{l\pi}=11.3\pm 0.5,
 \eeq
The error comes from $\mu$ dependence of the decay rate for the non-leptonic process. One result from the early measurement by CDF collaboration is \cite{Aaltonen:2008eu}
 \beq
 R^{\Lambda_c}_{l\pi}=16.6\pm 3.0{\rm (stat)}\pm 1.0{\rm (syst)}^{+2.6}_{-3.4}{\rm (PDG)}\pm 0.3{\rm (EBR)}.
 \eeq
Our fitted value from the semi- and non-leptonic processes gives
 \beq
 R^{\Lambda_c}_{l\pi}=12.6\pm 3.0.
 \eeq
One can find the consistency between theory and the data.

Another ratio is proposed to relate the baryon decay to the meson process in \cite{Leibovich:2003tw}. It is defined by
\beq
R_2^{\Lambda_c}=\frac{{\cal B}(\Lambda^0_b\to \Lambda^+_c \pi^-)}{{\cal B}(\bar B^0\to D^+\pi^-)}.
\eeq
The study of this ratio is helpful to understand the meson-baryon similarity. In the small velocity and heavy quark limit, $R_2=2$. The early experiment gives \cite{Abulencia:2006df}
 \beq
 \frac{f_{\Lambda_b^0}}{f_d}\frac{{\cal B}(\Lambda^0_b\to \Lambda^+_c \pi^-)}{{\cal B}(\bar B^0\to D^+\pi^-)}
  =0.82\pm 0.08({\rm stat})\pm 0.11({\rm syst})\pm 0.22({\rm BR})
 \eeq
We will discuss the production fraction $f_{\Lambda_b^0}$ in more detail in the part of $\Lambda^0_b\to \Lambda J/\psi$. The value of $f_{\Lambda_b^0}/f_d$ is chosen to be $0.458$. The CDF result is $R_2^{\Lambda_c}\cong 1.79\pm 0.33$. Our fitted value from the data of $\Lambda^0_b\to \Lambda^+_c \pi^-$ and $\bar B^0\to D^+\pi^-$  processes gives $R_2^{\Lambda_c}=1.95\pm 0.25$. In our model, the decay ratio of the process $\Lambda^0_b\to \Lambda^+_c \pi^-$ is ${\cal B}(\Lambda^0_b\to \Lambda^+_c \pi^-)=4.96\times 10^{-3}$. By use of the data for B meson ${\cal B}(\bar B^0\to D^+\pi^-)^{\rm expt}=(2.52\pm 0.13)\times 10^{-3}$, we obtain $R_2^{\Lambda_c}=1.97$. Our result accords with the experiment and the heavy quark symmetry relation very well. By comparison, the result in \cite{Leibovich:2003tw} is $R_2^{\Lambda_c}=1.6\frac{\tau_{\Lambda_b}}{\tau_{B^0}}=1.54$, which is smaller than ours and the data.

The $\Lambda_b^0$ decays can also be employed to test the factorization hypothesis. According to the QCD factorization, the processes $\Lambda_b^0\to\Lambda_c^+\pi^-(K^-)$ with one heavy and one light final states is factorizable, while the heavy-heavy processes $\Lambda_b^0\to\Lambda_c^+ D^-(D_s^-)$ are non-factorizable. If it is so, the theory prediction of QCDF approach will become worse when the final meson are heavier. We choose the four observed processes $\Lambda_b^0\to\Lambda_c^+\pi^-(K^-,D^-,D_s^-)$ for discussion. If the process $\Lambda_b^0\to\Lambda_c^+\pi^-$ is used to adjust the phenomenological parameters to fit the experiment. When the final meson is heavy $D^-$ or $D_s^-$ where QCD factorization is not applicable, the deviations of theory prediction from the experiment should occur and will be largest for $\Lambda_b^0\to\Lambda_c^+ D_s^-$. However, we don't see the deviations from Table \ref{tab:non11}. The consistency between the theory and the experiment data is nearly at the same accuracy for the four processes.

To make our point more clear, we use the relative ratio of the decay rates to reduce the model uncertainties in the baryon-to-baryon form factors. In order to test the factorization assumption, we define three ratios below
 \beq
 R_{\pi K}=\frac{{\cal B}(\Lambda^0_b\to\Lambda^+_c\pi^-)}
  {{\cal B}(\Lambda^0_b\to\Lambda^+_c K^-)}, ~~~~
 R_{\pi D}=\frac{{\cal B}(\Lambda^0_b\to\Lambda^+_c \pi^-)}
  {{\cal B}(\Lambda^0_b\to\Lambda^+_c D^-)}, ~~~~
 R_{\pi D_s}=\frac{{\cal B}(\Lambda^0_b\to\Lambda^+_c \pi^-)}
  {{\cal B}(\Lambda^0_b\to\Lambda^+_c D_s^-)}.
 \eeq
By calculations, the results for the ratios are given as
 \beq
 &&R_{\pi K}^{\rm th}=12.6\pm 1.2,    \qquad   ~~~R_{\pi K}^{\rm expt}=13.6\pm 1.6,  \non\\
 &&R_{\pi D}^{\rm th}=9.6\pm 0.9,     \qquad   ~~~~R_{\pi D}^{\rm expt}=10.6\pm 1.6,  \non\\
 &&R_{\pi Ds}^{\rm th}=0.40\pm 0.04,  \qquad   R_{\pi Ds}^{\rm expt}=0.45\pm 0.05.
 \eeq
The theory results are obtained by using the predictions given in Table \ref{tab:non11}. The central values
are given at $\mu=m_b$. The experimental values are our fitted results from the data.

To go further, we define the ratio of theory to experiment as $R'=R^{\rm th}/R^{\rm expt}$. Thus£¬
 \beq
 R'_{\pi K}=0.93\pm 0.14, ~~~~ R'_{\pi D}=0.91\pm 0.16, ~~~~R'_{\pi D_s}=0.89\pm 0.13.
 \eeq
Within the errors, the ratios $R'$ are consistent with 1. There is really a small trend for $R'$ to become smaller for heavier mesons. But the difference in the three ratios are so small that we can regard them to be equal. Thus, we can draw a conclusion that the factorization assumption for $\Lambda_b^0\to \Lambda_c^+ D^-(D_s^-)$ processes containing two heavy charmed mesons is still applicable. The mechanism of factorization cannot be explained by the color transparency argument or the perturbative framework. A test of factorization in the heavy-heavy $B$ meson decays is given in \cite{Chen:2005rp}. The conclusion from the $B$ meson system is similar to ours in the baryon case. Comparing the numerical results of \cite{Chen:2005rp} with the present  precise data from PDG, we can obtain another conclusion: the $N_c^{eff}=\infty$ prediction is not supported by the experiment. Thus, the large $N_c$ limit is not a justified mechanism of factorization. There must be some non-perturbative mechanism which prefer the factorization of a large-size charmed meson or baryon from a soft cloud.

It is interesting to compare the experimental data with the predictions within the heavy quark limit which are given in \cite{Ke:2007tg}. In that work, the effective coefficient is simply chosen as $a_1=1$ without the QCD corrections. The heavy-to-heavy baryon form factors are reduced to one Isgur-Wise function $\zeta(\omega)$ with $\omega=v\cdot v'$. At the zero-recoil point, $\zeta(1)=1$. At other momentum regions, the Isgur-Wise function can be approximated as a linear function described by a universal slope parameter $\rho^2\equiv -\frac{d\zeta(\omega)}{d\omega}|_{\omega=1}$. One can find that the results within the heavy quark limit accord with the present data very well. From the consistency, we obtain a conclusion that the $\Lambda^0_b\rightarrow \Lambda^+_c + M$ decay is governed by one universal slope parameter and a meson decay constant. This is the leading and dominant contribution. Other QCD corrections, no matter perturbative or non-perturbative, are perturbations near the stable point within the heavy quark limit.

\subsection{$\Lambda^0_b\rightarrow p + M$ decays}

For the non-leptonic decays $\Lambda^0_b\rightarrow p + M$, there are 8 processes which are similar to $\Lambda^0_b\rightarrow \Lambda_c^+ + M$ decays. But the branching fractions are smaller by two or three orders. The tree diagram contribution is proportional to $V_{ub}$ and thus suppressed by small CKM parameters. The charmless processes belong to the rare decays. But these processes are important in exploring the CP violation. As we will show below, the direct CP violation in some processes can be large, at the order of 10\%. We may call this phenomenon as "small ratio and large CP violation".

The theory predictions for the branching ratios of decays $\Lambda^0_b\rightarrow p + M$ are given in Table \ref{tab:non21}. The fractions of the four processes with final meson being light are at the order of $10^{-6}$. The processes of $\Lambda_b^0\rightarrow p \pi^-(\rho^-)$ are color-allowed tree diagram dominant. The processes of $\Lambda_b\rightarrow p K^-(K^{*-})$ are QCD penguin dominant. Although suppressed by $\alpha_s$, the $b\to s$ penguin is enhanced by CKM matrix elements $V_{cs}V_{cb}$. So the branching ratios of $\Lambda_b^0\rightarrow p K^-(K^{*-})$ decays are of the same order as the $\Lambda_b^0\rightarrow p \pi^-(\rho^-)$ decays. A  detailed discussion about the $\Lambda_b^0\rightarrow p K^-$ process in QCDF approach is given in \cite{Zhu:2016bra}. The processes $\Lambda_b\rightarrow p D_s^{(*)-}$ have only the color-allowed tree operator contribution and have the ratios of order of $10^{-5}$.  The processes $\Lambda_b^0\rightarrow p D^{(*)-}$ are color-allowed, but they are Cabibbo-suppressed. So the ratios are of the order of $10^{-7}$. Up to  now, only two processes $\Lambda_b^0\rightarrow p\pi^-$ and $p K^-$ are observed. The experiment provide an upper limit for $\Lambda_b\rightarrow p D_s^-$ which is close to the theory prediction.

\begin{table}
\caption{Branching ratios of $\Lambda^0_b\rightarrow p+ M$ decays.}\label{tab:non21}
\begin{center}
\begin{tabular}{l | c |c | c |c}\hline\hline
 ~~~Mode &~~~~~$\mu=m_b/2$~~~~~ & ~~~~~$\mu=m_b$~~~~~   & ~~~~~$\mu=2m_b$~~~~~& ~~~~Experiment
  \cite{Patrignani:2016xqp}~~ \\\hline
 $\Lambda^0_b\rightarrow p \pi^-$     & $4.57\times 10^{-6}$  & $4.30\times10^{-6}$
  & $4.11\times 10^{-6}$  & $(4.2\pm 0.8)\times 10^{-6}$  \\
 $\Lambda^0_b\rightarrow p \rho^-$    & $7.89\times 10^{-6}$  & $7.47\times10^{-6}$
  & $7.17\times 10^{-6}$  & $-$ \\
 $\Lambda^0_b\rightarrow p K^- $      & $3.15\times 10^{-6}$  & $2.17\times10^{-6}$
  & $1.70\times 10^{-6}$  & $(5.1\pm1.0)\times 10^{-6}$  \\
 $\Lambda^0_b\rightarrow p K^{*-}$    & $1.08\times 10^{-6}$  & $1.01\times10^{-6}$
  & $0.94\times 10^{-6}$  & $-$ \\
 $\Lambda^0_b\rightarrow p D^- $      & $6.65\times 10^{-7}$  & $6.29\times10^{-7}$
  & $6.04\times 10^{-7}$  & $-$ \\
 $\Lambda^0_b\rightarrow p D^{*-}$    & $6.91\times 10^{-7}$  & $6.54\times10^{-7}$
  & $6.28\times 10^{-7}$  & $-$   \\
 $\Lambda^0_b\rightarrow p D^-_{s}$   & $1.70\times 10^{-5}$  & $1.61\times10^{-5}$
  & $1.54\times 10^{-5}$  & $<4.8\times 10^{-4}$  \\
 $\Lambda^0_b\rightarrow p D^{*-}_{s}$&$1.48\times 10^{-5}$   & $1.41\times10^{-5}$
  & $1.35\times 10^{-5}$ & $-$ \\ \hline\hline
\end{tabular}
\end{center}
\end{table}

\begin{table}
\caption{ Up-down and CP asymmetries for $\Lambda^0_b\rightarrow p + M$ decays.}\label{tab:non22}
\begin{center}
\begin{tabular}{l |c|c|c|c|c|c}\hline\hline
~~~Mode & \multicolumn{3}{c|}{\tabincell{c}{$\alpha$}}
 & \multicolumn{3}{c}{$A_{CP}$} \\ \cline{2-7}
 & ~$\mu=m_b/2$~ & ~~$\mu=m_b$~~ & ~$\mu=2m_b$~ & ~$\mu=m_b/2$~ & $\mu=m_b$ & $\mu=2m_b$ \\ \hline
$\Lambda^0_b\to p \pi^-$    & $-0.966$  & -0.978 & $-0.984$ & $-3.74\times10^{-2}$ & $-3.37\times10^{-2}$
 & $-3.08\times10^{-2}$ \\
$\Lambda^0_b\to p \rho^-$   & $-0.810$  & -0.810 & $-0.810$ & $-3.44\times10^{-2}$ & $-3.19\times10^{-2}$
 & $-2.96\times10^{-2}$ \\
$\Lambda^0_b\to p K^-$      & $0.463$   & $0.270$    & $0.134$   & $0.081$   & $0.101$  & $0.114$ \\
$\Lambda^0_b\to p K^{*-}$   & $-0.790$  & $-0.790$  & $-0.790$   & $0.339$   & $0.311$  & $0.292$ \\
$\Lambda^0_b\to p D^- $     & $-0.995$  & $-0.995$  & $-0.995$   & $0$       & $0$      & $0$  \\
$\Lambda^0_b\to p D^{*-}$   & $-0.518$  & $-0.518$  & $-0.518$   & $0$       & $0$      & $0$  \\
$\Lambda^0_b\to p D^-_{s}$  & $-0.993$  & $-0.993$  & $-0.993$   & $0$       & $0$      & $0$  \\
$\Lambda^0_b\to p D^{*-}_{s}$  & $-0.489$   & $-0.489$  & $-0.489$   & $0$   & $0$      & $0$  \\ \hline\hline
\end{tabular}
\end{center}
\end{table}

The theory predictions for the up-down and direct CP asymmetries are given in Table \ref{tab:non22}. Similar to the $\Lambda^0_b\rightarrow \Lambda^+_c + M$ decays, nearly all the up-down asymmetries $\alpha$ are negative. There is one exception. The up-down asymmetry in the $\Lambda^0_b\to p K^-$ is positive, and the value is small about 0.3. The reason is due to a significant contribution from the $a_6$ term.
The absolute values of $\alpha$ are about 1 for most processes.  The two processes with final states
$D_{(s)}^{*-}$ have the up-down asymmetries about $0.5$. The direct CP violations are at the order of $10^{-2}$  in $\Lambda^0_b\to p\pi^-(\rho^-)$ decays. The predictions for the direct CP violations in $\Lambda^0_b\to p K^-(K^{*-})$ decays are large, about 0.1 or 0.3. We will discuss the large CP violation in more detail below.

The process of $\Lambda^0_b\to p\pi^-$ is is important in phenomenology, like the $\bar B^0\to \pi^+\pi^-$ in the $B$ meson system. This process is observed in experiment, and the branching ratio is measured to be $(4.2\pm 0.8)\times 10^{-6}$. Similar to the definition of $R_2^{\Lambda_c}$, the ratio of the baryon-to-meson decay rates for proton is defined by
 \beq
 R_2^p=\frac{{\cal B}(\Lambda^0_b\to p \pi^-)}{{\cal B}(\bar B^0\to \pi^+\pi^-)}.
 \eeq
From the experiment data, $R_2^p=0.82\pm 0.16$. That means the branching ratio of ${\cal B}(\Lambda^0_b\to p \pi^-)$ is smaller than the corresponding meson process. However, for the $\Lambda^0_b\to \Lambda_c^+\pi^-$, its branching ratio is larger than the corresponding meson decay rate and the the ratio of baryon-to-meson $R_2^{\Lambda_c}\approx 2$. In fact, the fractions of $\Lambda^0_b\to \Lambda_c^+\pi^-$ is nearly equal to the sum of two ratios of ${\cal B}({\bar B^0}\to D^+\pi^-)$ and ${\cal B}({\bar B^0}\to D^{*+}\pi^-)$. If this rule can be applied to proton case, we expect ${\cal B}(\Lambda^0_b\to p \pi^-)={\cal B}(\bar B^0\to \pi^+ \pi^-)+{\cal B}(\bar B^0\to \rho^+ \pi^-)$. But the experimental data shows that ${\cal B}(\Lambda^0_b\to p \pi^-)<{\cal B}(\bar B^0\to \pi^+ \pi^-)$. Why the branching ratio of the $\Lambda^0_b\to p \pi^-$ decay is small? One reason may be the small form factors $f_1(0)\cong g_1(0)\cong 0.13$. If it is so, the ratio of $\Lambda^0_b\to p l^-\bar\nu_l$ decay should be smaller than $\bar B^0\to \pi^+ l^-\bar\nu_l$. But the data tell us that ${\cal B}(\Lambda^0_b\to p l^-\bar\nu_l)\approx 3{\cal B}(\bar B^0\to \pi^+ l^-\bar\nu_l)$.
We can look at this problem from another ratio of semi-leptonic to non-leptonic decay rates.

Similar to the definition of $R^{\Lambda_c}_{l\pi}$, the ratio of semi-leptonic to non-leptonic decay rates for proton case is defined by
\beq
 R^p_{l\pi}=\frac{{\cal B}(\Lambda^0_b\to p l^-\bar{\nu}_l)}{{\cal B}(\Lambda^0_b\to p \pi^-)}.
\eeq
In our model, the result is $R^p_{l\pi}=93.5\pm 4.5$.  From the experimental data, the fitted result is  $R^p_{l\pi}=97.6\pm 30.2$ which accords with the theory very well. But, for the $\Lambda_c$ case, $R^{\Lambda_c}_{l\pi}=12.6\pm 3.0$. There is a factor of about 7 difference between the two ratios. Replacing the lepton pair $l\nu_l$ by a quark-anti-quark pair, the semi-leptonic process is changed to the non-leptonic process. The great difference between the $\Lambda_c$ and $p$ processes is difficult to understand. It's another result caused by the small branching ratio of $\Lambda^0_b\to p \pi^-$.

The ratio of pion to kaon decay rates is defined by
 \beq
 R_{\pi K}^p=\frac{{\cal B}(\Lambda^0_b\to p \pi^-)}{{\cal B}(\Lambda^0_b\to p K^-)}.
 \eeq
The LHCb collaboration reported a result $R_{\pi K}^p=0.86\pm 0.08\pm 0.05$ \cite{Aaij:2012as}. It is close to our fitted value $R_{\pi K}^p=0.82\pm 0.23$. In our theory,  the ratio is $R_{\pi K}^p=1.98\pm 0.69$. Theoretical uncertainties are large, as can be seen from the $\mu$ dependence of the branching ratio of $\Lambda^0_b\to p K^-$. A discrepancy between theory and the experiment can be found. But they can be consistent with $2\sigma$ deviations. In pQCD approach \cite{Lu:2009cm}, $R_{\pi K}^p=2.6^{+2.0}_{-0.5}$ which obviously disagrees with the data. In the generalized factorization approach \cite{Hsiao:2014mua}, $R_{\pi K}^p=0.84\pm 0.09$ which accords with the data.

Similarly for the ratio of $\rho$ to $K^*$ is
 \beq
 R_{\rho K^*}^p=\frac{{\cal B}(\Lambda^0_b\to p \rho^-)}{{\cal B}(\Lambda^0_b\to p K^{*-})}.
 \eeq
The ratio of $R_{\rho K^*}^p$ is suggested to test different factorization approach since the ratio is free of the hadronic uncertainties from the baryon-to-baryon form factors \cite{Hsiao:2014mua}.  In our theory, the prediction gives $R_{\rho K^*}^p=7.4\pm 0.9$.  In the generalized factorization approach (GFA) \cite{Hsiao:2014mua}, $R_{\rho K^*}^p=4.6\pm 0.5\pm 0.1$. There is obvious disagreement between different approaches. The reason can be explained by the importance of non-factorizable contributions in penguin dominated processes. The calculations of these non-factorizbale contributions contain large theory uncertainties in different factorization approaches, such as $\mu$-dependence, some non-perturbative effects etc.. The disagreement between different approaches will become more serious for direct CP violation.

Up to now, there is no confirmed direct CP violation in $\Lambda_b$ decays.  A recent measurement of CP violation in decays $\Lambda_b^0\to p \pi^-(K^-)$ comes from the CDF collaboration \cite{Aaltonen:2014vra}
 \beq
 &&A_{CP}(\Lambda_b^0\to p\pi^-)=+0.06\pm 0.07({\rm stat})\pm 0.03({\rm syst}), \non \\
 &&A_{CP}(\Lambda_b^0\to p K^-)=-0.10\pm 0.08({\rm stat})\pm 0.04({\rm syst}).
 \eeq
The central value of direct CP asymmetry for the decay $\Lambda_b^0\to p K^-$ is negative. Due to large errors in the data, we may say that the results are consistent with 0. About these two processes,  our predictions from the QCDF approach are
 \beq
 &&A_{CP}(\Lambda_b^0\to p \pi^-)= (-3.4\pm 0.4)\times 10^{-2}, \non\\
 &&A_{CP}(\Lambda_b^0\to p K^-)= (10.1\pm 2.0)\times 10^{-2}.
 \eeq
Because the direct CP violation comes from interference, it is more sensitive to detail of theory model than the branching ratio. In Table \ref{ACPp}, the direct CP violation for the four charmless processes $p\pi^-, ~p\rho^-, ~pK^-, ~pK^{*-}$ within different approaches are given. From the Table, one can see that our results are close to those in the generalized factorization approach, and differs from those in the pQCD approach.

\begin{table}
\caption{Direct CP asymmetries $A_{CP}$ (in units of $10^{-2}$) in different factorization approaches.}\label{ACPp}
\begin{center}
\begin{tabular}{lcccc} \hline\hline
 ~~Mode & ~~ QCDF (this work)~ &  ~~GFA \cite{Hsiao:2014mua} ~~ & ~~~~~pQCD \cite{Lu:2009cm}~~ & Experiment \cite{Aaltonen:2014vra}  \\\hline
 $\Lambda^0_b\rightarrow p\pi^-$  & $-3.4\pm 0.4$ & $-3.9\pm 0.2$ & $-31^{+43}_{-1}$ & $6\pm7\pm3$ \\\hline
 $\Lambda^0_b\rightarrow p\rho^-$ & $-3.2\pm 0.2$ & $-3.7\pm 0.3$ & $-$              & $-$         \\\hline
 $\Lambda^0_b\rightarrow p K^-$   & $10.1\pm 2.0$ & $5.8\pm 0.2$  & $-5^{+26}_{-5}$  & $-10\pm8\pm4$\\\hline
 $\Lambda^0_b\rightarrow pK^{*-}$ & $31.1\pm 2.8$ & $19.6\pm 1.4$ & $-$      & $-$ \\ \hline\hline
\end{tabular}
\end{center}
\end{table}

The decay of $\Lambda_b^0\to p K^{*-}$ is interesting. We find a very large direct CP violation in our approach as
 \beq
 A_{CP}(\Lambda_b^0\to p K^{*-})= (31.1\pm 2.8)\times 10^{-2}.
 \eeq
The predictions of direct CP violation in QCDF approach is usually small because the origin of strong phase is perturbative. So this large direct CP violation is out of expectation. This unusual phenomenon was first observed in \cite{Hsiao:2014mua}. The authors use the generalized factorization approach and obtain the result $A_{CP}(\Lambda_b^0\to p K^{*-})=(19.6\pm 1.0\pm 1.0)\times 10^{-2}$ which is smaller than ours but is
still large. The direct CP violation in this case comes from interference of tree contribution with $v_ua_1$ term and penguin contribution with $v_ca_4^c$ term. Penguin contribution is larger than the tree but their magnitudes are at the same order. The interference of a similar magnitude of tree and penguin contributions with different weak and strong phases is possible to produce a large CP violation. The processes $\Lambda_b^0\to p \pi^-(\rho^-)$ are tree dominated, and the CP violation is small. For the process $\Lambda_b^0\to p K^-$, the penguin contribution is enhanced by $a_6$ term. This leads to a larger branching ratio but a smaller CP asymmetry. In our approach, ${\cal B}(\Lambda_b^0\to p K^-)\approx 2{\cal B}(\Lambda_b^0\to p K^{*-})$ and $A_{CP}(\Lambda_b\to p K)\approx \frac{1}{3}A_{CP}(\Lambda_b^0\to p K^{*-})$. The process $\Lambda_b^0\to p K^{*-}$ is the only process with ratio of order $10^{-6}$ and large direct CP asymmetry. But we must stress that the prediction of CP violation in $\Lambda_b^0\to p K^{*-}$ is not stable. A small enhancement in the penguin contribution would modify the prediction of CP asymmetry.

The sign of direct CP violation is important since it represents whether $b$ quark is more possible to decay or the opposite. It is known that QCDF approach fails to explain the direct CP violation in $B^0\to \pi^+K^{(*)-}$. The present data provide a precise and confirmed result:
$A_{CP}(\bar B^0\to \pi^+K^{-})=-0.082\pm 0.006$, $A_{CP}(\bar B^0\to \pi^+K^{*-})=-0.22\pm 0.06$.
The direct CP violation is large and negative. However, the prediction of QCDF approach is small, only several percent and the sign is positive \cite{Beneke:2003zv}. How to explain a large and negative CP asymmetry is a difficult and unsolved puzzle in QCDF approach. In \cite{Beneke:2003zv}, the authors suggested a scenario (called by Scenario S3 (¡°universal annihilation¡±)) enhanced by weak annihilation. By choosing a phenomenological parameter of annihilation contribution and a proper strong phase, the direct CP violation can be changed to be negative. Since weak annihilation is non-pertubative, the importance of weak annihilation also implies the importance of non-perturbative effects on the strong phase.
We don't know what is the case in the heavy baryon system. The cental value of $A_{CP}(\Lambda_b^0\to p K^-)$ from CDF collaboration measurement is negative may be an indication.
Our prediction within QCDF approach is positive. Certainly, nothing is certain at present. We hope that the future experiment can provide some helps for us to think deeply about this question. So, the measurement of direct CP violation in $\Lambda_b^0\to p K^-$ and $\Lambda_b^0\to p K^{*-}$ decays is not only important to test different factorization approaches but also to explore the relation between the baryon and meson systems.

It seems that the results in the generalized factorization approach are more favorable \cite{Hsiao:2014mua} in phenomenology. But the generalized factorization approach is in principle a phenomenological method. To account for the non-factorizable corrections, a phenomenological color number $N_c^{eff}$ is introduced
and the effective coefficients for $b\to d$ and $b\to s$ transitions are different. The theory uncertainties caused by these treatments are difficult to estimate. The gluon momentum in the penguin loop is not determined. These conceptual problems are solved by QCD factorization. QCD factorization approach is rigorous in leading power of $1/m_b$. Beyond the leading power, the theory uncertainties is also not under control. Compared to the generalized factorization approach, the vertex corrections provide another source of strong phase in the QCDF approach. This may be the main reason that our predictions of CP violation for $\Lambda_b^0\to p K^-$ and $\Lambda_b^0\to p K^{*-}$ decays are larger than the ones in the generalized factorization approach. In phenomenology, the predictions of QCDF approach considering only the vertex and penguin corrections in this study should be consistent with those in the generalized factorization approach when $N_c^{eff}=3$.

\subsection{$\Lambda^0_b\rightarrow \Lambda + M$ decays}

There are fourteen processes for the class of $\Lambda^0_b\rightarrow \Lambda + M$. The theory predictions  and the experimental data for the branching ratios of $\Lambda^0_b\rightarrow \Lambda+ M$ decays are given in Table \ref{tab:non31}. The first eight processes which contain light meson are charmless modes. Their ratios are small, at the order of $10^{-8}$ to $10^{-6}$. Comparing these ratios with the $\Lambda^0_b\rightarrow p + M$ processes and the B meson data, the ratios are smaller by about one order or even two orders. Our theory predictions rely on the assumption of SU(3) symmetry relation for $\beta$ parameters $\beta_p=\beta_\Lambda$. Relaxing this restriction cannot produce a big enhancement because the numerical results are less sensitive to the variation of $\beta_\Lambda$. The processes with charmonium states $\eta_c$ and $J/\psi$ have the largest fractions of order of $10^{-4}$. The remained four processes with a final $D$ meson have ratios of $10^{-7}$ to $10^{-6}$. They have only the color-suppressed and Cabibbo-suppressed tree diagram contributions, so these processes have small fractions and no CP violation. The theory predictions for the up-down and direct CP asymmetries are given in Table \ref{tab:non32}.

\begin{table}
\caption{Branching ratios of $\Lambda^0_b\rightarrow \Lambda + M$ decays.}\label{tab:non31}
\begin{center}
\scalebox{1}[0.9]{
\begin{tabular}{l | c |c | c |c}\hline\hline
~~~Mode & ~~~~$\mu=m_b/2$~~~~ & ~~~~~~$\mu=m_b$~~~~~~ & ~~~~$\mu=2m_b$~~~~ & Experiment \\\hline
$\Lambda^0_b\to \Lambda \pi^0$   & $6.52\times 10^{-8}$ & $5.74\times10^{-8}$ & $5.26\times 10^{-8}$ & $-$ \\
$\Lambda^0_b\to \Lambda \rho^0$  & $1.13\times 10^{-7}$ & $9.75\times10^{-8}$ & $8.50\times 10^{-8}$ & $-$ \\
$\Lambda^0_b\to \Lambda K^0$     & $1.11\times 10^{-7}$ & $7.58\times10^{-8}$ & $5.84\times 10^{-8}$ & $-$ \\
$\Lambda^0_b\to \Lambda K^{*0}$  & $2.76\times 10^{-8}$ & $2.76\times10^{-8}$ & $2.59\times 10^{-8}$ & $-$ \\
$\Lambda^0_b\to \Lambda \eta$    & $6.37\times 10^{-7}$ & $4.39\times10^{-7}$ & $3.38\times 10^{-7}$
 & $(9.3^{+7.3}_{-5.3})\times 10^{-6}$ \cite{Aaij:2015eqa} \\
$\Lambda^0_b\to \Lambda \eta'$   & $6.75\times 10^{-6}$ & $4.03\times10^{-6}$ & $2.84\times 10^{-6}$
 & $<3.1\times10^{-6}$ \cite{Aaij:2015eqa} \\
$\Lambda^0_b\to \Lambda \omega$  & $2.08\times 10^{-8}$ & $1.13\times10^{-8}$ & $8.35\times 10^{-9}$
 & $-$ \\
$\Lambda^0_b\to \Lambda \phi$    & $6.93\times 10^{-7}$ & $6.33\times10^{-7}$ & $5.65\times 10^{-7}$
 & $(2.0\pm 0.5)\times 10^{-6}$ \cite{Patrignani:2016xqp} \\
$\Lambda^0_b\to \Lambda \eta_c$  & $2.80\times 10^{-4}$ & $2.47\times10^{-4}$ & $2.28\times 10^{-4}$
 & $-$ \\
$\Lambda^0_b\to \Lambda J/\psi$  & $5.34\times 10^{-4}$ & $4.67\times 10^{-4}$ & $4.38\times 10^{-4}$
 & $(5.8\pm0.8)\times10^{-5}/f_{\Lambda_b}$ \cite{Patrignani:2016xqp} \\
$\Lambda^0_b\to \Lambda D^0$     & $3.79\times 10^{-6}$ & $3.37\times10^{-6}$ & $3.18\times 10^{-6}$ & $-$ \\
$\Lambda^0_b\to \Lambda D^{*0}$  & $3.82\times 10^{-6}$ & $3.39\times10^{-6}$ & $3.20\times 10^{-6}$ & $-$ \\
$\Lambda^0_b\to \Lambda \bar{D}^0$  & $5.38\times 10^{-7}$ & $4.78\times10^{-7}$ & $4.51\times 10^{-7}$
 & $-$ \\
$\Lambda^0_b\to \Lambda \bar{D}^{*0}$ & $5.42\times 10^{-7}$ & $4.81\times10^{-7}$ & $4.54\times 10^{-7}$
 & $-$ \\ \hline\hline
\end{tabular}}
\end{center}
\end{table}

\begin{table}
\caption{ Up-down and CP asymmetries for $\Lambda^0_b\rightarrow \Lambda + M$ decays.}\label{tab:non32}
\begin{center}
\begin{tabular}{l |c|c|c|c|c|c}\hline\hline
~~~Mode &\multicolumn{3}{c|}{\tabincell{c}{$\alpha$}}&\multicolumn{3}{c}{$A_{CP}$}\\\cline{2-7}
& ~~$\mu=m_b/2$~~ & ~~$\mu=m_b$~~ & ~~$\mu=2m_b$~~ & ~~$\mu=m_b/2$~~ & ~~$\mu=m_b$~~ & ~~$\mu=2m_b$~~  \\\hline
$\Lambda^0_b\to\Lambda \pi^0$    & $-1$      & $-1$     & $-1$     & $0.331$    & $0.250$    & $0.202$  \\
$\Lambda^0_b\to\Lambda \rho^0$   & $-0.849$  & $-0.849$ & $-0.849$ & $0.328$    & $0.253$    & $0.210$  \\
$\Lambda^0_b\to\Lambda K^0$      & $0.599$   & $0.410$  & $0.282$  & $-0.224$   & $-0.206$   & $-0.189$ \\
$\Lambda^0_b\to\Lambda K^{*0}$   & $-0.828$  & $-0.828$ & $-0.828$ & $-0.325$   & $-0.251$   & $-0.210$ \\
$\Lambda^0_b\to\Lambda \eta$     & $0.433$   & $0.236$  & $0.116$  & $-0.028$   & $-0.034$   & $-0.038$ \\
$\Lambda^0_b\to\Lambda \eta'$    & $0.998$   & $0.991$  & $0.956$  & $0.008$    & $0.010$    & $0.011$  \\
$\Lambda^0_b\to\Lambda \omega$   & $-0.848$  & $-0.848$ & $-0.848$ & $0.600$    & $0.586$    & $0.408$  \\
$\Lambda^0_b\to\Lambda \phi$     & $-0.803$  & $-0.803$ & $-0.803$ & $0.020$    & $0.016$    & $0.013$  \\
$\Lambda^0_b\to\Lambda \eta_c$   & $-0.985$  & $-0.985$ & $-0.985$ & $0$        & $0$        & $0$ \\
$\Lambda^0_b\to\Lambda J/\psi$   & $-0.206$  & $-0.206$ & $-0.206$ & $0$        & $0$        & $0$ \\
$\Lambda^0_b\to\Lambda D^0$      & $-0.998$  & $-0.998$ & $-0.998$ & $0$        & $0$        & $0$ \\
$\Lambda^0_b\to\Lambda D^{*0}$   & $-0.539$  & $-0.539$ & $-0.539$ & $0$        & $0$        & $0$ \\
$\Lambda^0_b\to\Lambda \bar{D}^0$    & $-0.998$  & $-0.998$  & $-0.998$  & $0$  & $0$        & $0$ \\
$\Lambda^0_b\to\Lambda \bar{D}^{*0}$ & $-0.539$  & $-0.539$  & $-0.539$  & $0$  & $0$        & $0$ \\ \hline\hline
\end{tabular}
\end{center}
\end{table}

The $\Lambda^0_b\rightarrow \Lambda \pi^0(\rho^0)$ processes has no QCD penguin contributions. The $b\to s\bar u u$ transition is cancelled by $b\to s \bar d d$ contribution because the opposite sign for $\bar u u$ and $\bar d d$ components in $\pi^0(\rho^0)$. For the $\bar B^0\to \bar K^0\pi^0$ process, there is one extra term by the Fierz transformation, so that $b\to s \bar d d$ QCD penguin contribution is not canceled. The experimental data gives ${\cal B}(\bar B^0\to \bar K^0\pi^0)=(9.9\pm 0.5)\times 10^{-6}$ which is very large.  But for the baryon case, there is no QCD penguin contribution. This difference between the meson and baryon is due to a fact that the spectator in the baryon is a diquark and it is an anti-quark in the meson. The tree diagram is color-suppressed and is further suppressed by small CKM elements $V_{ub}V^{\ast}_{us}$. The electroweak penguin contribution is small  but cannot be neglected in this case.  The branching ratios are predicted to be very small, at the order of $10^{-7}$ or $10^{-8}$. They have large direct CP asymmetry, about 30\%, but difficult to measure in experiment.

The $\Lambda^0_b\rightarrow \Lambda  K^0(K^{*0})$ processes have no tree diagram contribution.
They are the pure penguin processes which is QCD penguin dominated. But they are $b\to d$ transition where the CKM elements $V_{tb}V^{\ast}_{td}$ is suppressed. For the $\Lambda^0_b\to \Lambda  K^{*0}$ process, only $a_4$ term contributes, the ratio is predicted to be very small, only at the order of $10^{-8}$. For the $\Lambda^0_b\to \Lambda K^0$, there is chirally-enhanced $a_6$ term, so the ratio is increased to be about $10^{-7}$. The direct CP violation is large for these two processes.

The $\Lambda^0_b\to \Lambda \eta(\eta')$ processes are important in phenomenology. They contain  information of $\eta-\eta'$ mixing and QCD anomaly related to $\eta'$. In this study, we don't consider the anomaly contribution to $\eta'$. The two processes $\Lambda^0_b\to \Lambda \eta(\eta')$ are $b\to s$ QCD penguin dominated.  The $a_6$ term is chirally-enhanced by $R_\eta$ or $R_{\eta'}$ defined in the Appendix \ref{function}. For the $\Lambda^0_b\to \Lambda \eta$ process, our approach gives the branching ratio ${\cal B}(\Lambda^0_b\to \Lambda \eta)=(3-6)\times 10^{-7}$ with large theoretical uncertainties. A recent measurement from the LHCb collaboration gives $(9.3^{+7.3}_{-5.3})\times 10^{-6}$. The experimental error is quite large. But it is certain that our theory prediction is smaller than the data. For the $\Lambda^0_b\to \Lambda \eta'$ process, our approach gives prediction as ${\cal B}(\Lambda^0_b\to \Lambda \eta')=(3-7)\times 10^{-6}$, which is about one order larger than the $\eta$ process. The LHCb data gives an upper limit ${\cal B}(\Lambda^0_b\to \Lambda \eta')<3.1\times10^{-6}$, which is close to the lower limit of our prediction. The further experiment may show some discrepancies between theory and experiment. The direct CP violation in these two processes are both small.

One can define a ratio of $\eta$ to $\eta'$ to reduce some model dependence. For this purpose, a ratio $R_{\eta\eta'}^{\Lambda}$ is defined by
\beq
R_{\eta\eta'}^{\Lambda}=\frac{{\cal B}(\Lambda^0_b\to \Lambda \eta)}{{\cal B}(\Lambda^0_b\to \Lambda \eta')}.
\eeq
In our approach, $R_{\eta\eta'}^{\Lambda}=0.11^{+0.12}_{-0.06}$. One early study used the generalized factorization approach and the results are \cite{Ahmady:2003jz}:  ${\cal B}(\Lambda^0_b\to \Lambda \eta)=11.47\times 10^{-6}$, ${\cal B}(\Lambda^0_b\to \Lambda \eta')=11.33\times 10^{-6}$, and $R_{\eta\eta'}^{\Lambda}=1.01$, for form factors calculated in QCD sum rules; ${\cal B}(\Lambda^0_b\to \Lambda \eta)=2.95\times 10^{-6}$, ${\cal B}(\Lambda^0_b\to \Lambda \eta')=3.24\times 10^{-6}$, and $R_{\eta\eta'}^{\Lambda}=0.91$, for form factors calculated in a pole model.
Another study also uses the generalized factorization approach \cite{Geng:2016gul}, and
the results are: ${\cal B}(\Lambda^0_b\to \Lambda \eta)=(1.47\pm 0.35)\times 10^{-6}$, ${\cal B}(\Lambda^0_b\to \Lambda \eta')=(1.83\pm 0.58)\times 10^{-6}$, and $R_{\eta\eta'}^{\Lambda}=0.80\pm 0.32$.
One can see a large difference in predictions between different approaches. The reason leads to the difference may be: (1) Anomaly contribution. In \cite{Geng:2016gul}, one effect of anomaly term is introduced in the $\eta(\eta')$ decay constants. (2) $a_6$ and $a_8$ contributions. In our study, we used the equation of motion, the $a_6$ and $a_8$ terms are enhanced by factor $R_{\eta'}=m_{\eta'}^2/(m_bm_s)=2.2$. Our prediction for the ratio ${\cal B}(\Lambda^0_b\to \Lambda \eta')$ is large. There is no enhancement for $\Lambda^0_b\to \Lambda \eta$, so the predicted ratio is small.

The $\Lambda^0_b\to \Lambda \omega$ process contains both the tree and penguin contributions. The tree is color-suppressed and CKM parameter suppressed. It seems that this process should be dominated by $b\to s$ transition QCD penguin. But the prediction of the ratio is very small, only at the order of $10^{-8}$. The reason is due to a destructive interference in the $a_3$, $a_5$, $a_9$ terms. This case is very similar to the cancellation of QCD penguin in $\Lambda^0_b\to \Lambda \pi^0(\rho)$ decays. The direct CP violation in $\Lambda^0_b\to \Lambda \omega$ is predicted to be quite large, about 60\%, but the small decay ratio makes it impossible to measure in experiment.

The process $\Lambda^0_b\to \Lambda \phi$ is interesting in both theory and experiment. In SM, the process can only be occurred through loop effects described by $b\to s \bar s s$ penguin diagrams. This flavor-changing-neutral-current (FCNC) transition is very sensitive to new physics effects. From experimental point of view, the measurement of its decay ratio, CP violation, and T-odd observable provide an important test of SM and different new physics models. The direct CP violation is predicted to be small, about $1-2\%$. The up-down asymmetry $\alpha$ is $-0.8$ in our approach. In experiment, this process has been observed. The measurement from the LHCb collaboration gives ${\cal B}(\Lambda^0_b \to \Lambda\phi)=(5.18 \pm 1.04 \pm 0.35^{+0.67}_{-0.62})\times 10^{-6}$ \cite{Aaij:2016zhm}. From the PDG on the web, 2017 updated result gives ${\cal B}(\Lambda^0_b \to \Lambda\phi)=(2.0\pm 0.5)\times 10^{-6}$ \cite{Patrignani:2016xqp}. The central value is lowered by a factor of 2 compared to the LHCb data. Our theory prediction is ${\cal B}(\Lambda^0_b \to \Lambda\phi)=(5-7)\times 10^{-7}$ which is smaller than the data. By comparison, the result in \cite{Geng:2016gul} using the generalized factorization approach gives ${\cal B}(\Lambda^0_b \to \Lambda\phi)=(3.53\pm 0.24)\times 10^{-6}$ when the number of color is chosen as $N_c^{eff}=2$.

Why our theory prediction is smaller than the data? One reason may be the small $\Lambda_b\to \Lambda$ form factors. By increasing the $\Lambda_b\to \Lambda$ form factors, the ratio of $\Lambda^0_b\to \Lambda \phi$ is increased. But the ratios of processes $\Lambda^0_b\to p \pi^-(K^-)$ will be larger than the data. Thus, this explanation is excluded. Another reason is the non-factorizable effects. In this study, we only consider the vertex and penguin corrections. There are other effects, such as hard spectator interactions, power corrections, etc..  According to the meson-baryon similarity, one can use the data of the meson process to extract the strong interaction information. All the non-factorizable effects are included in the effective coefficients. From Eq. (\ref{eq:relation}), the combined coefficient of $\Lambda^0_b\rightarrow \Lambda\phi$ is equal to the coefficient of the corresponding meson process $\bar{B}^0\rightarrow \bar{K}^0 \phi$. By use of the $\bar{B}^0\rightarrow \bar{K}^0 \phi$, the combined coefficient $\bar a$ can be obtained. Then, one can give prediction for the $\Lambda^0_b\rightarrow \Lambda\phi$ decay. The advantage of this method is that the theoretical uncertainties of the QCDF approach are reduced by the experiment data. This method has been adopted for $\Lambda^0_b\rightarrow pK^-$ process in \cite{Zhu:2016bra}. We want to note that this method is not rigorous for $\Lambda^0_b\rightarrow pK^-$ because the difference of chirally-enhanced term in the baryon and meson systems. The application of $\Lambda^0_b\rightarrow pK^-$ is based upon assumption that the chirally-enhanced contribution does not change the meson-baryon relation significantly. Table \ref{meson} gives the predictions of branching ratios of $\Lambda^0_b\rightarrow \Lambda \phi$ and $\Lambda^0_b\rightarrow p K^-$ by use of the meson-baryon similarity.

\begin{table}
\caption{ Estimations for the branching ratios of $\Lambda^0_b\rightarrow pK^-$ and $\Lambda^0_b\rightarrow \Lambda\phi$ processes by using the meson data.}\label{meson}
\begin{center}
\scalebox{1}[0.9]{
\begin{tabular}{l |c|c}\hline\hline
 ~~Mode & ~~~~~Theory~~~~~  & ~~~~~~~Experiment~~~~~~~ \\\hline
 $\Lambda^0_b\to p K^-$~~~~      & $6.67\times10^{-6}$   & $(5.1\pm1.0)\times 10^{-6}$  \\\hline
 $\Lambda^0_b\to\Lambda\phi$ & $1.76\times 10^{-6}$  & $(2.0\pm 0.5)\times 10^{-6}$ \\\hline\hline
\end{tabular}}
\end{center}
\end{table}

From Table \ref{meson}, we can find that the prediction of $\Lambda^0_b\rightarrow \Lambda\phi$ decay coincides with the experimental data very well. It verifies our speculation that the non-factorizable effects lead to the difference between the theory prediction of QCDF approach and the experimental data. However, it is not easy to improve the QCDF predictions because of technical difficulties. For example, the power corrections are non-perturbative in principle. The calculations is difficult and model dependent. The estimation of hard spectator interactions also requires some phenomenological parameters.

The $\Lambda^0_b\to \Lambda \eta_c(J/\psi)$ processes proceed via $b\to s\bar c c$ transitions at the quark level. The tree diagram is color suppressed but the CKM elements $V_{cb}V^{\ast}_{cs}$ are large. The QCD penguin contributions are important. Their ratios are predicted to be large, at the order of $10^{-4}$.
Because the $\Lambda^0_b\to \Lambda J/\psi$ process is more interesting in experiment. We discuss this process in more detail.

From PDG, one can find that the ratio of $\Lambda^0_b\to \Lambda J/\psi$ process is not given directly. The data gives a value of the ratio of $\Lambda^0_b\to \Lambda J/\psi$ multiplied by a ratio of $\Lambda_b^0$ production. This is because there is no an accepted measurement of the production rate of $\Lambda^0_b$ which is defined by $f_{\Lambda_b^0}\equiv {\cal B}(b \to \Lambda_b^0)$. In literature, the choice of $f_{\Lambda_b^0}$ is different and arbitrary. In this study, we take the averaged value from Heavy Flavor Averaging Group \cite{Amhis:2016xyh}. Some other production rates are also provided for reference. We introduce $f_u$, $f_d$, $f_s$, $f_{\rm baryon}$, $f_{\Lambda_b^0}$ as fractions of $B^+$, $B^0$, $B_s^0$, $b$ baryon, $\Lambda_b^0$. For CDF measurement, $f_{\Lambda_b^0}/(f_u+f_d)=0.229\pm 0.062$, $f_u=f_d=0.340\pm 0.021$, $f_s=0.101\pm 0.015$, $f_{\rm baryon}=0.218\pm 0.047$ when using the Tevatron data only. Then, we obtain
 \beq
 f_{\Lambda_b^0}=0.156\pm 0.045.
 \eeq
In the previous study \cite{Wei:2009np}, $f_{\Lambda_b^0}=0.1$. By use of the above value of $f_{\Lambda_b^0}=0.156\pm 0.045$, the experimental data for the branching ratio of $\Lambda^0_b\to \Lambda J/\psi$ process can be given to be
 \beq
 {\cal B}(\Lambda^0_b\to \Lambda J/\psi)=(3.72\pm 1.07)\times 10^{-4}.
 \eeq
Our theory prediction is ${\cal B}(\Lambda^0_b\to \Lambda J/\psi)=4.67\times 10^{-4}$ when $\mu=m_b$. It is consistent with the experimental data within one standard deviation. The consistency is based upon that we choose a large $a_2$ for calculations. Considering only vertex and penguin corrections in leading power of $1/m_b$, the obtained $a_2$ is small and insufficient to explain the data for the color-suppressed processes. In fact, for the process $\Lambda^0_b\to \Lambda J/\psi$ where the the non-factorizable effects are substantial, the theoretical uncertainties in QCDF approach is very large although the factorization is applicable. By comparison, the result in \cite{Cheng:1996cs} gives the ratio ${\cal B}(\Lambda^0_b\to \Lambda J/\psi)=1.6\times 10^{-4}$ which is smaller than ours by a factor of 3 but still consistent with the data.

The up-down asymmetry $\alpha$ is also an experimentally interested quantity. From PDG, the parameter $\alpha$ for $\Lambda^0_b\rightarrow J/\psi\Lambda$ is $\alpha=0.18\pm 0.13$ \cite{Patrignani:2016xqp}. A recent measurement from the CMS collaboration gives
$\alpha=0.14\pm 0.14 {\rm (stat)}\pm 0.10{\rm (syst)}$ \cite{Sirunyan:2018bfd}.
Both the results are consistent with 0. Our theory prediction is $\alpha=-0.206$.

\subsection{$\Lambda^0_b\rightarrow n + M$ decays}

Up to now, there is no any experimental data on the process of $\Lambda^0_b\rightarrow n + M$. One reason is the difficulty in detection of the neutron. Maybe the future experiment can overcome this difficulty to improve the study in this class of processes. The theory predictions for the branching ratios of decays $\Lambda^0_b\rightarrow n+ M$ are given in Table \ref{tab:non41}. The up-down and CP asymmetries are given in Table \ref{tab:non41}. We will discuss $\Lambda^0_b\rightarrow n + M$ decays similar to the $\Lambda^0_b\rightarrow \Lambda + M$ decays.

\begin{table}
\caption{Branching ratios of $\Lambda^0_b\rightarrow n + M$ decays.}\label{tab:non41}
\begin{center}
\begin{tabular}{l | c |c | c }\hline\hline
~~~Mode & ~~~~~~~~$\mu=m_b/2$~~~~~~~~ & ~~~~~~~~$\mu=m_b$~~~~~~~~ & ~~~~~~~~$\mu=2m_b$~~~~~~~~ \\ \hline
$\Lambda^0_b\to n \pi^0$       & $1.34\times 10^{-7}$ & $1.14\times10^{-7}$ & $1.03\times 10^{-7}$ \\
$\Lambda^0_b\to n \rho^0$      & $2.13\times 10^{-7}$ & $1.89\times10^{-7}$ & $1.75\times 10^{-7}$ \\
$\Lambda^0_b\to n \bar K^0$    & $3.02\times 10^{-6}$ & $2.01\times10^{-6}$ & $1.52\times 10^{-6}$ \\
$\Lambda^0_b\to n \bar K^{*0}$ & $8.83\times 10^{-7}$ & $8.04\times10^{-7}$ & $7.16\times 10^{-7}$ \\
$\Lambda^0_b\to n \eta$        & $2.76\times 10^{-8}$ & $2.46\times10^{-8}$ & $2.39\times 10^{-8}$ \\
$\Lambda^0_b\to n \eta'$       & $8.32\times 10^{-8}$ & $5.02\times10^{-8}$ & $3.41\times 10^{-8}$ \\
$\Lambda^0_b\to n \omega$      & $8.89\times 10^{-8}$ & $8.85\times10^{-8}$ & $9.09\times 10^{-8}$ \\
$\Lambda^0_b\to n \phi$        & $3.03\times 10^{-9}$ & $2.29\times10^{-9}$ & $1.90\times 10^{-9}$ \\
$\Lambda^0_b\to n \eta_c$      & $1.59\times 10^{-5}$ & $1.43\times10^{-5}$ & $1.33\times 10^{-5}$ \\
$\Lambda^0_b\to n J/\psi$~~    & $2.36\times 10^{-5}$ & $2.06\times10^{-5}$ & $1.93\times 10^{-5}$ \\
$\Lambda^0_b\to n D^0$         & $7.26\times 10^{-5}$ & $6.45\times10^{-5}$ & $6.09\times 10^{-5}$ \\
$\Lambda^0_b\to n D^{*0}$      & $7.54\times 10^{-5}$ & $6.70\times10^{-5}$ & $6.32\times 10^{-5}$ \\
$\Lambda^0_b\to n \bar{D}^0$   & $2.93\times 10^{-8}$ & $2.60\times10^{-8}$ & $2.46\times 10^{-8}$ \\
$\Lambda^0_b\to n\bar{D}^{*0}$ & $3.05\times 10^{-8}$ & $2.71\times10^{-8}$ & $2.55\times 10^{-8}$ \\\hline\hline
\end{tabular}
\end{center}
\end{table}

\begin{table}
\caption{ Up-down and CP asymmetries for $\Lambda^0_b\rightarrow n+ M$ decays.}\label{tab:non42}
\begin{center}
\begin{tabular}{l |c|c|c|c|c|c}\hline\hline
 ~~~Mode &\multicolumn{3}{c|}{\tabincell{c}{$\alpha$}}&\multicolumn{3}{c}{$A_{CP}$}\\\cline{2-7}
 & ~~$\mu=m_b/2$~~ & ~~$\mu=m_b$~~ & ~~$\mu=2m_b$~~ & ~~$\mu=m_b/2$~~ &~~$\mu=m_b$~~ &~~$\mu=2m_b$~~
  \\\hline
$\Lambda^0_b\to n \pi^0$        & $-0.740$  & $-0.816$ & $-0.857$ & $0.227$   & $0.222$   & $0.202$  \\
$\Lambda^0_b\to n \rho^0$       & $-0.810$  & $-0.810$ & $-0.810$ & $0.338$   & $0.294$   & $0.253$  \\
$\Lambda^0_b\to n \bar K^0$     & $ 0.542$  & $ 0.376$ & $ 0.264$ & $0.011$   & $0.010$   & $0.009$  \\
$\Lambda^0_b\to n \bar K^{*0}$  & $-0.790$  & $-0.790$ & $-0.790$ & $0.016$   & $0.012$   & $0.010$  \\
$\Lambda^0_b\to n \eta$         & $-0.333$  & $-0.552$ & $-0.676$ & $-0.482$  & $-0.431$  & $-0.378$ \\
$\Lambda^0_b\to n \eta'$        & $-0.693$  & $-0.694$ & $-0.685$ & $ 0.529$  & $ 0.364$  & $ 0.194$ \\
$\Lambda^0_b\to n \omega$       & $-0.809$  & $-0.809$ & $-0.809$ & $-0.497$  & $-0.422$  & $-0.357$ \\
$\Lambda^0_b\to n \phi$         & $-0.776$  & $-0.776$ & $-0.776$ & $0$       & $0$       & $0$      \\
$\Lambda^0_b\to n \eta_c$       & $-0.964$  & $-0.964$ & $-0.964$ & $-0.033$  & $-0.017$  & $-0.007$ \\
$\Lambda^0_b\to n J/\psi$~~     & $-0.206$  & $-0.206$ & $-0.206$ & $0.017$   & $0.0126$  & $0.010$  \\
$\Lambda^0_b\to n D^0$          & $-0.995$  & $-0.995$ & $-0.995$ & $0$       & $0$       & $0$     \\
$\Lambda^0_b\to n D^{*0}$       & $-0.519$  & $-0.519$ & $-0.519$ & $0$       & $0$       & $0$     \\
$\Lambda^0_b\to n \bar{D}^0$    & $-0.995$  & $-0.995$ & $-0.995$ & $0$       & $0$       & $0$     \\
$\Lambda^0_b\to n\bar{D}^{*0}$  & $-0.519$  & $-0.519$ & $-0.519$ & $0$       & $0$       & $0$     \\ \hline\hline
\end{tabular}
\end{center}
\end{table}

Unlike the $\Lambda^0_b\to \Lambda \pi^0(\rho^0)$ processes where QCD penguin contributions cancel, $\Lambda^0_b\to n \pi^0(\rho^0)$ processes contain both the tree and penguin contributions. The tree diagram is color-suppressed and the CKM elements is $V_{ub}V^{\ast}_{ud}$. The QCD penguin is $b\to d$ transition which is suppressed by $V_{cb}V^{\ast}_{cd}$ or
$V_{ub}V^{\ast}_{ud}$. The tree and the penguin contribution are at the same order. The predicted branching ratios are at the order of $10^{-7}$. The direct CP violation is very large for these two processes, about 20-30\%. Considering the meson decay $\bar B^0\to \pi^0\pi^0$, the predicted ratio in the QCDF approach is also of order $10^{-7}$ but the data is about $10^{-6}$. The non-factorizable effects must be important in $\Lambda^0_b\to n \pi^0(\rho^0)$ processes. The measurement of $\Lambda^0_b\to n \pi^0(\rho^0)$ can test the effects of non-factorizable contributions.

The $\Lambda^0_b\to n  \bar K^0(\bar K^{*0})$ processes have no tree diagram contribution. Similar to $\Lambda^0_b\to \Lambda \phi$, they are the pure penguin processes dominated by QCD penguin. At the quark level, penguin diagram proceeds via $b\to s\bar d d$ transition where the CKM elements $V_{tb}V^{\ast}_{ts}$ is not suppressed. The ratios are predicted to be large, at the order of about $10^{-6}$. Explicitly, they are
 \beq
 {\cal B}(\Lambda^0_b\to n  \bar K^0)=(2.0\pm 0.5)\times 10^{-6}, \qquad
 {\cal B}(\Lambda^0_b\to n  \bar K^{*0})=(7.9\pm 0.8)\times 10^{-7}.
 \eeq
The $\Lambda^0_b\to n  \bar K^0$ process is expected to be observed in future experiment. One the contrary, due to the large decay ratio, the  direct CP violations in $\Lambda^0_b\to n  \bar K^0(\bar K^{*0})$ processes are both small, only about 1\%.

The $\Lambda^0_b\to n \eta(\eta')$ processes also provide information of the $\eta-\eta'$ mixing. But the penguin contributions proceed via $b\to d$ transitions which are suppressed by small CKM elements. So the ratios of these two processes are very small, only at the order of $10^{-8}$. The direct  CP violation is predicted to be about 40\%, but difficult to measure. Similarly, $\Lambda^0_b\to n \omega(\phi)$ processes are $b\to d$ transitions, and the branching ratios are small.

\begin{table}
\caption{ Branching ratios ${\cal B}$ and direct CP asymmetries $A_{CP}$ in GFA and our approach for the charmless prcoesses.}\label{tab:non51}
\begin{center}
\begin{tabular}{l |c|c|c|c}\hline\hline
 ~~~Mode & \multicolumn{2}{c|}{\tabincell{c}{${\cal B}\times 10^6$}}
  & \multicolumn{2}{c}{$A_{CP}\times 10^2$}\\ \cline{2-5}
  & ~~~~~~GFA \cite{Hsiao:2017tif}~~~~~~ & ~~~~~This work~~~~~
  & ~~~~~~GFA \cite{Hsiao:2017tif}~~~~~~ & ~~~~~This work~~~ \\\hline
 $\Lambda^0_b\to p \pi^-$        & $4.25^{+1.04}_{-0.48}$   & $4.30^{+0.27}_{-0.19}$
                                 & $-3.9\pm 0.4$            & $-3.4\pm 0.3$           \\
 $\Lambda^0_b\to p \rho^-$       & $11.03^{+2.72}_{-1.25}$  & $7.47^{+0.42}_{-0.30}$
                                 & $-3.8\pm 0.4$            & $-3.2\pm 0.2$           \\
 $\Lambda^0_b\to p K^-$          & $4.49^{+0.84}_{-0.39}$   & $2.17^{+0.98}_{-0.47}$
                                 & $6.7\pm 0.3$             & $10.1^{+1.3}_{-2.0}$    \\
 $\Lambda^0_b\to p K^{*-}$       & $2.86^{+0.62}_{-0.29}$   & $1.01\pm 0.07$
                                 & $19.7\pm 1.4$            & $31.1^{+2.8}_{-1.9}$    \\
 $\Lambda^0_b\to\Lambda \pi^0$   & $(3.4^{+0.8}_{-0.4})\times 10^{-2}$
  & $(5.74^{+0.78}_{-0.48})\times 10^{-2}$    & $0.0$       & $25.0^{+8.1}_{-4.8}$    \\
 $\Lambda^0_b\to\Lambda \rho^0$  & $(9.5^{+3.0}_{-1.3})\times 10^{-2}$
  & $(9.75^{+1.55}_{-1.25})\times 10^{-2}$ & $2.3^{+0.7}_{-0.8}$  & $25.3^{+7.5}_{-4.3}$  \\
 $\Lambda^0_b\to\Lambda K^0$     & $(9.4^{+2.3}_{-3.8})\times 10^{-3}$
  & $(7.58^{+3.52}_{-1.74})\times 10^{-2}$    & $0.2^{+0.1}_{-0.0}$  & $-20.6\pm 1.7$  \\
 $\Lambda^0_b\to\Lambda K^{*0}$  & $(9.2^{+4.7}_{-2.0})\times 10^{-2}$
  & $(2.76^{+0.0}_{-0.17})\times 10^{-2}$     & $1.3\pm 0.1$   & $-25.1^{+4.1}_{-7.4}$ \\
 $\Lambda^0_b\to\Lambda \eta$    & $1.59^{+0.38}_{-0.17}$   & $0.44^{+0.20}_{-0.10}$
                                 & $0.4\pm 0.2$             & $-3.4^{+0.6}_{-0.4}$     \\
 $\Lambda^0_b\to\Lambda \eta'$   & $1.90^{+0.68}_{-0.23}$   & $4.03^{+2.72}_{-1.19}$
                                 & $1.6\pm 0.1$             & $1.0^{+0.1}_{-0.2}$      \\
 $\Lambda^0_b\to\Lambda \omega$  & $0.71^{+1.59}_{-0.70}$   & $(1.1^{+1.0}_{-0.3})\times 10^{-2}$
                                 & $3.6^{+4.8}_{-4.0}$      & $58.6^{+1.4}_{-17.8}$    \\
 $\Lambda^0_b\to\Lambda \phi$    & $1.77^{+1.65}_{-1.68}$   & $0.63^{+0.06}_{-0.07}$
                                 & $1.4^{+0.7}_{-0.1}$      & $1.6^{+0.4}_{-0.3}$      \\
 $\Lambda^0_b\to n \pi^0$        & $0.10\pm 0.03$           & $0.11^{+0.02}_{-0.01}$
                                 & $8.0^{+1.2}_{-1.4}$      & $22.2^{+0.5}_{-2.0}$     \\
 $\Lambda^0_b\to n \rho^0$       & $0.18\pm 0.09$           & $0.19^{+0.02}_{-0.01}$
                                 & $14.0\pm 1.8$            & $29.4^{+4.4}_{-4.1}$     \\
 $\Lambda^0_b\to n \bar K^0$     & $4.61^{+1.31}_{-0.58}$   & $2.01^{+1.01}_{-0.49}$
                                 & $ 1.1\pm 0.0$            & $1.0\pm 0.01$            \\
 $\Lambda^0_b\to n \bar K^{*0}$  & $3.09^{+1.57}_{-0.67}$   & $0.80\pm 0.08$
                                 & $1.3\pm 0.1$             & $1.2^{+0.4}_{-0.2}$      \\
 $\Lambda^0_b\to n \eta$         & $(6.9^{+2.7}_{-2.4})\times 10^{-2}$
  & $(2.46^{+0.30}_{-0.07})\times 10^{-2}$  & $-16.8\pm 2.1$  & $-43.1^{+5.3}_{-5.1}$  \\
 $\Lambda^0_b\to n \eta'$        & $(4.2\pm 1.8)\times 10^{-2}$
  & $(5.02^{+3.30}_{-1.59})\times 10^{-2}$  & $-15.7^{+4.0}_{-5.6}$  & $36.4^{+16.5}_{-17.0}$  \\
 $\Lambda^0_b\to n \omega$       & $0.22^{+0.16}_{-0.10}$   & $(8.85^{+0.04}_{-0.0})\times 10^{-2}$
                                 & $-18.2^{+24.4}_{-4.2}$   & $-42.2^{+6.5}_{-7.5}$    \\
 $\Lambda^0_b\to n \phi$         & $0.02^{+0.17}_{-0.02}$
  & $(2.29^{+0.84}_{-0.39})\times 10^{-3}$  & $-8.8^{+7.4}_{-5.1}$      & $0$   \\ \hline\hline
\end{tabular}
\end{center}
\end{table}

The $\Lambda^0_b\to n \eta_c(J/\psi)$ processes proceed via $b\to d\bar c c$ transitions at the quark level. The tree diagram is color suppressed and the CKM elements $V_{cb}V^{\ast}_{cd}$ are suppressed. The predicted branching ratios are at the order of $10^{-5}$, which is smaller than ratios of $\Lambda^0_b\to \Lambda \eta_c(J/\psi)$ decays by one order. The CP violation is small, too. The processes $\Lambda^0_b\to n D^0(D^{*0})$ have the color-suppressed tree diagram contribution.  The branching ratios are orders of $10^{-5}$. The processes of $\Lambda^0_b\to n \bar D^0(\bar D^{*0})$ are further suppressed by small CKM element. The branching ratios are orders of $10^{-9}$ and direct CP violation is 0.

In \cite{Hsiao:2017tif}, the authors provide predictions of branching ratios and direct CP asymmetries for
20 charmless processes in the generalized factorization approach (GFA). We compare their results with ours in Table \ref{tab:non51}. For the errors of their results, we only list the error from non-factorizable effects or the largest error due to limit of space. From Table \ref{tab:non51}, most predictions in the two approaches are consistent within the theoretical uncertainties. There are some exceptions. The difference in $\Lambda^0_b\to\Lambda \eta(\eta')$ processes has been explained in the above subsection. Our prediction for the ratio of $\Lambda^0_b\to\Lambda \omega$ decay is small. But the errors of GFA result is large and the two approaches are consistent. For the direct CP violation, nearly all of our predictions are larger than the results of GFA. In some processes with small ratios, the difference becomes very obvious.

\section{Conclusions and discussions}

In this study, we provide a comprehensive study of the semi-leptonic and non-leptonic decays of $\Lambda_b^0$. Compared to our previous analysis, there are several improvements. The baryon-baryon form factors are calculated in the covariant light-front approach where the quantities of $f_3$ and $g_3$ can be evaluated. Different ratios and asymmetries in the six semi-leptonic processes are studied. The two-body non-leptonic decays are analyzed beyond the tree operator contribution. The penguin diagram contributions including the QCD and electroweak operators are taken into account.  We calculate the non-leptonic decays of $\Lambda^0_b$ into a baryon plus a s-wave meson (pseudoscalar or vector) including 44 processes in total within the framework of QCD factorzation approach. For some processes, our calculations are given for the first time up to our knowledge. Among the 44 processes, there are about 9 processes observed in experiment. Compared to the precise and large amount of data for the $B$ meson from PDG, the experimental results for $\Lambda_b^0$ are very few. The weak decays of $\Lambda_b^0$ provide an important place to explore CP violation and QCD dynamics in the baryon environment. We hope that this work can promote the study of $\Lambda_b$ and provide a reference for future experiments.

For the semi-leptonic processes, the theory predictions are in accord with the experiment. This accordance verifies the diquark approximation. The semi-leptonic decays with tau lepton are predicted to be at the same order as the light lepton process. The ratios of tau to electron or muon decays provide a test of theory models in SM. We extract the CKM parameter $|V_{ub}|$ from the data of $\Lambda_b^0\to p\mu^-\bar \nu_{\mu}$ by use of our model.

For the non-leptonic decays $\Lambda_b^0\to \Lambda_c^+ D_{(s)}^{(*)-}$ where the final states are both heavy, factorization hypothesis works very well. But in QCDF, these processes are not factorizable. We test the factorization assumption by use of several relative ratios and don't find deviations.
The mechanism of factorization should be beyond the "color transparency" argument and the perturbative framework. The large $N_c$ limit is also not a justified mechanism of factorization. There must be some non-perturbative mechanisms which prefer the factorization of a large-size charmed meson from a soft background.

The charmless non-leptonic decays are interesting in both theory and experiment. The branching ratios of the observed $\Lambda_b^0$ decays are at the order of $10^{-6}$. By comparison, the corresponding $B$ meson decays have the ratios of order of $10^{-5}$. This fact implies that the ratios of the $\Lambda_b^0$ decays are smaller than those of the $B$ meson by about one order. Because the data for the $\Lambda_b^0$ and B meson decays in the semi-leptonic and charmful non-leptonic processes are quite similar, the difference occurred in the charmless non-leptonic processes seems to be a problem. From the theoretical point of view, the baryon-to-baryon transition form factors have to be adjusted to be small, about 0.1. The heavy-to-light form factors for the $B$ meson are about 0.3. A natural question arises: why the heavy-to-light baryon form factors are smaller than the heavy-to-light meson form factors by a factor of 2 or 3? With the diquark picture, it is difficult to understand this question.

According to the numerical results, we list the processes with large branching ratios which may be observed in the future experiment: $\Lambda_c^+\rho^-$, $\Lambda_c^+K^{*-}$, $\Lambda_c^+ D^{*-}$, $\Lambda_c^+ D_s^{*-}$, $p\rho^-$, $pK^{*-}$, $pD_s^-$, $pD_s^{*-}$, $\Lambda\eta'$, $\Lambda\eta_c$, $\Lambda D^0$, $\Lambda D^{*0}$, $n\bar K^0$, $n\bar K^{*0}$, $n \eta_c$, $n J/\psi$, $n D^0$, $n D^{*0}$.

The direct CP violations in the processes of $p K$ and $pK^*$ are predicted to be large. The values are about 10\% and 30\%, respectively. The $pK^*$ process are most promising. This phenomenon was first observed in \cite{Hsiao:2014mua} by use of the generalized factorization approach. Their prediction of direct CP asymmetry is 20\%. Our prediction is larger than theirs. In QCDF approach, the vertex corrections provide another source of strong phase. The large CP violation is caused by the interference of tree and penguin contributions. The $pK^*$ process is a rare case that the tree and the dominant QCD penguin contributions  have the same magnitude and contain different weak and strong phases.

We compare our results with the predictions given in the generalized factorization approach. We find that most results of the two approaches are consistent within the theoretical errors. This is not accidental.
Our results should be close to the predictions in generalized factorization approach when $N_c=3$.
QCD factorization solves some conceptual problems in the generalized factorization and develops a more rigorous method. We stress that we neglect some non-factorizable effects in our calculations, such as the hard spectator scattering, weak annihilation etc.. These effects are important in phenomenology. When the data becomes more precise, these effects should be taken into account.

Under the diquark approximation, the baryon is similar to meson. The $\Lambda\phi$ process can be used to test the meson-baryon similarity. Replace a diquark with an antiquark, $\Lambda^0_b\rightarrow \Lambda\phi$ process is changed to $\bar{B}^0\rightarrow \bar{K}^0 \phi$. At the quark level, the QCD dynamics for the two processes are same. By use of the data of $\bar{B}^0\rightarrow \bar{K}^0 \phi$, we can extract the combined coefficient and then predict the ratio of $\Lambda^0_b\rightarrow \Lambda\phi$. The prediction by this method coincides with the experiment very well.

Conventional wisdom is that baryon system is more complicated than the meson. This opinion is based upon the three-quark picture for a baryon. The complication can be seen clearly in an analysis of $\Lambda_b\to p\pi(K)$ process in the perturbative QCD approach \cite{Lu:2009cm}.  There are more than 100 Feynmann diagrams even at the tree level. However, our study may provide another picture: the baryon is as simple as a meson. The bridge to relate the baryon and meson is the diquark. This study, in particular in decays of $\Lambda^0_b\rightarrow \Lambda\phi$, and many previous studies verify the effectiveness of the diquark assumption.  With the diquark approximation, the study of heavy baryon may be developed to a similar stage as the $B$ meson physics.

\appendix

\section{The conventional light-front approach}

In the conventional light-front approach, a baryon $\Lambda_Q$ with total momentum $P$ and spin $S=1/2$ is composed of a quark $q_1$ and a scalar diquark can be written as
\begin{eqnarray}
|\Lambda_Q(P,S,S_z)\rangle&=&\int\{d^3p_1\}\{d^3p_2\}2(2\pi)^2\delta^3(\widetilde{P}-\widetilde{p_1}
-\widetilde{p_2})\nonumber\\
&\times&\sum_{\lambda_1}\Psi^{SS_z}(\widetilde{p_1},\widetilde{p_2},\lambda_1)
C^{\alpha}_{\beta\gamma}F^{bc}|Q_{\alpha}(p_1,\lambda_1)[q^{\beta}_{1b}q^{\gamma}_{2c}](p_2)\rangle,
\end{eqnarray}
where Q represent b, c, s u, d, [$q_1q_2$] represents [$ud$], $\lambda$ denotes helicity.  $p_1$, $p_2$ are the on-mass-shell light-front momenta defined by
\begin{eqnarray}
\widetilde{p}=(p^{+},p_{\perp}),~~~~~~~~~p_{\perp}=(p^1,p^2),~~~~~~~~~p^-=\frac{m^2+p_{\perp}^2}{p^+},
\end{eqnarray}
and
\begin{eqnarray}
&&\{d^3p\}\equiv\frac{dp^+d^2p_{\perp}}{2(\pi)^3}, \qquad \delta^3(\widetilde{p})=\delta(p^+)\delta^2(p_{\perp}),\non\\
&&|Q(p_1,\lambda_1)[q_1q_2](p_2)\rangle=b^{\dag}_{\lambda_1}(p_1)a^{\dag}(p_2)|0\rangle,\nonumber\\
&&\left[a(p'),a^{\dag}(p)\right]=2(2\pi)^3\delta^3(\widetilde{p}'-\widetilde{p}),\nonumber\\
&&\{d_{\lambda'}(p'),d^{\dag}_{\lambda}(p)\}=2(2\pi)^3\delta^3(\widetilde{p}'-\widetilde{p})
\delta_{\lambda'\lambda}.
\end{eqnarray}

The coefficient $C^{\alpha}_{\beta\gamma}$ is a normalized color factor and $F^{bc}$ is a normalized flavor coefficient. They satisfy
\begin{eqnarray}
&&C^{\alpha}_{\beta\gamma}F^{bc}C^{\alpha'}_{\beta'\gamma'}F^{b'c'}\langle Q_{\alpha'}(p_1',\lambda_1')[q^{\beta'}_{1b'}q^{\gamma'}_{2c'}](p'_2)\mid Q_{\alpha}(p_1,\lambda_1)[q^{\beta}_{1b}q^{\gamma}_{2c}](p_2)\rangle \nonumber\\
&&~=2^2(2\pi)^6\delta^3(\widetilde{p}'_1-\widetilde{p}_1)\delta^3(\widetilde{p}'_2-\widetilde{p}_2)
\delta_{\lambda'_1\lambda_1}.
\end{eqnarray}

The intrinsic variables $(x_i,k_{i\perp})$ with $i=1,2$ are
\begin{eqnarray}
&&p_1^{+}=x_1P^+, \qquad p_2^+=x_2P^+, \qquad x_1+x_2=1,\nonumber\\
&&p_{1\perp}=x_1P_{\perp}+k_{1\perp}, \qquad p_{2\perp}=x_2P_{\perp}+k_{2\perp},
 \qquad k_{\perp}=-k_{1\perp}=k_{2\perp},
\end{eqnarray}
where $x_i$ with $0<x_1,x_2<1$ are the light-front momentum fractions. The variables $(x_i,k_{i\perp})$ are independent of the total momentum of the hadron and thus are Lorentz-invariant variables. The invariant mass square $M_0^2$ is defined as
\begin{eqnarray}
M^2_0=\frac{k^2_{1\perp}+m_1^2}{x_1}+\frac{k^2_{2\perp}+m_2^2}{x_2}.
\end{eqnarray}

We define the internal momenta as
\begin{eqnarray}
k_i=(k^-_i,k^+_i,k_{i\perp})=(e_i-k_{iz},e_i+k_{iz},k_{i\perp})=(\frac{m_i^2+k^2_{i\perp}}{x_iM_0},x_iM_0,k_{i\perp}).
\end{eqnarray}
Then, it is easy to obtain
\begin{eqnarray}
M_0&=&e_1+e_2,\nonumber\\
e_i&=&\frac{x_iM_0}{2}+\frac{m_i^2+k_{i\perp}^2}{2x_iM_0}=\sqrt{m_i^2+k_{i\perp}^2+k^2_{iz}},\nonumber\\
k_{iz}&=&\frac{x_iM_0}{2}-\frac{m^2_i+k^2_{i\perp}}{2x_iM_0}.
\end{eqnarray}
where $e_i$ denotes the energy of the i-th constituent. $k_{i\perp}$ and $k_{iz}$ constitute a momentum vector $\overrightarrow{k}_i=(k_{i\perp},k_{iz})$ and correspond to the components in the transverse and z directions respectively.

The momentum-space function $\Psi^{SS_z}$ in Eq. (A1) is expressed as
\begin{eqnarray}
\Psi^{SS_z}(\widetilde{p}_1,\widetilde{p}_2,\lambda_1)=\langle\lambda_1\mid\mathcal{R}^{\dag}_M(x_1,k_{1\perp},m_1)\mid s_1\rangle\langle 00;\frac{1}{2}s_1\mid\frac{1}{2}S_z\rangle\phi(x,k_{\perp}),
\end{eqnarray}
where $\phi(x,k_{\perp})$ is the light-front wave function which describes the momentum distribution of the constituents in the bound state with $x=x_2$, $k_{\perp}=k_{2\perp}$; $\langle 00;\frac{1}{2}s_1|\frac{1}{2}S_z\rangle$ is the corresponding Clebsch-Gordan coefficient with spin $s=s_z=0$ for the scalar diquark; $\langle\lambda_1\mid\mathcal{R}^{\dag}_{M}(x_1,k_{1\perp},m_1)\mid s_1\rangle $ is the well-known Melosh transformation matrix element which transforms the conventional spin states in the instant form into the light-front helicity eigenstates,
 \begin{eqnarray}
 \langle\lambda_1\mid \mathcal{R}^{\dag}_M(x_1,k_{1\perp},m_1)\mid s_1\rangle &=&\frac{\bar{u}(k_1,\lambda_1)u_D(k_1,s_1)}{2m_1}\nonumber\\
 &=&\frac{(m_1+x_1M_0)\delta_{\lambda_1s_1}+i\overrightarrow{\sigma}_{\lambda_1s_1}\cdot\overrightarrow{k}_{1\perp}\times
 \overrightarrow{n}}{\sqrt{(m_1+x_1M_0)^2+k_{1\perp}^2}},
 \end{eqnarray}
where $u_{(D)}$ denotes  a Dirac spinor in the  the light-front (instant) form and $\widetilde{n}=(0,0,1)$ is a unit vector in the $z$ direction. In practice, it is more convenient to use the covariant form
\begin{eqnarray}
&&\langle\lambda_1\mid\mathcal{R}^{\dag}(x_1,k_{1\perp},m_1)\mid s_1\rangle\langle00;\frac{1}{2}s_1\mid\frac{1}{2}S_z\rangle\nonumber\\
&&=\frac{1}{\sqrt{2M_0(e_1+m_1)}}\bar{u}(p_1,\lambda_1)\Gamma u(\bar{P},S_z),
\end{eqnarray}
where $\Gamma=1$ for scalar diquark.

The heavy baryon state is normalized as
\begin{eqnarray}
\langle\Lambda(P',S',S_z')\mid\Lambda(P,S,S_z)\rangle=2(2\pi)^3P^+\delta^3(\widetilde{P}'-\widetilde{P})\delta_{S'S}\delta_{S_z'S_z}.
\end{eqnarray}
Thus, the light-front wave function satisfies the constraint
\begin{eqnarray}
\int\frac{dxd^2k_{\perp}}{2(2\pi^3)}\mid\phi(x,k_{\perp})\mid^2=1.
\end{eqnarray}


\section{ The coefficient $a_i$ in QCDF approach}

Here, we give the results for the coefficients $a_i$ at next-to-leading order in $\alpha_s$. From \cite{Beneke:2001ev}, their formulae are given by
\begin{eqnarray}
 a_1&=&C_1+\frac{C_2}{N_c}\left[1+\frac{C_F\alpha_s}{4\pi}V_M\right],\nonumber\\[8pt]
 a_2&=&C_2+\frac{C_1}{N_c}\left[1+\frac{C_F\alpha_s}{4\pi}V_M\right],\nonumber\\[8pt]
 a_3&=&C_3+\frac{C_4}{N_c}\left[1+\frac{C_F\alpha_s}{4\pi}V_M\right],\nonumber\\[8pt]
 a_4^q&=&C_4+\frac{C_3}{N_c}\left[1+\frac{C_F\alpha_s}{4\pi}V_M\right]+\frac{C_F\alpha_s}{4\pi}
 \frac{P^q_{M,2}}{N_c},\nonumber\\[8pt]
 a_5&=&C_5+\frac{C_6}{N_c}\left[1+\frac{C_F\alpha_s}{4\pi}(-V'_M)\right],\nonumber\\[8pt]
 a_6^q&=&C_6+\frac{C_5}{N_c}\left(1-6\frac{C_F\alpha_s}{4\pi}\right)+\frac{C_F\alpha_s}{4\pi}
 \frac{P^q_{M,3}}{N_c},\nonumber\\[8pt]
  a_7&=&C_7+\frac{C_8}{N_c}\left[1+\frac{C_F\alpha_s}{4\pi}(-V'_M)\right],\nonumber\\[8pt]
 a_8^q&=&C_8+\frac{C_7}{N_c}\left(1-6\frac{C_F\alpha_s}{4\pi}\right)+\frac{\alpha}{9\pi}
 \frac{P^{q,EW}_{M,3}}{N_c},\nonumber\\[8pt]
 a_9&=&C_9+\frac{C_{10}}{N_c}\left[1+\frac{C_F\alpha_s}{4\pi}(-V'_M)\right],\nonumber\\[8pt]
 a_{10}^q&=&C_{10}+\frac{C_9}{N_c}\left[1+\frac{C_F\alpha_s}{4\pi}V_M\right]+\frac{\alpha}{9\pi}
 \frac{P^{q,EW}_{M,2}}{N_c}.
\end{eqnarray}
where $C_i\equiv C_i(\mu)$, $\alpha_s\equiv\alpha_s(\mu)$, $C_F=(N^2_c-1)/(2N_c)$, and $N_c=3$.

The vertex corrections are given by
\begin{eqnarray}
V_M&=&12\ln\frac{m_b}{\mu}-18+\int_0^1dx ~g(x)\Phi_M(x),\nonumber\\
V'_M&=&12\ln\frac{m_b}{\mu}-6+\int_0^1dx ~g(1-x)\Phi_M(x),\nonumber\\
g(x)&=&3\left(\frac{1-2x}{1-x}\ln x-i\pi\right)\nonumber\\
&&+\left[2\text{Li}_2(x)-\ln^2x+\frac{2\ln x}{1-x}-(3+2i\pi)\ln x-(x\leftrightarrow1-x)\right],\nonumber
\end{eqnarray}
where $\phi_M(x)=6x(1-x)$ is the leading-twist light-cone distribution amplitudes. The asymptotic form of the twist-2 distribution amplitude is adopted. A discussion on the non-asymptotic form of the pion distribution amplitude can be found in \cite{Chang:2013pq}. For the asymptotic form, we have $\int_0^1dx~g(x)\phi_M(x)=-\frac{1}{2}-3i\pi$.

The penguin contributions are given by
\begin{eqnarray*}
P^q_{M,2}&=&C_1\left[\frac{4}{3}\ln\frac{m_b}{\mu}+\frac{2}{3}-G_M(s_q)\right]
+C_3\left[\frac{8}{3}\ln
\frac{m_b}{\mu}+\frac{4}{3}-G_M(0)-G_M(1)\right]\\
&&+(C_4+C_6)\left[\frac{4n_f}{3}\ln\frac{m_b}{\mu}-(n_f-2)G_M(0)-G_M(s_c)-G_M(1)\right]\\
&&-2C^{eff}_{8g}\int^1_0\frac{dx}{1-x}\phi_M(x),\\
P^{q,EW}_{M,2}&=&(C_1+N_cC_2)\left[\frac{4}{3}\ln\frac{m_b}{\mu}+\frac{2}{3}
-G_M(s_q)\right]-3C^{eff}_{7\gamma}\int_0^1\frac{1}{1-x}\phi_M(x),
\end{eqnarray*}
where $n_f=5$ is the number of light quark flavors, and $s_u=0$, $s_c=(m_c/m_b)^2$ are mass ratios involved in the penguin diagrams.
The function $G_M(s)$ is given by
\begin{eqnarray*}
G_K(s)&=&\int_0^1dx~G(s-i\epsilon,1-x)\phi_M(x),\\
G(s,x)&=&-4\int_0^1du~u(1-u)\ln[s-u(1-u)x]\\
&=&\frac{2(12s+5x-3x\ln s)}{9x}-\frac{4\sqrt{4s-x}(2s+x)}{3x^{3/2}}\arctan\sqrt{\frac{x}{4s-x}}~,
\end{eqnarray*}
and
\begin{eqnarray*}
G_M(s_c)&=&\frac{5}{3}-\frac{2}{3}\ln s_c+\frac{32}{3}s_c+16s_c^2-\frac{2}{3}\sqrt{1-4s_c}(1+2s_c+24s_c^2)\\
&&\times(2\text{arctanh}\sqrt{1-4s_c}-i\pi)
+12s_c^2\left[1-\frac{4}{3}s_c\right]\times(2\text{arctanh}\sqrt{1-4s_c}-i\pi)^2,\\
G_M(0)&=&\frac{5}{3}+\frac{2i\pi}{3},\\
G_M(1)&=&\frac{85}{3}-6\sqrt{3}\pi+\frac{4\pi^2}{9}.
\end{eqnarray*}
where $\int\frac{dx}{1-x}\phi_M(x)=3$.

The twist-3 terms are
\begin{eqnarray*}
P^q_{M,3}&=&C_1\left[\frac{4}{3}\ln\frac{m_b}{\mu}+\frac{2}{3}-\hat{G}_M(s_q)\right]
+C_3\left[\frac{8}{3}\ln\frac{m_b}{\mu}+\frac{4}{3}-\hat{G}_M(0)-\hat{G}_M(1)\right]\\
&&+(C_4+C_6)\left[\frac{4n_f}{3}\ln\frac{m_b}{\mu}-(n_f-2)\hat{G}_M(0)-\hat{G}_M(s_c)
-\hat{G}_M(1)\right]-2C_{8g}^{eff},\\
P^{q,EW}_{M,3}&=&(C_1+N_cC_2)\left[\frac{4}{3}\ln\frac{m_b}{\mu}+\frac{2}{3}
-\hat{G}_M(s_q)\right]-3C^{eff}_{7\gamma},
\end{eqnarray*}
where
$$\hat{G}_M(s)=\int^1_0dx~G(s-i\epsilon,1-x)\phi_q^M(x).$$

The asymptotic twist-3 distribution amplitude is $\phi_q^M(x)=1$. We have
\begin{eqnarray*}
\hat{G}_M(s_c)&=&\frac{16}{9}(1-3s_c)-\frac{2}{3}\left[\ln s_c+(1-4s_c)^{3/2}(2\text{arctanh}\sqrt{1-4s_c}-i\pi)\right],\\
\hat{G}_M(0)&=&\frac{16}{9}+\frac{2\pi}{3}i,\\
\hat{G}_M(1)&=&\frac{2\pi}{\sqrt{3}}-\frac{32}{9}.
\end{eqnarray*}

The numerical values of the Wilson coefficients are taken from \cite{Beneke:2001ev} and are collected in Table \ref{tab:ciapp}.

\begin{table}
\caption{The Wilson coefficients $C_i$ at different scale $\mu$.}\label{tab:ciapp}
\begin{center}
\begin{tabular}{c|c|c|c|c|c|c}\hline\hline
~~~~~~~~$\mu$~~~~~~~~ & ~~~~$C_1$~~~~ & ~~~~$C_2$~~~~ & ~~~~$C_3$~~~~ & ~~~~$C_4$~~~~ & ~~~~$C_5$~~~~
 & ~~~~$C_6$~~~~ \\ \hline
$\mu=m_b/2$ & 1.185    & $-0.387$ & 0.018    & $-0.038$ & 0.010    & $-0.053$ \\
$\mu=m_b$   & 1.117    & $-0.268$ & 0.012    & $-0.027$ & 0.008    & $-0.034$ \\
$\mu=2m_b$  & 1.074    & $-0.181$ & 0.008    & $-0.019$ & 0.006    & $-0.022$ \\\hline\hline
~~$\mu$~~   & $C_7/\alpha$        & $C_8/\alpha$        & $C_9/\alpha$
            & $C_{10}/\alpha$     & $C^{eff}_{7\gamma}$ & $C^{eff}_{8g}$      \\\hline
$\mu=m_b/2$ & $-0.012$ & 0.045    & $-1.358$ & 0.418    & $-0.364$ & $-0.169$ \\
$\mu=m_b$   & $-0.001$ & 0.029    & $-1.276$ & 0.288    & $-0.318$ & $-0.151$ \\
$\mu=2m_b$  & $ 0.018$ & 0.019    & $-1.212$ & 0.193    & $-0.281$ & $-0.316$ \\\hline\hline
\end{tabular}
\end{center}
\end{table}

\section{$\lambda$ functions for different decay modes}\label{function}

\noindent (1)~~ $\Lambda^0_b\rightarrow\Lambda^+_c+M$ processes

\vspace{0.8cm}

In $\Lambda^0_b\rightarrow\Lambda^+_c\pi^-$,
$$\lambda=\frac{G_F}{\sqrt{2}}f_{\pi}V_{cb}V^{\ast}_{ud}a_1.$$
\vspace{0.8cm}

In $\Lambda^0_b\rightarrow\Lambda^+_c\rho^-$,
$$\lambda=\frac{G_F}{\sqrt{2}}f_{\rho}V_{cb}V^{\ast}_{ud}a_1.$$
\vspace{0.8cm}

In $\Lambda^0_b\rightarrow\Lambda^+_c K^-$,
$$\lambda=\frac{G_F}{\sqrt{2}}f_K V_{cb}V^{\ast}_{us}a_1.$$
\vspace{0.8cm}

In $\Lambda^0_b\rightarrow\Lambda^+_c K^{\ast-}$,
$$\lambda=\frac{G_F}{\sqrt{2}}f_{K^{\ast}}V_{cb}V^{\ast}_{us}a_1.$$
\vspace{0.8cm}

In $\Lambda^0_b\rightarrow\Lambda^+_c D^-$,

A term:
\begin{eqnarray*}
\lambda&=&\frac{G_F}{\sqrt{2}}f_D\left[V_{cb}V^{\ast}_{cd}a_1+V_{ub}V^{\ast}_{ud}
(a_4^u+a_{10}^u)+V_{cb}V^{\ast}_{cd}(a_4^c+a_{10}^c)\right.\\
&&\left.+R_{D^-}(V_{ub}V^{\ast}_{ud}(a_6^u+a_{8}^u)+V_{cb}V^{\ast}_{cd}(a_6^c+a_{8}^c))\right],
\end{eqnarray*}

B term:
\begin{eqnarray*}
\lambda&=&\frac{G_F}{\sqrt{2}}f_D\left[V_{cb}V^{\ast}_{cd}a_1+V_{ub}V^{\ast}_{ud}
(a_4^u+a_{10}^u)+V_{cb}V^{\ast}_{cd}(a_4^c+a_{10}^c)\right.\\
&&\left.-R_{D^-}(V_{ub}V^{\ast}_{ud}(a_6^u+a_{8}^u)+V_{cb}V^{\ast}_{cd}(a_6^c+a_{8}^c))\right],
\end{eqnarray*}
with $ R_{D^-}=\frac{2m^2_{D^-}}{(m_c+m_d)m_b}.$
\vspace{0.8cm}

In $\Lambda^0_b\rightarrow\Lambda^+_c D^{\ast-}$,
\begin{eqnarray*}
\lambda&=&\frac{G_F}{\sqrt{2}}f_{D^{\ast}}\left[V_{cb}V^{\ast}_{cd}a_1+V_{ub}
V^{\ast}_{ud}(a_4^u+a_{10}^u)+V_{cb}V^{\ast}_{cd}(a_4^c+a_{10}^c)\right].
\end{eqnarray*}
\vspace{0.8cm}

In $\Lambda^0_b\rightarrow\Lambda^+_c D_s^-$,

A term:
\begin{eqnarray*}
\lambda&=&\frac{G_F}{\sqrt{2}}f_{D_s}\left[V_{cb}V^{\ast}_{cs}a_1+V_{ub}V^{\ast}_{us}
(a_4^u+a_{10}^u)+V_{cb}V^{\ast}_{cs}(a_4^c+a_{10}^c)\right.\\
&&\left.+R_{D_s^-}(V_{ub}V^{\ast}_{us}(a_6^u+a_{8}^u)+V_{cb}V^{\ast}_{cs}(a_6^c+a_{8}^c))\right],
\end{eqnarray*}

B term:
\begin{eqnarray*}
\lambda&=&\frac{G_F}{\sqrt{2}}f_{D_s}\left[V_{cb}V^{\ast}_{cs}a_1+V_{ub}V^{\ast}_{us}
(a_4^u+a_{10}^u)+V_{cb}V^{\ast}_{cs}(a_4^c+a_{10}^c)\right.\\
&&\left.-R_{D_s^-}(V_{ub}V^{\ast}_{us}(a_6^u+a_{8}^u)+V_{cb}V^{\ast}_{cs}(a_6^c+a_{8}^c))\right],
\end{eqnarray*}
with $ R_{D_s^-}=\frac{2m^2_{D_s^-}}{(m_c+m_s)m_b}$.
\vspace{0.8cm}

In $\Lambda^0_b\rightarrow\Lambda^+_c D_s^{\ast-}$,
\begin{eqnarray*}
\lambda&=&\frac{G_F}{\sqrt{2}}f_{D_s^{\ast}}\left[V_{cb}V^{\ast}_{cs}a_1+V_{ub}V^{\ast}_{us}
(a_4^u+a_{10}^u)+V_{cb}V^{\ast}_{cs}(a_4^c+a_{10}^c)\right].
\end{eqnarray*}

\vspace{1.5cm}

\noindent (2)~~ $\Lambda^0_b\rightarrow p+M$ processes
\vspace{0.8cm}

In $\Lambda^0_b\rightarrow p \pi^-$,

A term:
\begin{eqnarray*}
\lambda&=&\frac{G_F}{\sqrt{2}}f_{\pi}\left[V_{ub}V^{\ast}_{ud}a_1+V_{ub}V^{\ast}_{ud}
(a_4^u+a_{10}^u)+V_{cb}V^{\ast}_{cd}(a_4^c+a_{10}^c)\right.\\
&&+R_{\pi^-}\left.(V_{ub}V^{\ast}_{ud}(a_6^u+a_{8}^u)+V_{cb}V^{\ast}_{cd}(a_6^c+a_{8}^c))\right],
\end{eqnarray*}

B term:
\begin{eqnarray*}
\lambda&=&\frac{G_F}{\sqrt{2}}f_{\pi}\left[V_{ub}V^{\ast}_{ud}a_1+V_{ub}V^{\ast}_{ud}
(a_4^u+a_{10}^u)+V_{cb}V^{\ast}_{cd}(a_4^c+a_{10}^c)\right.\\
&&-R_{\pi^-}\left.(V_{ub}V^{\ast}_{ud}(a_6^u+a_{8}^u)+V_{cb}V^{\ast}_{cd}(a_6^c+a_{8}^c))\right],
\end{eqnarray*}
with $ R_{\pi^-}=\frac{2m^2_{\pi^-}}{(m_u+m_d)m_b}$.
\vspace{0.8cm}

In $\Lambda^0_b\rightarrow p\rho^-$,
$$\lambda=\frac{G_F}{\sqrt{2}}f_{\rho}\left[V_{ub}V^{\ast}_{ud}a_1+V_{ub}V^{\ast}_{ud}
(a_4^u+a_{10}^u)+V_{cb}V^{\ast}_{cd}(a_4^c+a_{10}^c)\right].$$
\vspace{0.8cm}

In $\Lambda^0_b\rightarrow p K^-$,

A term:
\begin{eqnarray*}
\lambda&=&\frac{G_F}{\sqrt{2}}f_K\left[V_{ub}V^{\ast}_{us}a_1+V_{ub}V^{\ast}_{us}
(a_4^u+a_{10}^u)+V_{cb}V^{\ast}_{cs}(a_4^c+a_{10}^c)\right.\\
&&+R_{K^-}\left.(V_{ub}V^{\ast}_{us}(a_6^u+a_{8}^u)+V_{cb}V^{\ast}_{cs}(a_6^c+a_{8}^c))\right],
\end{eqnarray*}

B term:
\begin{eqnarray*}
\lambda&=&\frac{G_F}{\sqrt{2}}f_K\left[V_{ub}V^{\ast}_{us}a_1+V_{ub}V^{\ast}_{us}
(a_4^u+a_{10}^u)+V_{cb}V^{\ast}_{cs}(a_4^c+a_{10}^c)\right.\\
&&-R_{K^-}\left.(V_{ub}V^{\ast}_{us}(a_6^u+a_{8}^u)+V_{cb}V^{\ast}_{cs}(a_6^c+a_{8}^c))\right],
\end{eqnarray*}
with $ R_{K^-}=\frac{2m^2_{K^-}}{(m_u+m_s)m_b}$.
\vspace{0.8cm}

In $\Lambda^0_b\rightarrow p K^{\ast-}$,
$$\lambda=\frac{G_F}{\sqrt{2}}f_{K^{\ast}}\left[V_{ub}V^{\ast}_{us}a_1+V_{ub}V^{\ast}_{us}
(a_4^u+a_{10}^u)+V_{cb}V^{\ast}_{cs}(a_4^c+a_{10}^c)\right].$$
\vspace{0.8cm}

In $\Lambda^0_b\rightarrow p D^-$,
$$\lambda=\frac{G_F}{\sqrt{2}}f_{D}V_{ub}V^{\ast}_{cd}a_1.$$
\vspace{0.8cm}

In $\Lambda^0_b\rightarrow p D^{\ast-}$,
$$\lambda=\frac{G_F}{\sqrt{2}}f_{D^{\ast}}V_{ub}V^{\ast}_{cd}a_1.$$
\vspace{0.8cm}

In $\Lambda^0_b\rightarrow p D_s^-$,
$$\lambda=\frac{G_F}{\sqrt{2}}f_{D_s}V_{ub}V^{\ast}_{cs}a_1.$$
\vspace{0.8cm}

In $\Lambda^0_b\rightarrow p D_s^{\ast-}$,
$$\lambda=\frac{G_F}{\sqrt{2}}f_{D_s^{\ast}}V_{ub}V^{\ast}_{cs}a_1.$$
\vspace{1.5cm}

\noindent (3)~~ $\Lambda^0_b\rightarrow \Lambda+M$ processes
\vspace{0.8cm}

In $\Lambda^0_b\rightarrow\Lambda \pi^0$,
$$\lambda=\frac{G_F}{\sqrt{2}}f^{d}_{\pi^0}\left[V_{ub}V^{\ast}_{us}(-a_2)-V_{tb}V^{\ast}_{ts}
\left(\frac{3}{2}a_7-\frac{3}{2}a_9\right)\right].$$
\vspace{0.8cm}

In $\Lambda^0_b\rightarrow\Lambda \rho^0$,
$$\lambda=\frac{G_F}{\sqrt{2}}f^{d}_{\rho^0}\left[V_{ub}V^{\ast}_{us}(-a_2)-V_{tb}V^{\ast}_{ts}
\left(-\frac{3}{2}a_7-\frac{3}{2}a_9\right)\right],$$
with $f^{d}_{\rho^0}=\frac{f_{\rho}}{\sqrt 2}$.
\vspace{0.8cm}

In $\Lambda^0_b\rightarrow\Lambda K^0$,

A term:
\begin{eqnarray*}
\lambda&=&\frac{G_F}{\sqrt{2}}f_K\left[V_{ub}V^{\ast}_{ud}\left(a_4^u-\frac{1}{2}a_{10}^u
 +R_{K^0}\left(a_6^u-\frac{1}{2}a_8^u\right)\right)\right. \\
  &&\left. +V_{cb}V^{\ast}_{cd}\left(a_4^c-\frac{1}{2}a_{10}^c
 +R_{K^0}\left(a_6^c-\frac{1}{2}a_8^c\right)\right)  \right].
\end{eqnarray*}

B term:
\begin{eqnarray*}
\lambda&=&\frac{G_F}{\sqrt{2}}f_K\left[V_{ub}V^{\ast}_{ud}\left(a_4^u-\frac{1}{2}a_{10}^u
 -R_{K^0}\left(a_6^u-\frac{1}{2}a_8^u\right)\right)\right. \\
  &&\left. +V_{cb}V^{\ast}_{cd}\left(a_4^c-\frac{1}{2}a_{10}^c
 -R_{K^0}\left(a_6^c-\frac{1}{2}a_8^c\right)\right)  \right].
\end{eqnarray*}
with $R_{K^0}=\frac{2m_{K^0}^2}{(m_s+m_d)m_b}$.

\vspace{0.8cm}

In $\Lambda^0_b\rightarrow\Lambda K^{*0}$,
\begin{eqnarray*}
\lambda=\frac{G_F}{\sqrt{2}}f_{K^*}\left[V_{ub}V^{\ast}_{ud}\left(a_4^u-\frac{1}{2}a_{10}^u\right)
 +V_{cb}V^{\ast}_{cd}\left(a_4^c-\frac{1}{2}a_{10}^c\right)\right].
\end{eqnarray*}

\vspace{0.8cm}

In $\Lambda^0_b\rightarrow\Lambda \eta$,

A term:
\begin{eqnarray*}
\lambda&=&\frac{G_F}{\sqrt{2}}f^u_{\eta}\left\{V_{ub}V^{\ast}_{us}a_2+V_{ub}V^{\ast}_{us}
\left[\left(2a_3-2a_5-\frac{1}{2}a_7+\frac{1}{2}a_9\right)+\frac{f^s_{\eta}}{f^u_{\eta}}
\left(a_3+a_4^u-a_5+\frac{1}{2}a_7\right.\right.\right.\\
&&\left.\left. -\frac{1}{2}a_9-\frac{1}{2}a_{10}^u\right)\right]  +V_{cb}V^{\ast}_{cs}\left[\left(2a_3-2a_5-\frac{1}{2}a_7+\frac{1}{2}a_9\right)+\frac{f^s_{\eta}}
{f^u_{\eta}}\left(a_3+a_4^c-a_5+\frac{1}{2}a_7\right.\right.\\
&&\left.\left.\left.-\frac{1}{2}a_9-\frac{1}{2}a_{10}^c\right)\right]\right\} +\frac{G_F}{\sqrt{2}}R_{\eta}f^u_{\eta}\left\{V_{ub}V^{\ast}_{us}
\frac{f^s_{\eta}}{f^u_{\eta}}\left(a_6^u-\frac{1}{2}a_8^u\right)+V_{cb}V^{\ast}_{cs}
\frac{f^s_{\eta}}{f^d_{\eta}}\left(a_6^c-\frac{1}{2}a_8^c\right)\right\},
\end{eqnarray*}

B term:
\begin{eqnarray*}
\lambda&=&\frac{G_F}{\sqrt{2}}f^u_{\eta}\left\{V_{ub}V^{\ast}_{us}a_2+V_{ub}V^{\ast}_{us}
\left[\left(2a_3-2a_5-\frac{1}{2}a_7+\frac{1}{2}a_9\right)+\frac{f^s_{\eta}}{f^u_{\eta}}
\left(a_3+a_4^u-a_5+\frac{1}{2}a_7\right.\right.\right.\\
&&\left.\left. -\frac{1}{2}a_9-\frac{1}{2}a_{10}^u\right)\right]  +V_{cb}V^{\ast}_{cs}\left[\left(2a_3-2a_5-\frac{1}{2}a_7+\frac{1}{2}a_9\right)+\frac{f^s_{\eta}}
{f^u_{\eta}}\left(a_3+a_4^c-a_5+\frac{1}{2}a_7\right.\right.\\
&&\left.\left.\left.-\frac{1}{2}a_9-\frac{1}{2}a_{10}^c\right)\right]\right\} -\frac{G_F}{\sqrt{2}}R_{\eta}f^u_{\eta}\left\{V_{ub}V^{\ast}_{us}
\frac{f^s_{\eta}}{f^u_{\eta}}\left(a_6^u-\frac{1}{2}a_8^u\right)+V_{cb}V^{\ast}_{cs}
\frac{f^s_{\eta}}{f^d_{\eta}}\left(a_6^c-\frac{1}{2}a_8^c\right)\right\},
\end{eqnarray*}
with  $R_{\eta}=\frac{2m^2_{\eta}}{(m_s+m_s)m_b}$.
\vspace{0.8cm}

In $\Lambda^0_b\rightarrow\Lambda \eta'$,

A term:
\begin{eqnarray*}
\lambda&=&\frac{G_F}{\sqrt{2}}f^u_{\eta'}\left\{V_{ub}V^{\ast}_{us}a_2+V_{ub}V^{\ast}_{us}
\left[\left(2a_3-2a_5-\frac{1}{2}a_7+\frac{1}{2}a_9\right)+\frac{f^s_{\eta'}}{f^u_{\eta'}}
\left(a_3+a_4^u-a_5+\frac{1}{2}a_7\right.\right.\right.\\
&&\left.\left. -\frac{1}{2}a_9-\frac{1}{2}a_{10}^u\right)\right]  +V_{cb}V^{\ast}_{cs}\left[\left(2a_3-2a_5-\frac{1}{2}a_7+\frac{1}{2}a_9\right)+\frac{f^s_{\eta'}}
{f^u_{\eta'}}\left(a_3+a_4^c-a_5+\frac{1}{2}a_7\right.\right.\\
&&\left.\left.\left.-\frac{1}{2}a_9-\frac{1}{2}a_{10}^c\right)\right]\right\} +\frac{G_F}{\sqrt{2}}R_{\eta'}f^u_{\eta'}\left\{V_{ub}V^{\ast}_{us}
\frac{f^s_{\eta'}}{f^u_{\eta'}}\left(a_6^u-\frac{1}{2}a_8^u\right)+V_{cb}V^{\ast}_{cs}
\frac{f^s_{\eta'}}{f^d_{\eta'}}\left(a_6^c-\frac{1}{2}a_8^c\right)\right\},
\end{eqnarray*}

B term:
\begin{eqnarray*}
\lambda&=&\frac{G_F}{\sqrt{2}}f^u_{\eta'}\left\{V_{ub}V^{\ast}_{us}a_2+V_{ub}V^{\ast}_{us}
\left[\left(2a_3-2a_5-\frac{1}{2}a_7+\frac{1}{2}a_9\right)+\frac{f^s_{\eta'}}{f^u_{\eta'}}
\left(a_3+a_4^u-a_5+\frac{1}{2}a_7\right.\right.\right.\\
&&\left.\left. -\frac{1}{2}a_9-\frac{1}{2}a_{10}^u\right)\right]  +V_{cb}V^{\ast}_{cs}\left[\left(2a_3-2a_5-\frac{1}{2}a_7+\frac{1}{2}a_9\right)+\frac{f^s_{\eta'}}
{f^u_{\eta'}}\left(a_3+a_4^c-a_5+\frac{1}{2}a_7\right.\right.\\
&&\left.\left.\left.-\frac{1}{2}a_9-\frac{1}{2}a_{10}^c\right)\right]\right\} -\frac{G_F}{\sqrt{2}}R_{\eta'}f^u_{\eta'}\left\{V_{ub}V^{\ast}_{us}
\frac{f^s_{\eta'}}{f^u_{\eta'}}\left(a_6^u-\frac{1}{2}a_8^u\right)+V_{cb}V^{\ast}_{cs}
\frac{f^s_{\eta'}}{f^d_{\eta'}}\left(a_6^c-\frac{1}{2}a_8^c\right)\right\},
\end{eqnarray*}
with  $ R_{\eta'}=\frac{2m^2_{\eta'}}{(m_s+m_s)m_b}$.
\vspace{0.8cm}

In $\Lambda^0_b\rightarrow\Lambda \eta_c$,
$$\lambda=\frac{G_F}{\sqrt{2}}f_{\eta_c}\left[V_{cb}V^{\ast}_{cs}a_2-V_{tb}V^{\ast}_{ts}
(a_3-a_5-a_7+a_9)\right].$$
\vspace{0.8cm}

In $\Lambda^0_b\rightarrow\Lambda J/\psi$,
$$\lambda=\frac{G_F}{\sqrt{2}}f_{J/\psi}\left[V_{cb}V^{\ast}_{cs}a_2-V_{tb}V^{\ast}_{ts}
(a_3+a_5+a_7+a_9)\right].$$
\vspace{0.8cm}

In $\Lambda^0_b\rightarrow\Lambda \omega$,
$$\lambda=\frac{G_F}{\sqrt{2}}f^d_{\omega}\left[V_{ub}V^{\ast}_{us}a_2-V_{tb}V^{\ast}_{ts}
\left(2a_3+2a_5+\frac{1}{2}a_7+\frac{1}{2}a_9\right)\right],$$
with $f^d_{\omega}=\frac{f_{\omega}}{\sqrt 2}$.
\vspace{0.8cm}

In $\Lambda^0_b\rightarrow\Lambda \phi$,
\begin{eqnarray*}
\lambda&=&\frac{G_F}{\sqrt{2}}f_{\phi}\left[V_{ub}V^{\ast}_{us}\left(a_3+a^u_4+a_5-\frac{1}{2}a_7
 -\frac{1}{2}a_9-\frac{1}{2}a^u_{10}\right)\right.\\
 &&+\left.V_{cb}V^{\ast}_{cs}\left(a_3+a^c_4+a_5-\frac{1}{2}a_7-\frac{1}{2}a_9
 -\frac{1}{2}a^c_{10}\right)\right].
\end{eqnarray*}
\vspace{0.8cm}

In $\Lambda^0_b\rightarrow\Lambda D^0$,
$$\lambda=\frac{G_F}{\sqrt{2}}f_{D}V_{cb}V^{\ast}_{us}a_2.$$
\vspace{0.8cm}

In $\Lambda^0_b\rightarrow\Lambda D^{\ast 0}$,
$$\lambda=\frac{G_F}{\sqrt{2}}f_{D^{\ast}}V_{cb}V^{\ast}_{us}a_2.$$
\vspace{0.8cm}

In $\Lambda^0_b\rightarrow\Lambda \bar{D}^0$,
$$\lambda=\frac{G_F}{\sqrt{2}}f_{D}V_{ub}V^{\ast}_{cs}a_2.$$
\vspace{0.8cm}

In $\Lambda^0_b\rightarrow\Lambda \bar{D}^{\ast 0}$,
$$\lambda=\frac{G_F}{\sqrt{2}}f_{D^{\ast}}V_{ub}V^{\ast}_{cs}a_2.$$

\vspace{1.5cm}

\noindent (4)~~ $\Lambda^0_b\rightarrow n+M$ processes
\vspace{1.0cm}

In $\Lambda^0_b\rightarrow n \pi^0$,

A term:
\begin{eqnarray*}
\lambda&=&\frac{G_F}{\sqrt{2}}f^d_{\pi^0}\left\{V_{ub}V^{\ast}_{ud}(-a_2)+V_{ub}V^{\ast}_{ud}
 \left(a_4^u+\frac{3}{2}a_7-\frac{3}{2}a_9-\frac{1}{2}a^u_{10}\right)
 +V_{cb}V^{\ast}_{cd}\left(a_4^c+\frac{3}{2}a_7\right.\right. \\
 &&\left.\left. -\frac{3}{2}a_9-\frac{1}{2}a^c_{10}\right)
 +R_{\pi^0}\left[V_{ub}V^{\ast}_{ ud}\left(a^u_6-\frac{1}{2}a^u_8\right)
 +V_{cb}V^{\ast}_{cd}\left(a^c_6-\frac{1}{2}a^c_8\right)\right]\right\},
\end{eqnarray*}

B term:
\begin{eqnarray*}
\lambda&=&\frac{G_F}{\sqrt{2}}f^d_{\pi^0}\left\{V_{ub}V^{\ast}_{ud}(-a_2)+V_{ub}V^{\ast}_{ud}
 \left(a_4^u+\frac{3}{2}a_7-\frac{3}{2}a_9-\frac{1}{2}a^u_{10}\right)
 +V_{cb}V^{\ast}_{cd}\left(a_4^c+\frac{3}{2}a_7\right.\right. \\
 &&\left.\left. -\frac{3}{2}a_9-\frac{1}{2}a^c_{10}\right)
 -R_{\pi^0}\left[V_{ub}V^{\ast}_{ ud}\left(a^u_6-\frac{1}{2}a^u_8\right)
 +V_{cb}V^{\ast}_{cd}\left(a^c_6-\frac{1}{2}a^c_8\right)\right]\right\},
\end{eqnarray*}
with $f^d_{\pi^0}=\frac{f_{\pi}}{\sqrt 2}$ and $R_{\pi^0}=\frac{2m^2_{\pi^0}}{(m_d+m_d)m_b}$.
\vspace{0.8cm}

In $\Lambda^0_b\rightarrow n \rho^0$,
\begin{eqnarray*}
\lambda&=&\frac{G_F}{\sqrt{2}}f^d_{\rho^0}\left[V_{ub}V^{\ast}_{ud}(-a_2)+V_{ub}V^{\ast}_{ud}
 \left(a_4^u-\frac{3}{2}a_7-\frac{3}{2}a_9-\frac{1}{2}a^u_{10}\right)\right.\\
 &&\left.+V_{cb}V^{\ast}_{cd}\left(a_4^c-\frac{3}{2}a_7-\frac{3}{2}a_9-\frac{1}{2}a^c_{10}\right)\right].
\end{eqnarray*}
\vspace{0.8cm}

In $\Lambda^0_b\rightarrow n \bar K^0$,

A term:
\begin{eqnarray*}
\lambda&=&\frac{G_F}{\sqrt{2}}f_K\left[V_{ub}V^{\ast}_{us}\left(a_4^u-\frac{1}{2}a_{10}^u
 +R_{K^0}\left(a_6^u-\frac{1}{2}a_8^u\right)\right)\right. \\
  &&\left. +V_{cb}V^{\ast}_{cs}\left(a_4^c-\frac{1}{2}a_{10}^c
 +R_{K^0}\left(a_6^c-\frac{1}{2}a_8^c\right)\right)  \right].
\end{eqnarray*}

B term:
\begin{eqnarray*}
\lambda&=&\frac{G_F}{\sqrt{2}}f_K\left[V_{ub}V^{\ast}_{us}\left(a_4^u-\frac{1}{2}a_{10}^u
 -R_{K^0}\left(a_6^u-\frac{1}{2}a_8^u\right)\right)\right. \\
  &&\left. +V_{cb}V^{\ast}_{cs}\left(a_4^c-\frac{1}{2}a_{10}^c
 -R_{K^0}\left(a_6^c-\frac{1}{2}a_8^c\right)\right)  \right].
\end{eqnarray*}
with $R_{K^0}=\frac{2m_{K^0}^2}{(m_s+m_d)m_b}$.

\vspace{0.8cm}

In $\Lambda^0_b\rightarrow n \bar K^{*0}$,
\begin{eqnarray*}
\lambda=\frac{G_F}{\sqrt{2}}f_{K^*}\left[V_{ub}V^{\ast}_{us}\left(a_4^u-\frac{1}{2}a_{10}^u\right)
 +V_{cb}V^{\ast}_{cs}\left(a_4^c-\frac{1}{2}a_{10}^c\right)\right].
\end{eqnarray*}

\vspace{0.8cm}

In $\Lambda^0_b\rightarrow n \eta$,

A term:
\begin{eqnarray*}
\lambda&=&\frac{G_F}{\sqrt{2}}f^u_{\eta}\left\{V_{ub}V^{\ast}_{ud}a_2+V_{ub}V^{\ast}_{ud}
 \left[\left(2a_3+a_4^u-2a_5-\frac{1}{2}a_7+\frac{1}{2}a_9-\frac{1}{2}a^u_{10}\right)\right.\right.\\
 &&\left.\left.+\frac{f^s_{\eta}}{f^u_{\eta}}\left(a_3-a_5+\frac{1}{2}a_7-\frac{1}{2}a_9\right)\right]  +V_{cb}V^{\ast}_{cd}\left[\left(2a_3+a_4^c-2a_5-\frac{1}{2}a_7 \right.\right.\right.\\
 &&\left.\left.\left.+\frac{1}{2}a_9-\frac{1}{2}a^c_{10}\right)
 +\frac{f^s_{\eta}}{f^u_{\eta}}\left(a_3-a_5+\frac{1}{2}a_7
 -\frac{1}{2}a_9\right)\right]\right\}  \\
 &&+\frac{G_F}{\sqrt{2}}R_{\eta}f^u_{\eta}\left(1-\frac{f^u_{\eta}}{f^s_{\eta}}\right)
 \left\{V_{ub}V^{\ast}_{ud}\left(a_6^u-\frac{1}{2}a_8^u\right)
 +V_{cb}V^{\ast}_{cd}\left(a_6^c-\frac{1}{2}a_8^c\right)\right\},
\end{eqnarray*}

B term:
\begin{eqnarray*}
\lambda&=&\frac{G_F}{\sqrt{2}}f^u_{\eta}\left\{V_{ub}V^{\ast}_{ud}a_2+V_{ub}V^{\ast}_{ud}
 \left[\left(2a_3+a_4^u-2a_5-\frac{1}{2}a_7+\frac{1}{2}a_9-\frac{1}{2}a^u_{10}\right)\right.\right.\\
 &&\left.\left.+\frac{f^s_{\eta}}{f^u_{\eta}}\left(a_3-a_5+\frac{1}{2}a_7-\frac{1}{2}a_9\right)\right]  +V_{cb}V^{\ast}_{cd}\left[\left(2a_3+a_4^c-2a_5-\frac{1}{2}a_7 \right.\right.\right.\\
 &&\left.\left.\left.+\frac{1}{2}a_9-\frac{1}{2}a^c_{10}\right)+\frac{f^s_{\eta}}
 {f^u_{\eta}}\left(a_3-a_5+\frac{1}{2}a_7-\frac{1}{2}a_9\right)\right]\right\}  \\
 &&-\frac{G_F}{\sqrt{2}}R_{\eta}f^u_{\eta}\left(1-\frac{f^u_{\eta}}{f^s_{\eta}}\right)
 \left\{V_{ub}V^{\ast}_{ud}\left(a_6^u-\frac{1}{2}a_8^u\right)
 +V_{cb}V^{\ast}_{cd}\left(a_6^c-\frac{1}{2}a_8^c\right)\right\},
\end{eqnarray*}
Here, we adopt a treatment for $\eta(\eta')$ matrix elements from \cite{Ali:1998eb}.

\vspace{0.8cm}

In $\Lambda^0_b\rightarrow n \eta'$,

A term:
\begin{eqnarray*}
\lambda&=&\frac{G_F}{\sqrt{2}}f^u_{\eta'}\left\{V_{ub}V^{\ast}_{ud}a_2+V_{ub}V^{\ast}_{ud}
 \left[\left(2a_3+a_4^u-2a_5-\frac{1}{2}a_7+\frac{1}{2}a_9-\frac{1}{2}a^u_{10}\right)\right.\right.\\
 &&\left.\left.+\frac{f^s_{\eta'}}{f^u_{\eta'}}\left(a_3-a_5+\frac{1}{2}a_7-\frac{1}{2}a_9\right)\right]  +V_{cb}V^{\ast}_{cd}\left[\left(2a_3+a_4^c-2a_5-\frac{1}{2}a_7 \right.\right.\right.\\
 &&\left.\left.\left.+\frac{1}{2}a_9-\frac{1}{2}a^c_{10}\right)
 +\frac{f^s_{\eta'}}{f^u_{\eta'}}\left(a_3-a_5+\frac{1}{2}a_7
 -\frac{1}{2}a_9\right)\right]\right\}  \\
 &&+\frac{G_F}{\sqrt{2}}R_{\eta'}f^u_{\eta'}\left(1-\frac{f^u_{\eta'}}{f^s_{\eta'}}\right)
 \left\{V_{ub}V^{\ast}_{ud}\left(a_6^u-\frac{1}{2}a_8^u\right)
 +V_{cb}V^{\ast}_{cd}\left(a_6^c-\frac{1}{2}a_8^c\right)\right\},
\end{eqnarray*}

B term:
\begin{eqnarray*}
\lambda&=&\frac{G_F}{\sqrt{2}}f^u_{\eta'}\left\{V_{ub}V^{\ast}_{ud}a_2+V_{ub}V^{\ast}_{ud}
 \left[\left(2a_3+a_4^u-2a_5-\frac{1}{2}a_7+\frac{1}{2}a_9-\frac{1}{2}a^u_{10}\right)\right.\right.\\
 &&\left.\left.+\frac{f^s_{\eta'}}{f^u_{\eta'}}\left(a_3-a_5+\frac{1}{2}a_7-\frac{1}{2}a_9\right)\right]  +V_{cb}V^{\ast}_{cd}\left[\left(2a_3+a_4^c-2a_5-\frac{1}{2}a_7 \right.\right.\right.\\
 &&\left.\left.\left.+\frac{1}{2}a_9-\frac{1}{2}a^c_{10}\right)
 +\frac{f^s_{\eta'}}{f^u_{\eta'}}\left(a_3-a_5+\frac{1}{2}a_7
 -\frac{1}{2}a_9\right)\right]\right\}  \\
 &&-\frac{G_F}{\sqrt{2}}R_{\eta'}f^u_{\eta'}\left(1-\frac{f^u_{\eta'}}{f^s_{\eta'}}\right)
 \left\{V_{ub}V^{\ast}_{ud}\left(a_6^u-\frac{1}{2}a_8^u\right)
 +V_{cb}V^{\ast}_{cd}\left(a_6^c-\frac{1}{2}a_8^c\right)\right\}.
\end{eqnarray*}

\vspace{0.8cm}

In $\Lambda^0_b\rightarrow n \eta_c$,
$$\lambda=\frac{G_F}{\sqrt{2}}f_{\eta_c}\left[V_{cb}V^{\ast}_{cd}a_2-V_{tb}V^{\ast}_{td}
(a_3-a_5-a_7+a_9)\right].$$
\vspace{0.8cm}

In $\Lambda^0_b\rightarrow n J/\psi$,
$$\lambda=\frac{G_F}{\sqrt{2}}f_{J/\psi}\left[V_{cb}V^{\ast}_{cd}a_2-V_{tb}V^{\ast}_{td}
(a_3+a_5+a_7+a_9)\right].$$
\vspace{0.8cm}

In $\Lambda^0_b\rightarrow n \omega$,
\begin{eqnarray*}
 \lambda&=&\frac{G_F}{\sqrt{2}}f^u_{\omega}\left[V_{ub}V^{\ast}_{ud}a_2+V_{ub}V^{\ast}_{ud}
 \left(2a_3+a^u_4+2a_5+\frac{1}{2}a_7+\frac{1}{2}a_9-\frac{1}{2}a^u_{10}\right)\right.\\
 &&\left.+V_{cb}V^{\ast}_{cd}\left(2a_3+a^c_4+2a_5+\frac{1}{2}a_7
 +\frac{1}{2}a_9-\frac{1}{2}a^c_{10}\right)\right].
\end{eqnarray*}
\vspace{0.8cm}

In $\Lambda^0_b\rightarrow n \phi$,
$$\lambda=\frac{G_F}{\sqrt{2}}f_{\phi}\left[-V_{tb}V^{\ast}_{td}\left(a_3+a_5-\frac{1}{2}
a_7-\frac{1}{2}a_9\right)\right].$$
\vspace{0.8cm}

In $\Lambda^0_b\rightarrow n \bar{D}^0$,
$$\lambda=\frac{G_F}{\sqrt{2}}f_{D}V_{ub}V^{\ast}_{cd}a_2.$$
\vspace{0.8cm}

In $\Lambda^0_b\rightarrow n \bar{D}^{\ast 0}$,
$$\lambda=\frac{G_F}{\sqrt{2}}f_{D^{\ast}}V_{ub}V^{\ast}_{cd}a_2.$$
\vspace{0.8cm}

In $\Lambda^0_b\rightarrow n D^0$,
$$\lambda=\frac{G_F}{\sqrt{2}}f_{D}V_{cb}V^{\ast}_{ud}a_2.$$
\vspace{0.8cm}

In $\Lambda^0_b\rightarrow n D^{\ast 0}$
$$\lambda=\frac{G_F}{\sqrt{2}}f_{D^{\ast}}V_{cb}V^{\ast}_{ud}a_2.$$
\vspace{1.5cm}

\section*{Acknowledgement}

This work is supported by the National Natural Science Foundation
of China (NNSFC) under the contract Nos. 11175091, 11375128.

\end{document}